\documentclass[11pt, a4paper, pdftex]{article}
\usepackage[utf8]{inputenc}		
\usepackage[T1]{fontenc}

\usepackage{amsmath}
\usepackage{amssymb}
\usepackage{setspace}
\usepackage{latexsym}
\usepackage{fancyhdr}
\usepackage{lastpage}
\usepackage{times}

\usepackage[pdftex]{graphicx}

\usepackage[english]{babel}

\usepackage{authblk}

\graphicspath{ {les_figures/}}

\newcommand{\pa}{\paragraph{}}

\def\og{\leavevmode\raise.3ex\hbox{$\scriptscriptstyle\langle\!\langle$~}}
\def\fg{\leavevmode\raise.3ex\hbox{$\scriptscriptstyle\rangle\!\rangle$}}

\begin{document}

\title{Revisiting proximity effect using broadband signals}
\author[1, 2]{Laurent Millot \thanks{Electronic address: l.millot@ens-louis-lumiere.fr; corresponding author}}
\author[2]{Mohammed Elliq}
\author[2]{Manuel Lopes}
\author[1, 2]{Gérard Pelé}
\author[1, 2]{Dominique Lambert}
\affil[1]{Institut d'esthétique, des arts et technologies - UMR 8153 (CNRS/Université Paris 1/MENRT), 
Université Paris 1, 47 rue des bergers, Paris, 75015, France}
\affil[2]{ENS Louis-Lumière, Noisy-le-Grand, 7 allée du promontoire, 93160, France}

\date{01/07/2007}

\maketitle

	\begin{abstract}
Experiments studying mainly proximity effect are presented. Pink noise and music were used as stimuli and a combo guitar amplifier as source to test several microphones: omnidirectional and directional. We plot in-axis levels and spectral balances as functions of x, the distance to the source. Proximity effect was found for omnidirectional microphones. In-axis level curves show that 1/x law seems poorly valid. Spectral balance evolutions depend on microphones and moreover on stimuli: bigger decreases of low frequencies with pink noise; larger increases of other frequencies with music. For a naked loudspeaker, we found similar in-axis level curves under and above the cut-off frequency and propose an explanation. Listening equalized music recordings will help to demonstrate proximity effect for tested microphones.

Paper 7106 presented at the 122th Convention of the Audio Engineering Society, Wien, 2007
	\end{abstract}
	

\section{Introduction}
\pa Classical description of microphone behavior even to build multichannel recording setups, using signal processing \cite{Remy:04} or not \cite{Williams:04}, is given by a quite simple description based on the directivity function $D(\theta)$ given by $D(\theta)=m+(1-m).\cos(\theta)$. And this directivity function can also be used to design real-time recording digital simulators \cite{Braasch:05}.

\pa Some works \cite{Torio:98} investigate the inner description of microphone, introducing electroacoustical analogies and description of the transduction process,  but still rely on the same assumption for the modeling of the excitation of the back or side excitation of the microphone: a version of the original signal modulated by a $d.\cos(\theta)$ factor where $\theta$ is the incidence angle and $d$ the apparent path length difference.

\pa In textbooks for sound engineering students dealing with recording setups, one may encounter also the same simple model used to explain which would be the theoretical result of the recording using a given microphone, a given stereophonic recording setup or even a multichannel one.

\pa In a previous paper \cite{Millot:06}, we show all the restrictive assumptions needed to find again this model, notably only valid for a monochromatic excitation. So we propose a broadband physical model and its digi\-tal version  constituting an extension of the classical directivity-based model and also demonstrating that such models assume that there is no microphone at the recording location. Indeed, these models only use a complex mixing of the pressure fields for three close points and the physical presence of the microphone is even a problem for the model!

\pa But, we also pointed out the fact that the proximity effect, relying on the spherical waves assumption, is clearly not valid and that we do not have sufficient experimental knowledge of the physical phenomena in the proximity of the source to be able to derive a satisfying model.

\pa In this paper, we review some of the experimental results we collected using broadband stimuli (pink noise as "classical" reference and a piece of music) and 4 different microphones giving access to 6 directivities. The results given in this paper illustrate the gap existing between the $1/x$ law assumption and the measurements, varying with the microphone nature, and also give some information about the mean distance over which one may be able to consider a wave model for pressure field. We plotted the variation of the mean in-axis level and of the global spectral balance according to the distance from the source.

\pa To try to access some more information about the proximity effect, we also consider the "simpler" case of the behavior of a naked loudspeaker under and above its cut-off frequency to review the assumption of destructive interferences between back and front waves as suggested in literature. Given the experimental results, we propose an alternative explanation for this phenomenon which rejects the wave assumption.      

\pa Finally, considering all the collected information, we discuss how to enhance the knowledge about the proximity effect, we propose some clues for other measurements campaigns and, above all, a competitive physical model for proximity phenomena excluding waves: the acoustical flows paradigm.

\section{Experimental setups}
\pa All experiments have been done in a small anechoic chamber using broadband signals, pink noise or a piece of jazz music (30~s of "One", Ahmad Jamal, \textit{Digital Works}, Atlantic, 1985), to test the behavior of different microphones using a guitar combo loudspeaker (Roland Cube 15) as source or a naked loudspeaker (Celestion G12M Greenback). We also recorded pure sinusoids to test the short-circuit phenomenon in the case of the naked loudspeaker. 

\begin{figure}[h!t]
\begin{center}
\includegraphics[width=10cm]{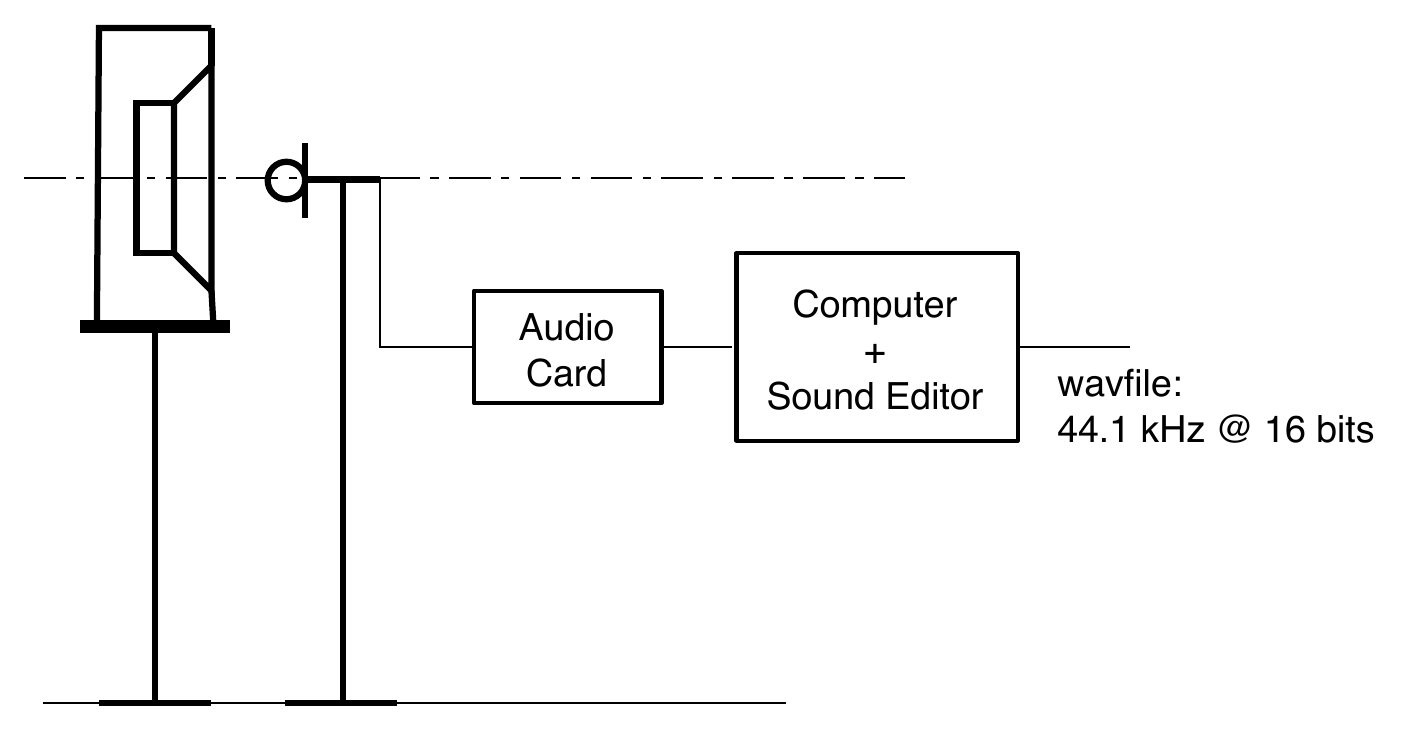}
\end{center}
\caption[1]{Experimental setup to test different microphones.}
\label{fig:Man1}
\end{figure}

\pa Figure \ref{fig:Man1} is a scheme of the first experimental device: a loudspeaker as a source, a given microphone put in the axis of the source with distance varying from the closest possible one from the source to one meter, a CD player with an amplifier to diffuse the stimuli, an USB external audio card (Lexicon Omega), a laptop computer (Apple PowerBook) using a sound editor (Audacity 1.2.3) to record each measurement. 

\pa Four microphones have been tested with both pink noise and musical stimuli: Audio-Technica AT2020 cardioid microphone (large diaphragm);  Behringer C-2 cardioid microphone (small diaphragm); ECM8000 Behringer omnidirectional microphone (small diaphragm); Neumann U89i microphone (large diaphragm) with bidirectional, cardioid and omnidirectional directivities.

\pa Maximal excitations levels were important, from 100 to 122~dB SPL just  in front off the grid of the combo loudspeaker, to be able to get some usable recordings at a distance of 1~m from the source.

\pa All broadband measurements were analyzed using the IDS analyzer \cite{Millot:04} which provides 10 listenable subband signals, the mean level in dB FS for the stimulus and also its mean IDS balance: the relative global weight of each subband for a given measurement.

\begin{figure}[h!t]
\begin{center}
\includegraphics[width=10cm]{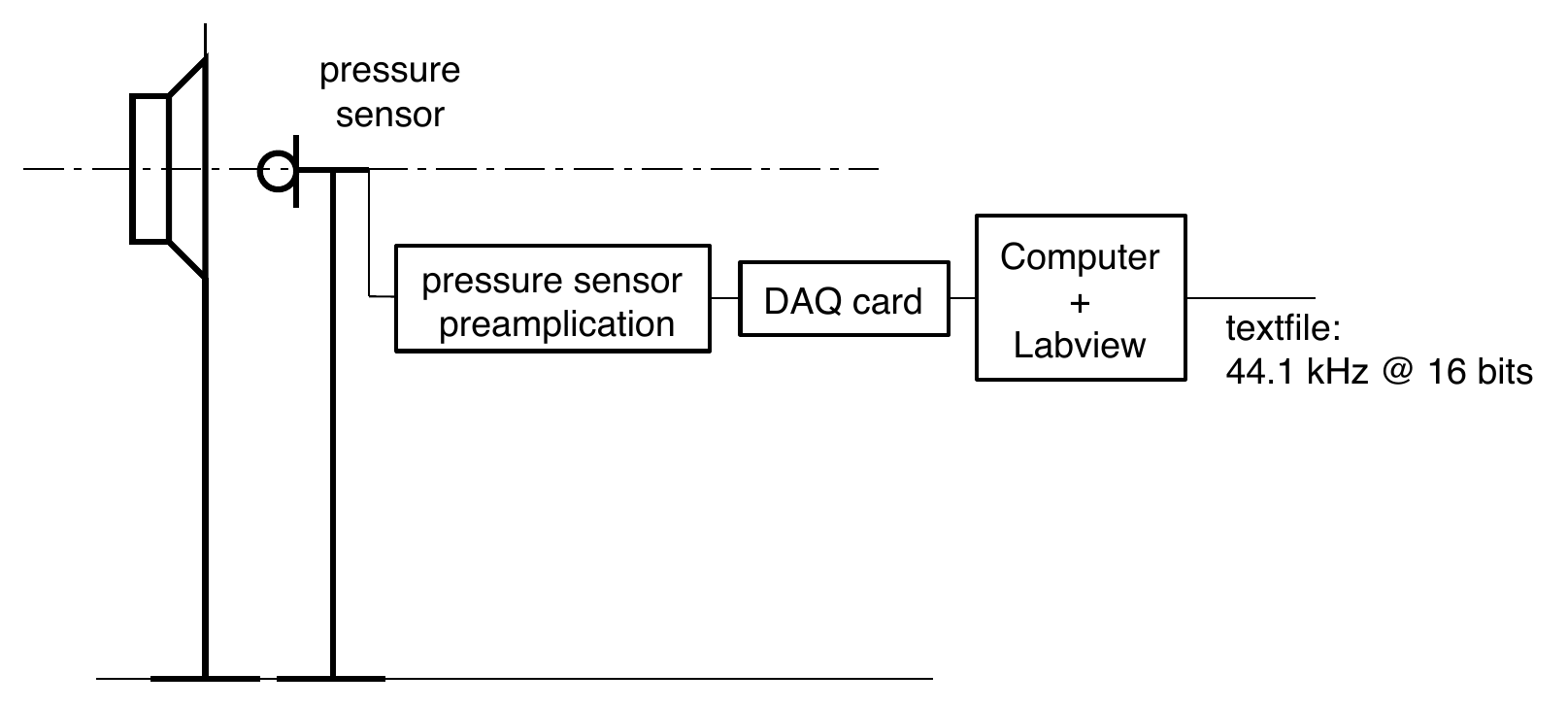}
\end{center}
\caption[2]{Experimental setup to study the short-circuit naked loudspeaker phenomenon.}
\label{fig:Man2}
\end{figure}

\pa To study the short-circuit phenomenon for a naked loudspeaker, that is to say the theoretical difference of behavior under and over the cut-off frequency, we have used the second experimental setup given in figure \ref{fig:Man2}.

\pa Compared to the first experimental device, we have only changed the following elements for the se\-cond one: instead of a microphone we used a pressure sensor (Endevco 8507C-2) with its 136 signal conditioner; a DAQ card (PCI-6036E from National Instrument) with its BNC interface (BNC 2110) ; a PC desktop computer using LabView software (version 7.1). We recorded the signals according to the distance from the source for two sinusoids: one under the cut-off frequency; the other above it. We also use both broadband stimuli but the results contain some noise because our pressure sensor has a too tiny sensitivity as it is designed for huge pressure levels (up to 178~dB SPL) while the encountered levels where quickly smaller than 100~dB SPL even with a 120 or 130~dB SPL mean level measured just in front of the loudspeaker.

\pa All level indications were measured using a sonometer model SP-dBMeter (Sphynx company).

\section{Influence of microphone mature}
\subsection{Mean level vs distance to the source}
\pa The mean level at a distance of 1~m from the sound source is taken as 0~dB reference so, for all the in-axis level curves, a reduction of the distance from the source is associated with an amplification of the mean level. In the following, we note $x$ the distance from the microphone to the sound source.

\pa We have chosen the longest distance from the microphone to the source (1~m) as reference because waves derivation assumes that we are far away from any source. So, while the wave assumption is valid, the difference between the theoretical $1/x$ spherical wave law for the level and the measurements would be small. But, as soon as this gap becomes significant, we can assume that the phenomenon can not be approximated any more by a wave model and that another model or explanation is then needed.

\pa In Fig. 3 to 8, we have systematically plotted the evolution of the mean level amplification for each microphone, or for each directivity in the case of the Behringer ECM8000 microphone, for pink noise and musical stimuli. The pink noise measurements are presented to build a bridge with the  measurements one may find in the literature, even if they are often done for monochromatic excitation rather than pink noise: pink noise is often one of the stimuli used to study the response of loudspeakers so we thought it would be interesting to use pink noise as "classical" stimulus. But, we are much more interested by the results collected with music stimulus because they would be much more representative of usual recording situations, even if all experiments were performed in an anechoic chamber. Moreover, using music as stimulus, we can access the transitory and dynamic response of the microphone for example when listening or studying associated recording (or IDS subbands decomposition). 

\pa The comparison of the mean levels curves for pink noise and music are, by the way, quite interesting because in most of the cases both curves are significantly different. And one may be able to note that from microphone to the other the curve can also vary significantly.

\begin{figure}[h!t]
\begin{center}
\includegraphics[width=7.5cm]{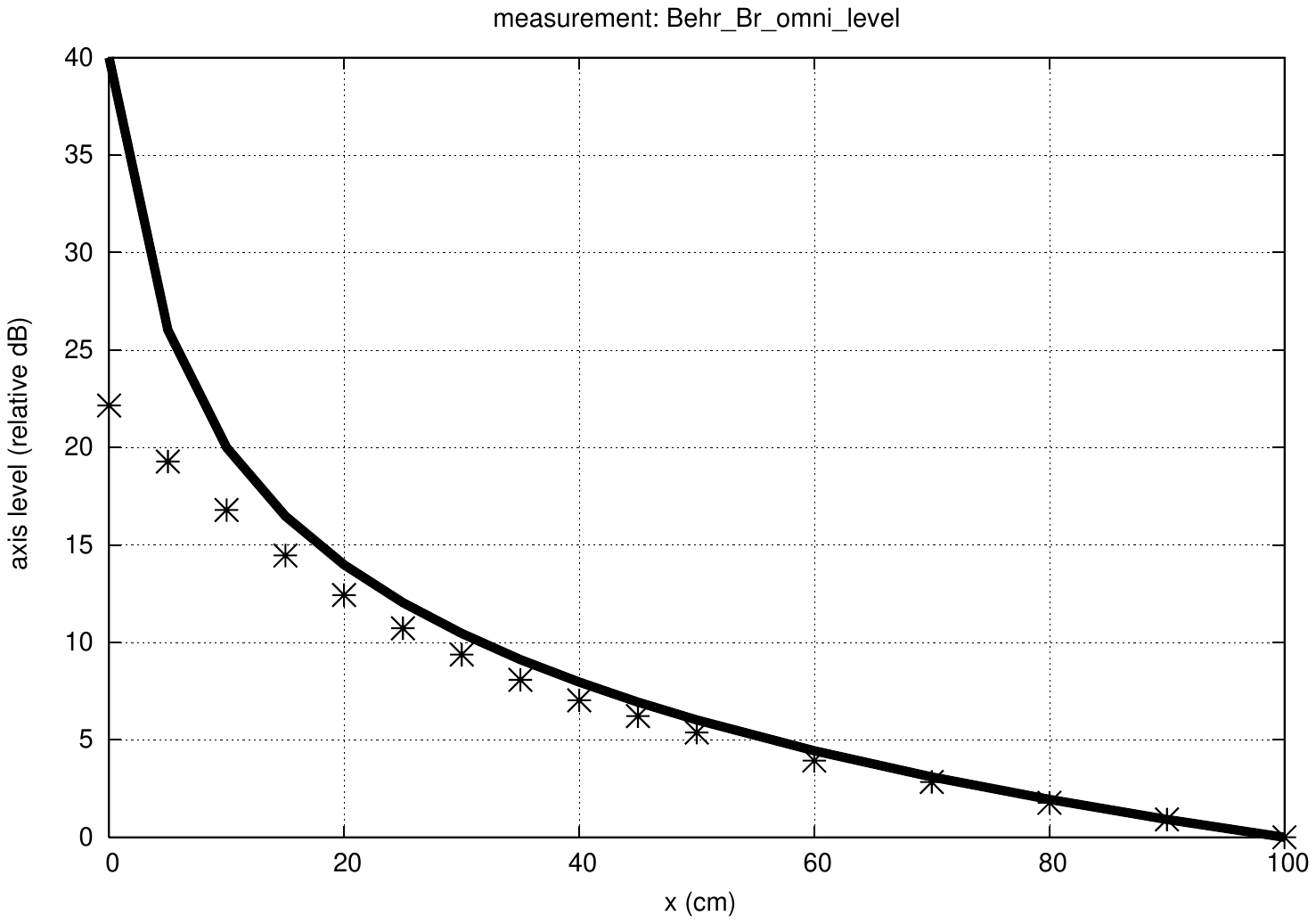}
\includegraphics[width=7.5cm]{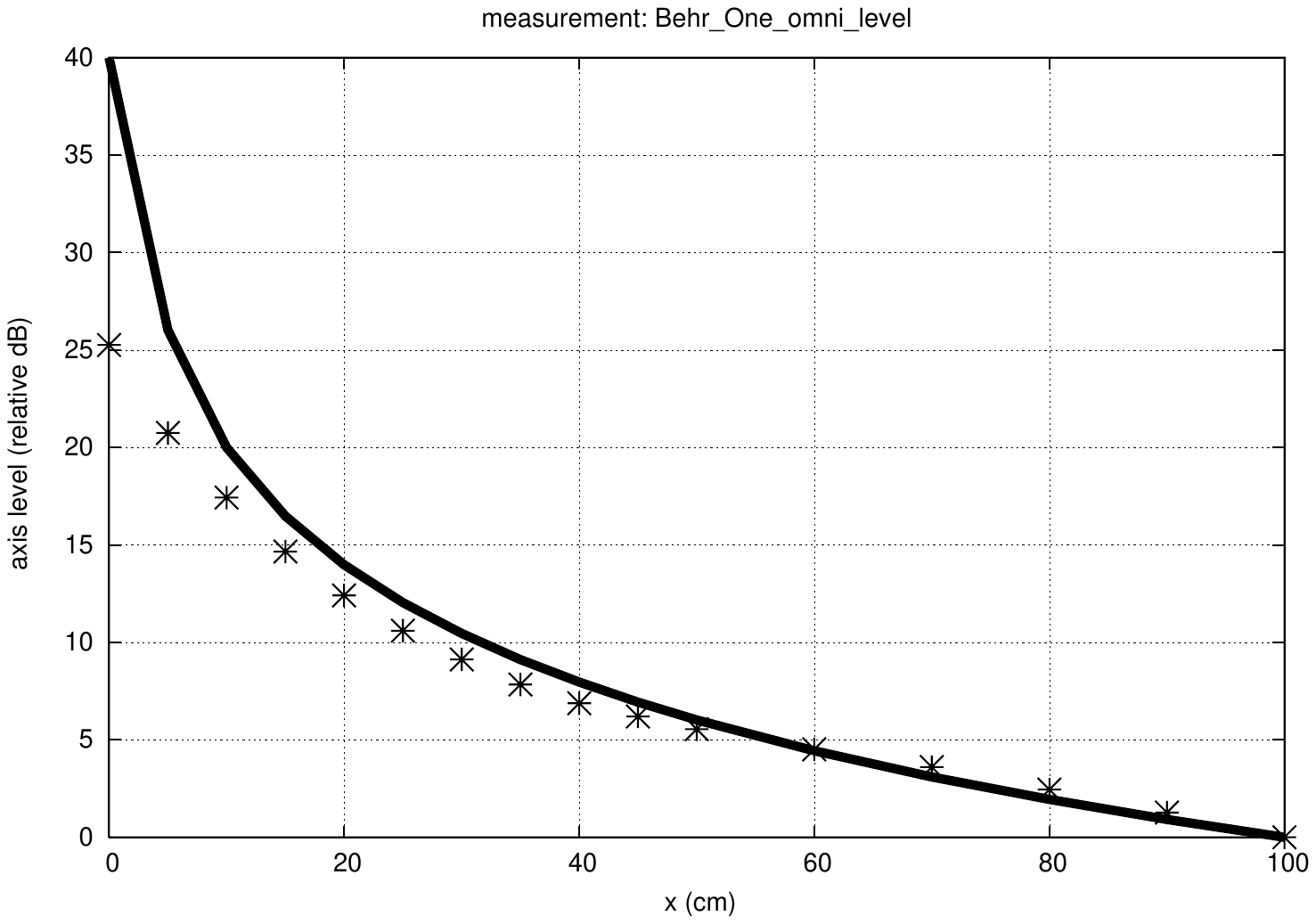}
\end{center}
\caption[3]{Mean level as a function of $x$ for a Behringer  ECM800 microphone with omnidirectional directivity: experimental points (crosses); theoretical amplification (solid line). Excitations:  pink noise (left) and music (right).}
\end{figure}

\pa Pink noise (116/117~dB SPL at $x=0$~cm) and music (109-116 dB SPL, mean level of 113/114 dB SPL at $x=0$~cm) curves are similar for the Behringer ECM8000 omnidirectional microphone (see Fig. 3) which is presented and used as a pressure sensor (with a small diaphragm) rather than a microphone: it is said to be designed for acoustical measurements and so would be as flat and neutral as possible. So, if we assume that this microphone is a pressure sensor, the mean level curves associated with this microphone would give the information about the "real" acoustical pressure field. And, the similarity of mean level amplification curves for pink noise and music stimuli gives some validity to this assumption.

\pa Considering the Behringer ECM8000 microphone as a pressure sensor, we can note that the gap between the $1/x$ wave law and the measurements becomes significant when $x$ is lower than 50~cm. So, waves assumptions would be reconsidered for distances lower than 50~cm, because we must remind the fact that the mean level at the source was important and we will suggest an alternative explanation for acoustical phenomenon,  at least valid in the proximity of the source, for which, the minimum distance over which (spherical) wave assumption is valid would vary with the level: it may increase with the level.

\pa At the source, the measured mean level amplification is from 18~dB (pink noise) to 15~dB (music) lower than the one predicted by the $1/x$ theoretical wave law which is quite significant. And from $x=0$~cm to the \textit{a priori} "wave validity limit" of $x=50$~cm, the mean level amplification decrease is equal to 20~dB for music and 17~dB for pink noise. It is also inte\-resting to note that the measured mean level amplification is always lower than the one predicted by the $1/x$ law for $x\leq 50$~cm: so the decrease of the mean level when $x$ increases is quite important and not explained by the wave assumption. This means that we have another physical phenomenon at least under the "wave validity limit" which seems to be well approximated by spherical waves after this "wave validity limit". But it does not mean that physical phenomenon, after "wave validity limit", is a wave: pressure mimics spherical wave as soon as the distance from the source is  sufficient; waves are just a fair approximation far away from the source...
  
\pa Fig. 4 to 6 correspond to the measurements made with the Neumann U89i microphone with three directivities: omnidirectional (figure 4), cardioid (figure 5) and bidirectional (figure 6). As all measurements for a given distance $x$ were performed only by changing the directivity, the excitation levels at $x=0$~cm were the same for each stimulus: 102~dB SPL for pink noise and 90 to 100~dB SPL for music with a mean level of 94/95~dB SPL.  

\pa This microphone has a large "dual-diaphragm" and, as its directivity can be changed, we can not consider it as a pressure sensor which explains the fact that the mean level curves found for this microphone are rather different from the ones found for the ECM8000 microphone. 

\pa Due to the size of its suspension, it was not possible to put it at a distance to the loudspeaker lower than 5~cm but we can point out that the gap between theory and measurements is around  6~dB  at distance $x=5$~cm for the ECM800 microphone,  assumed to give  a realistic pressure field.
    
\pa We can note that this gap varies according to the nature of the stimulus and according to the directivity, even in the case of the omnidirectional directivity. And the theoretical waves curves and the experimental points differ significantly according to the directivity.

\pa It is also interesting to underline the fact that these differences are bigger for music stimulus for cardioid and bidirectional directivity, and also bigger for the omnidirectional directivity for distances $x$ in the range 50 to 100~cm, that is to say the \textit{a priori} range of validity for the spherical wave assumption.
    
\begin{figure}[h!t]
\begin{center}
\includegraphics[width=7.5cm]{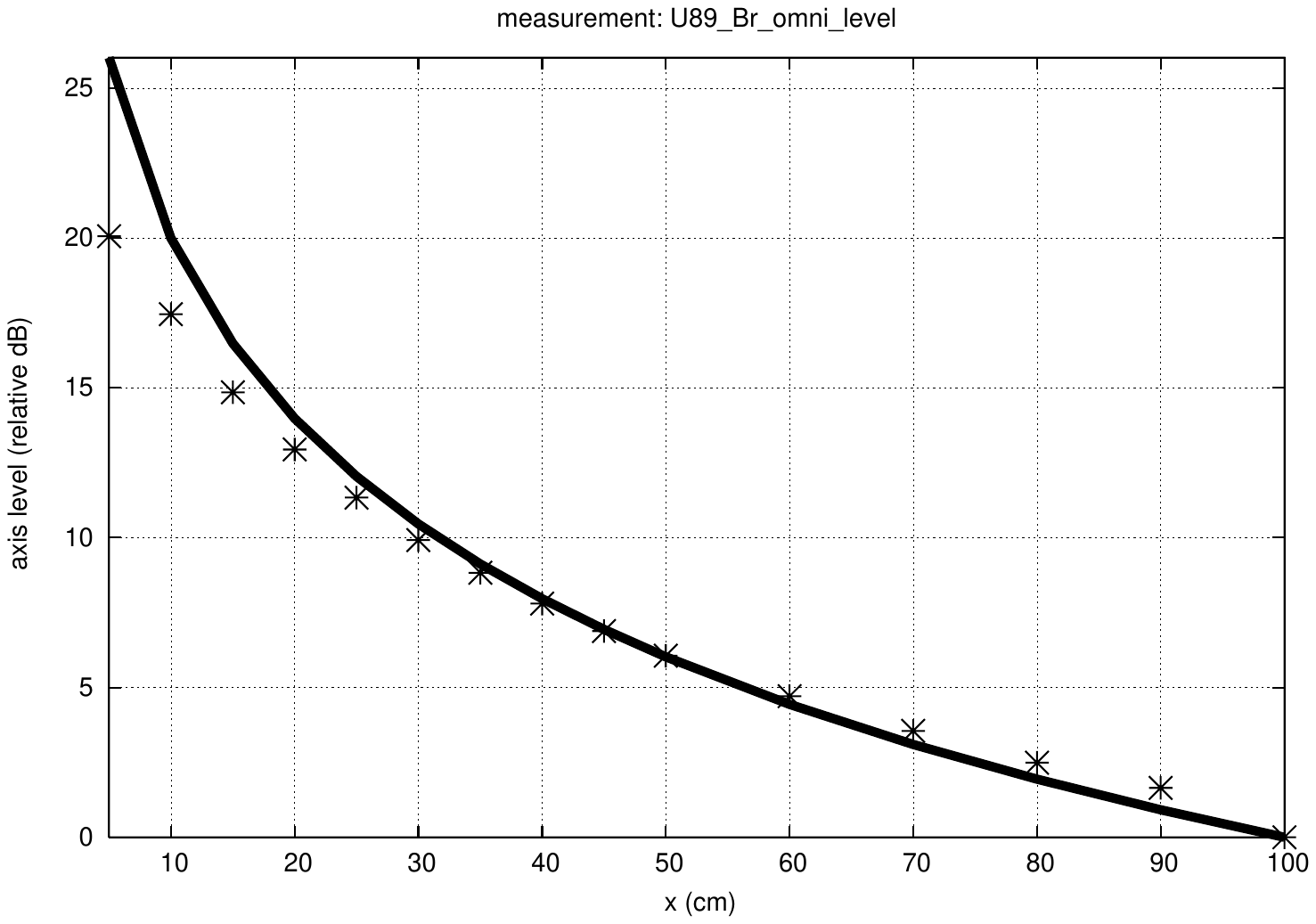}
\includegraphics[width=7.5cm]{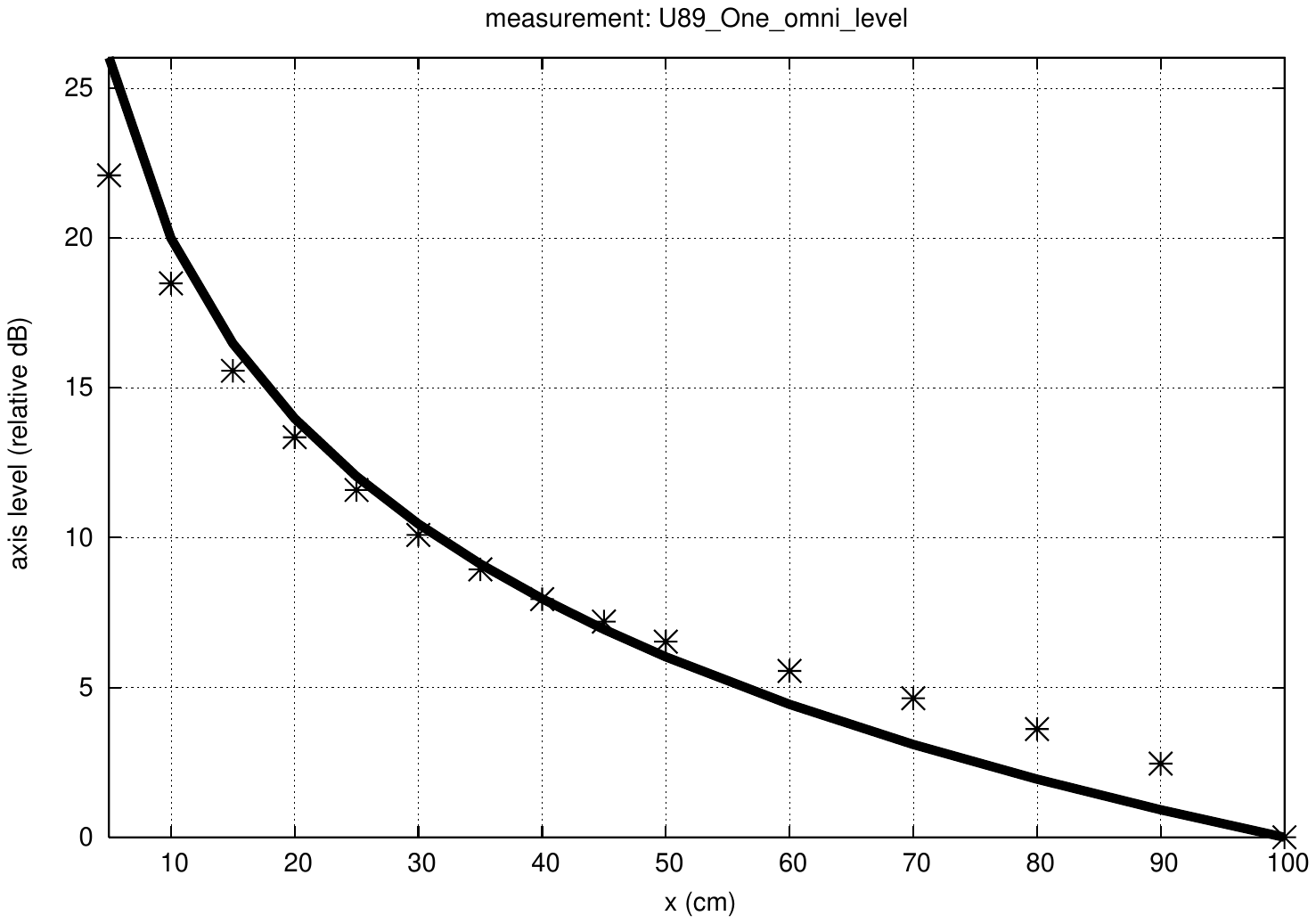}
\end{center}
\caption[4]{Mean level as a function of $x$ for a U89i Neumann microphone with omnidirectional directivity: experimental points (crosses); theoretical amplification (solid line). Excitations:  pink noise (left) and music (right).}
\end{figure}

\pa One can note that we refind main characteristics of omnidirectional and bidirectional measurements, with some attenuation (or averaging?) when considering the cardioid measurements, which would be understandable if this microphone is built with an omnidirectional capsule and a bidirectional one. In the omnidirectional case, the level amplification is bigger in the 50-100~cm range and lower in the 0-50~cm one. In the bidirectional case, the level amplification is bigger in the whole range but much more important in the 0-50~cm (at least for the music stimulus). For music stimulus, the difference is almost a boost of 10~dB at $x=5$~cm in the case of the bidirectional directivity. 

\begin{figure}[h!t]
\begin{center}
\includegraphics[width=7.5cm]{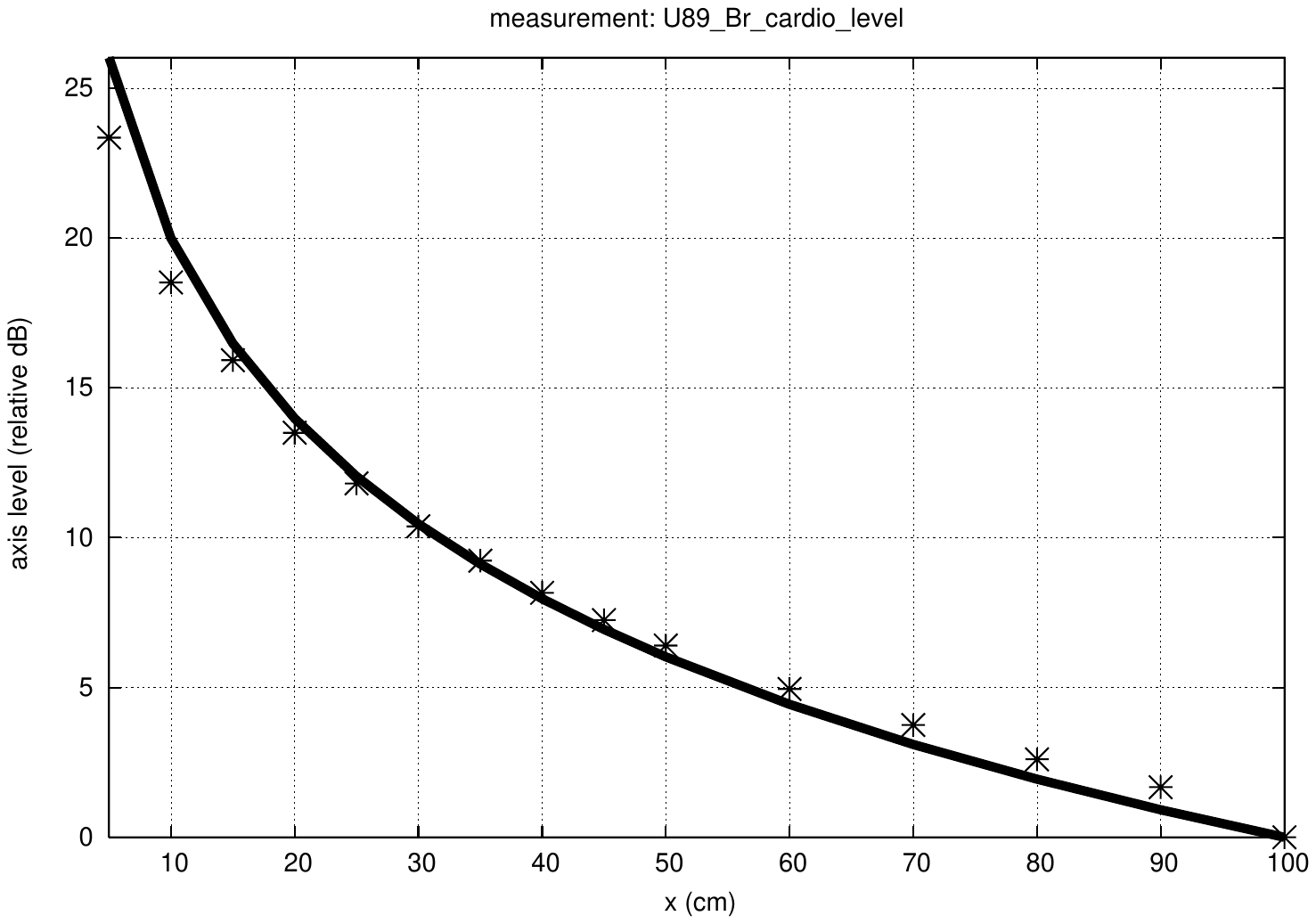}
\includegraphics[width=7.5cm]{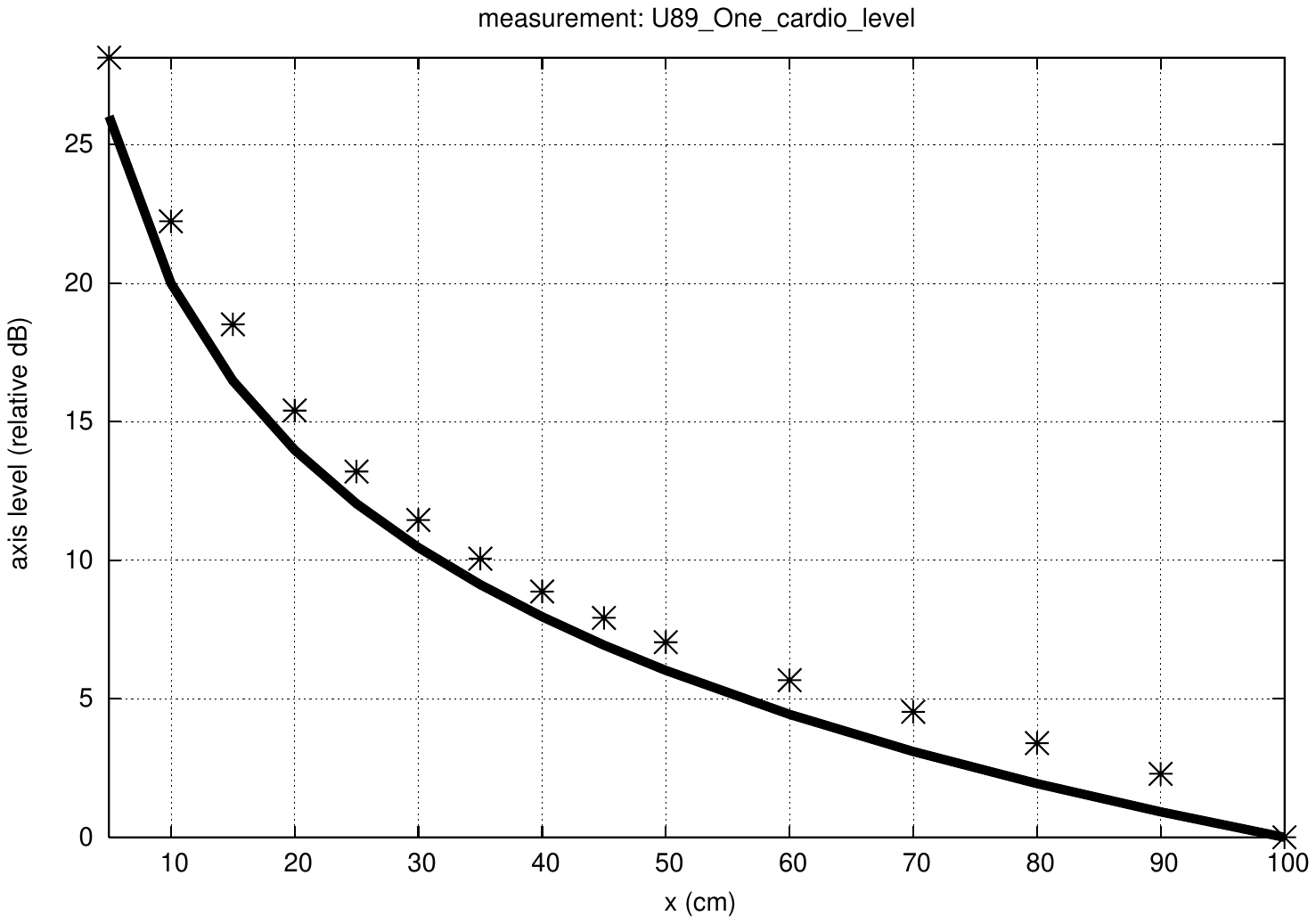}
\end{center}
\caption[5]{Mean level as a function of $x$ for a U89i Neumann microphone with cardioid directivity: experimental points (crosses); theoretical amplification (solid line). Excitations:  pink noise (left) and music (right).}
\end{figure}

\pa Studying the level amplification and taking the ECM8000 curves as reference, we can assume that we have a proximity effect, for the level, even in the case of the omnidirectional directivity. But this level proximity effect (compared to the reference) is big for the cardioid directivity and much more bigger for the bidirectional directivity: a boost of around 12~dB compared to the reference in the case of the music stimulus.

\begin{figure}[h!t]
\begin{center}
\includegraphics[width=7.5cm]{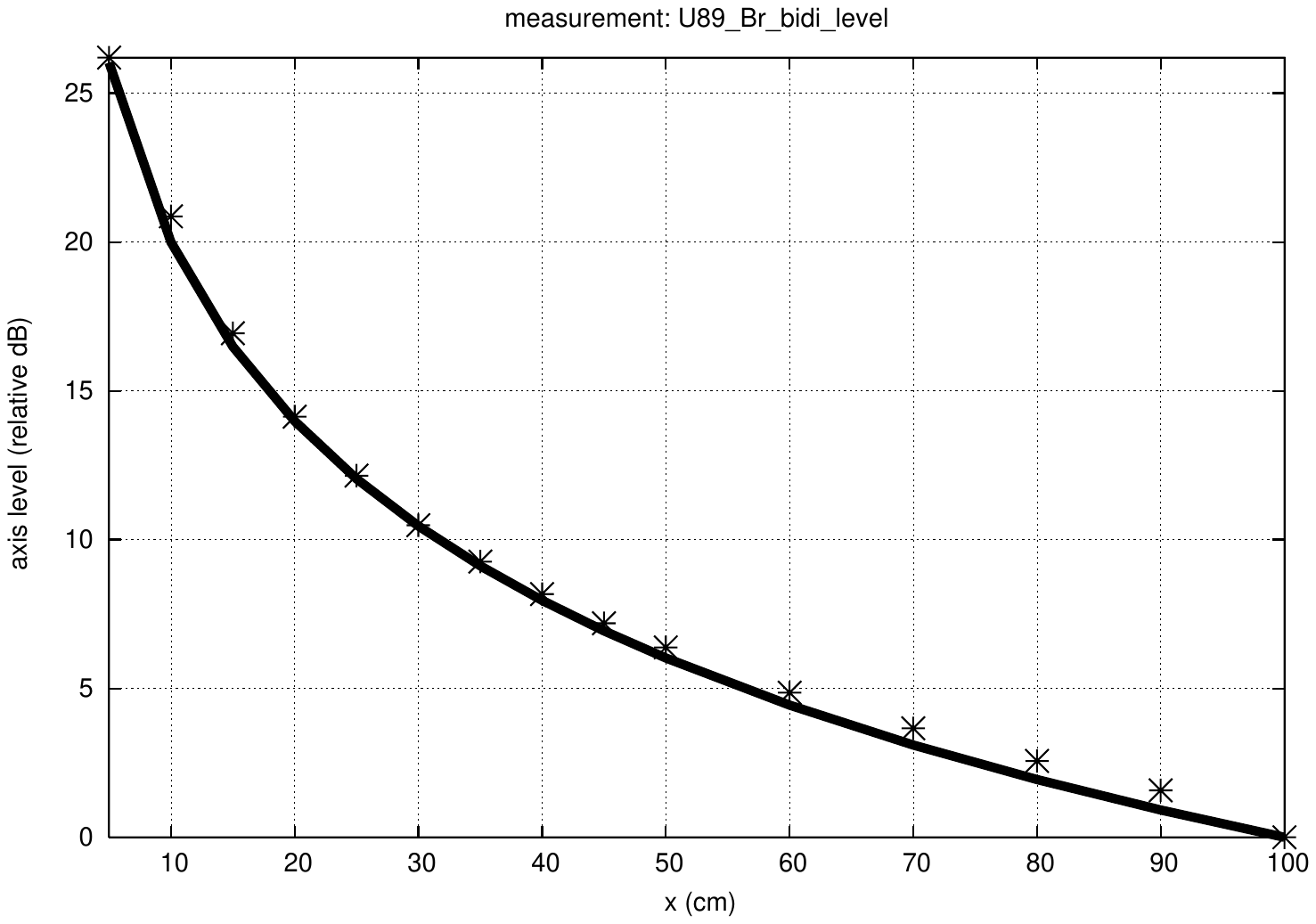}
\includegraphics[width=7.5cm]{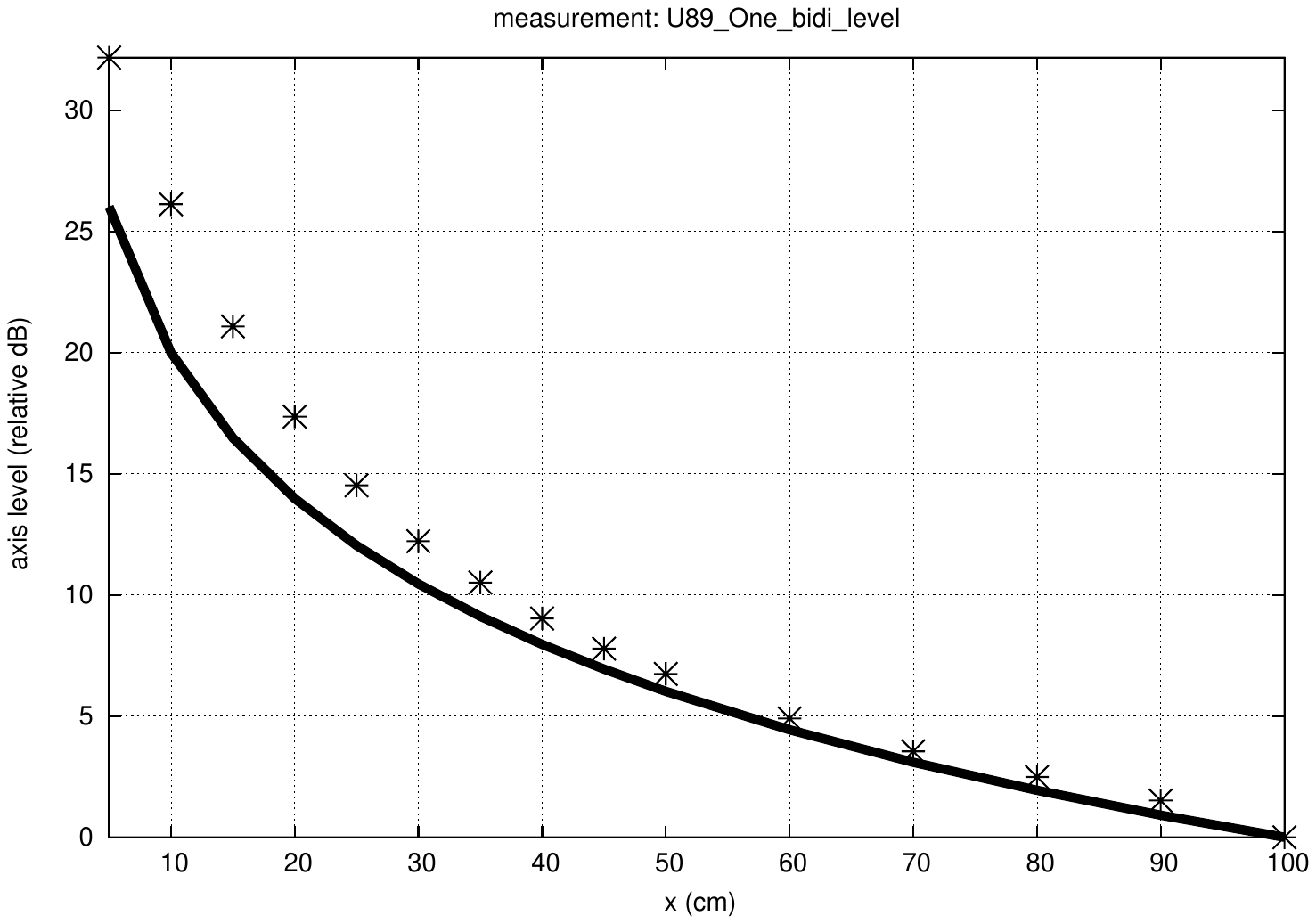}
\end{center}
\caption[6]{Mean level as a function of $x$ for a U89i Neumann microphone with bidirectional directivity: experimental points (crosses); theoretical amplification (solid line). Excitations:  pink noise (left) and music (right).}
\end{figure}

\pa As we still encounter a small boost for the omnidirectional directivity
(around 2~dB), we can think about the question of the influence of the diaphragm size as the ECM8000 microphone has a small diaphragm and the U89i, a large dual one and also a huge grid.

\pa With the U89i microphone, we can consider that we have a boost of level, much more important with music stimulus, for the 50-100~cm range. It would be interesting to make measurements in a bigger anechoic chamber and bigger distance in order to determine how the experimental level amplification curve would vary.

\pa Fig. 7 gives the results for the Audio-Technica AT2020 cardioid microphone which has a large dia\-phragm and also a huge shield grid. Measurements were done from $x=0$ to 100~cm for both stimuli.

\begin{figure}[h!t]
\begin{center}
\includegraphics[width=7.5cm]{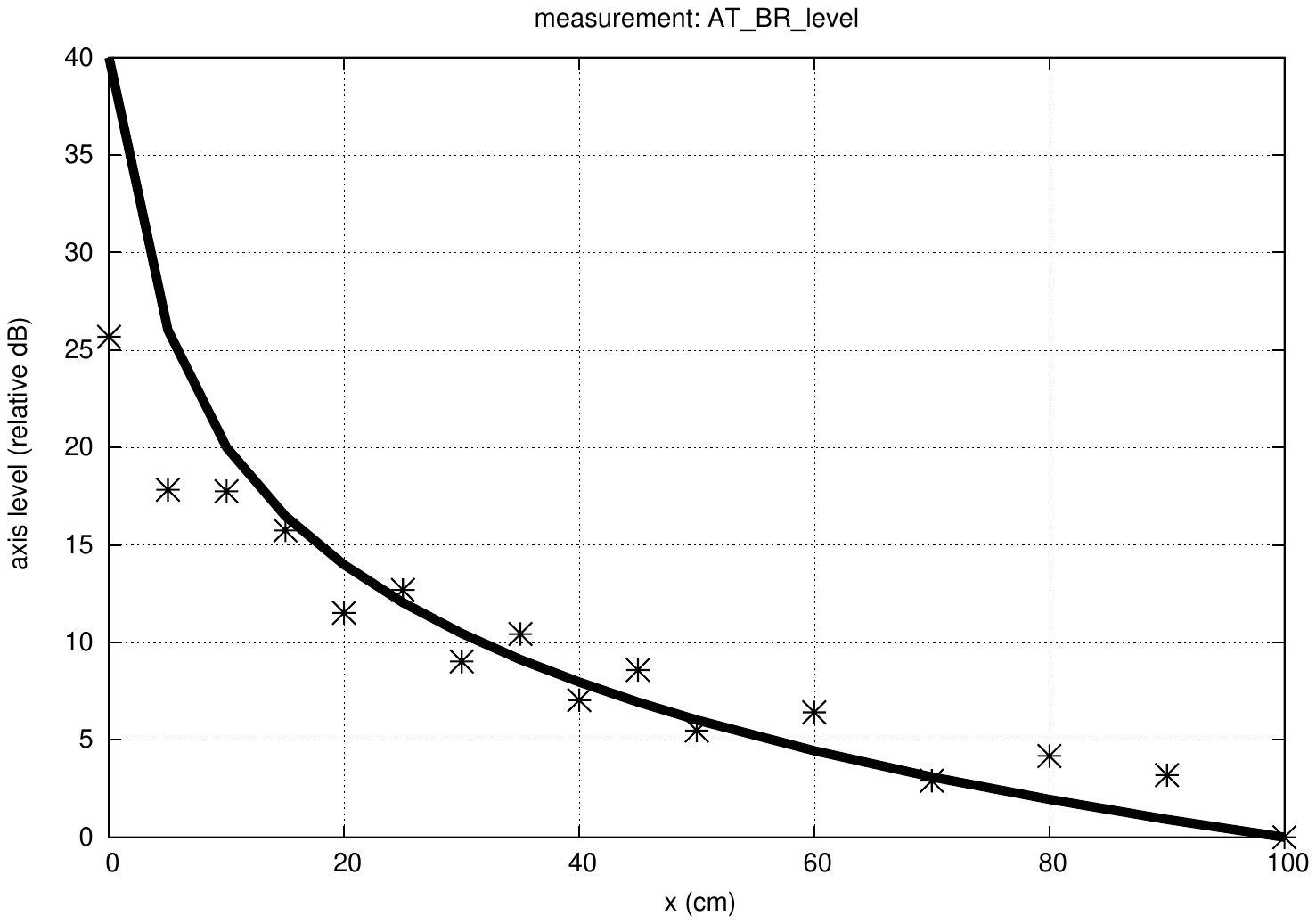}
\includegraphics[width=7.5cm]{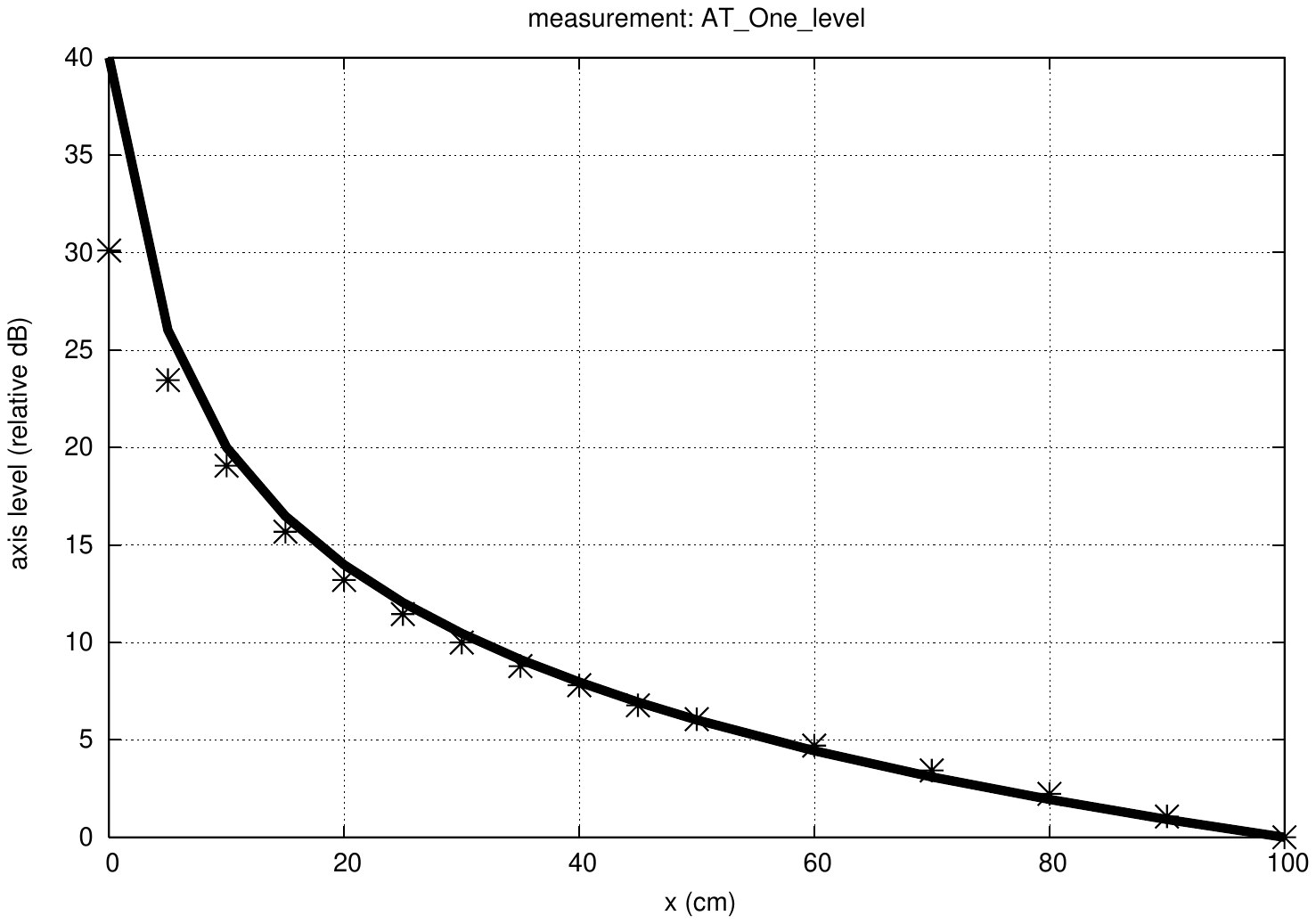}
\end{center}
\caption[7]{Mean level as a function of $x$ for a Audio-Technica AT2020 cardioid microphone: experimental points (crosses); theoretical amplification (solid line). Excitations:  pink noise (left) and music (right).}
\end{figure}

\pa We can observe that the gap between $1/x$ law and measurements points are significantly different when we consider either the case of the pink noise stimulus or the one of the music stimulus. Indeed, the gap is rather small, compared to the one for the chosen reference (ECM8000 microphone), for music stimulus at least for distances $x\geq 10$~cm while the gap is significant when using pink noise as we have measurements points under and over the theoretical curve with significant differences for most of the points. 

\pa We can think that this microphone may have been designed in order to follow as precisely as possible the wave theoretical curve for musical recordings. And, this fact gives some weight to the question of the usefulness of pink noise and, above all, pure  sinusoids stimuli compared to musical ones even if the choice of the musical stimuli need to be argued. In fact, using music for measurements may give some more insight about the behavior of the microphone.  

\pa Compared to the reference curves of Fig. 3, curves of Fig. 7 also point out a proximity effect due to the use of the AT2020 microphone and this effect (boost at 0~cm, points quite close to the wave curve for instance) is significant, but perhaps more obvious and informative in the case of the music stimulus. And, compared to the U89i cases, this effect consequence is a bigger proximity to the wave curves for distances $x\geq 10$~cm.

\pa Fig. 8 presents the case of the Behringer C-2 cardioid microphone (small diaphragm) for distances $x$ varying from 0 to 1~m.

\begin{figure}[h!]
\begin{center}
\includegraphics[width=7.5cm]{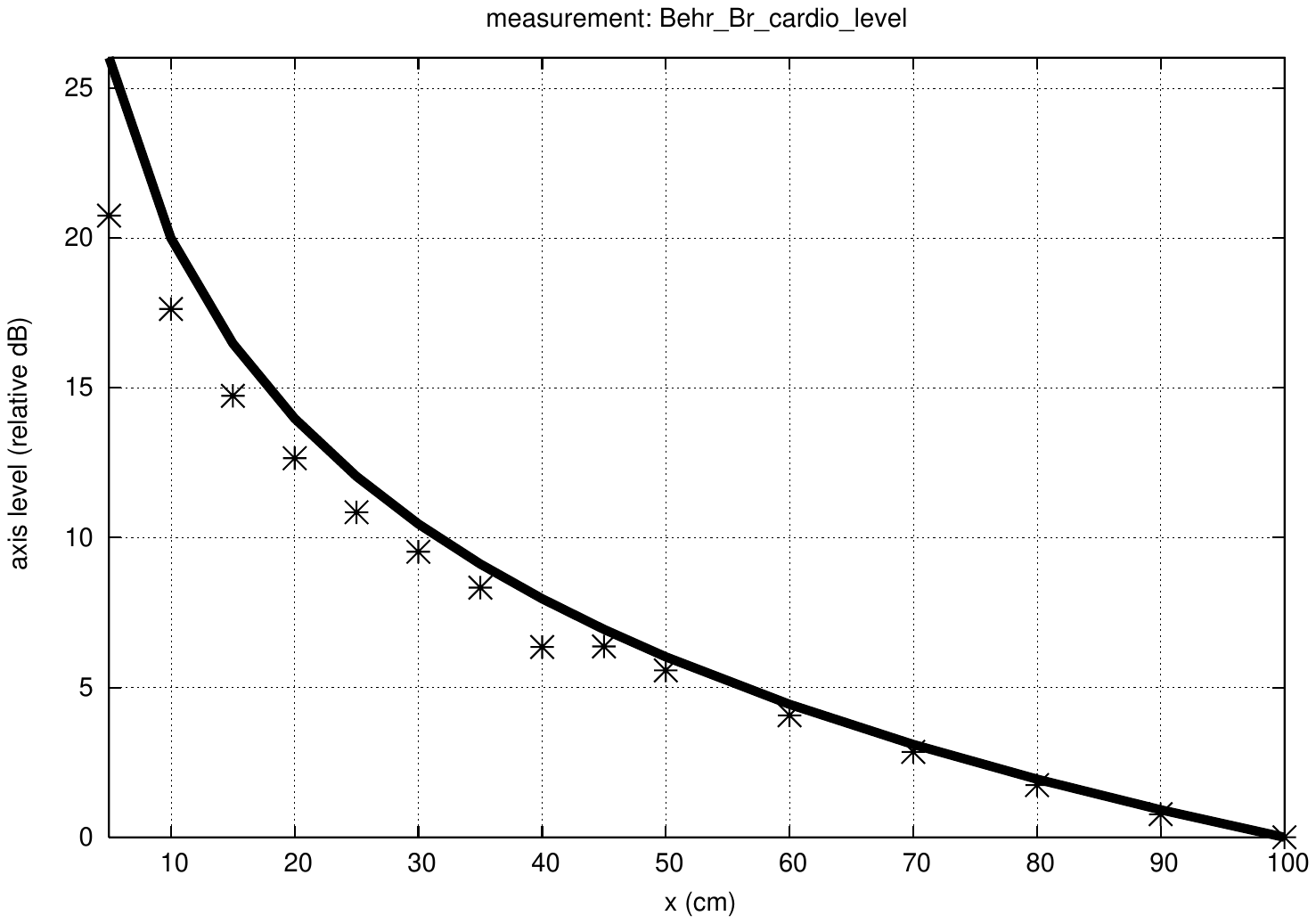}
\includegraphics[width=7.5cm]{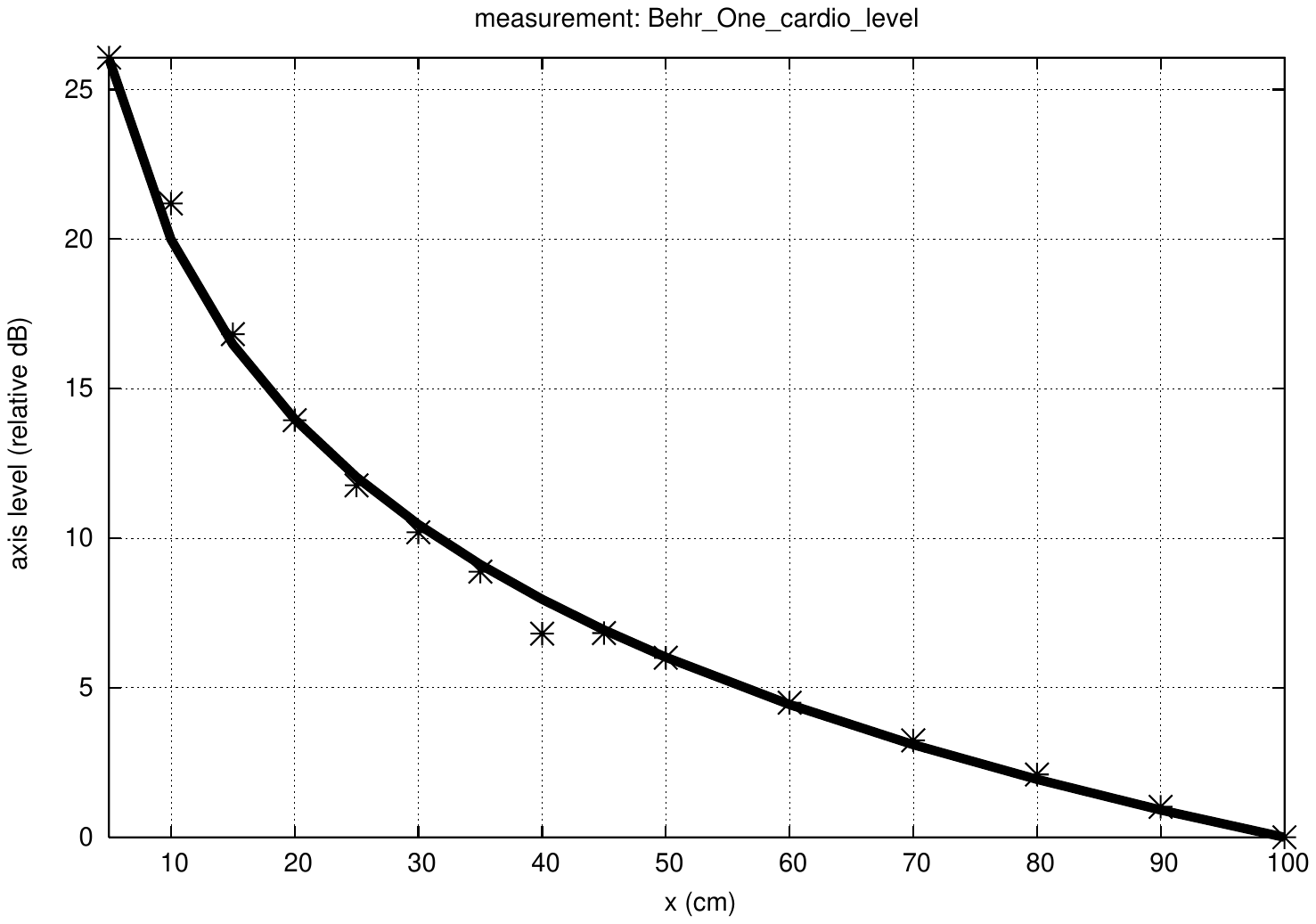}
\end{center}
\caption[8]{Mean level as a function of $x$ for a Behringer C-2 cardioid microphone: experimental points (crosses); theoretical amplification (solid line). Excitations:  pink noise (left) and music (right).}
\end{figure}

\pa Comparison of the curves for the pink noise stimu\-lus case and music one show that both measurements are rather different. For the pink noise stimu\-lus there are some significant differences between wave curve and experimental points for distances $x\leq 50$~cm, with a 40~cm point which may be discarded as the difference is quite stronger than the others for both stimuli. But, for music stimulus, the agreement between wave curve and experimental points is almost perfect. 

\pa In fact, we refind the question of the optimization of the design of the microphone to follow the wave level theoretical curve when recording music and, again, the question of the usefulness of pink noise measurements if one desire to access the behavior of the microphone during a recording. It seems that using musical stimuli to perform measurements could be very informative even when considering the mean or global behavior of the microphone for a piece of music.

\pa As the curves for the C-2 microphone are much closer to the wave ones than the curves we found for the ECM8000 microphone, proposed by the same company, we can assume that the C-2 microphone exhibits a proximity effect for the mean pressure level. 

\pa At this point we only have reviewed the variation of the mean level according to the microphone (or its directivity) and the distance from the source. But, the proximity effect as described in literature (see for instance \cite{Williams:04}) also concerns the spectral behavior of a microphone, and notably of a directional one.

\subsection{IDS balance vs distance from the source}
\pa The 100~cm level is also taken as 0~dB reference for all subband weight curves and, as soon as we would have reached the area where spherical wave model would be a fair model, the weight of each subband would not vary any more, in theory, but significantly in practice. The distance over which the weight presents only some "residual" variations may be considered as the "validity limit" distance for wave approximation.

\subsubsection{Omnidirectional directivity}
\pa In Fig. 9 to 18, we present the evolution of the subband weight according to the distance from the source $x$ for all the 10 subbands considered in the IDS analysis \cite{Millot:04} using the ECM800 microphone, that is to say the so-called pressure sensor giving access to the "real" pressure field.

\pa The 10 subbands considered in the IDS analysis are the following: 
\begin{itemize}
\item subband 1: 0-50~Hz;
\item subband 2: 50-200~Hz;
\item subband 3: 200-400~Hz;
\item subband 4: 400-800~Hz;
\item subband 5: 800-1200~Hz;
\item subband 6: 1.2-1.8~kHz;
\item subband 7: 1.8-3~kHz;
\item subband 8: 3-6~kHz;
\item subband 9: 6-15~kHz;
\item subband 10: 15-22.05~kHz (Nyquist frequency for recordings with 44.1~kHz sampling frequency).
\end{itemize}

\pa We must precise the fact that the global weight of each subband is calculated as the sum of the squares of all the samples of a subband, divided by the sum of the squares of the samples of the global signal associated with the recording. These relative weights, giving the global spectral balance for the studied recording, are then expressed in relative dB. And for all IDS evolution curves presented here we plotted the difference of the relative weight for a given distance and of the relative weight of the same subband for the 1~m location, the reference location. 

\pa As explained for the study of the mean level amplification curves, we consider that the wave assumption should be, at least, valid for the 1~m location. 

\begin{figure}[h!t]
\begin{center}
\includegraphics[width=7.5cm]{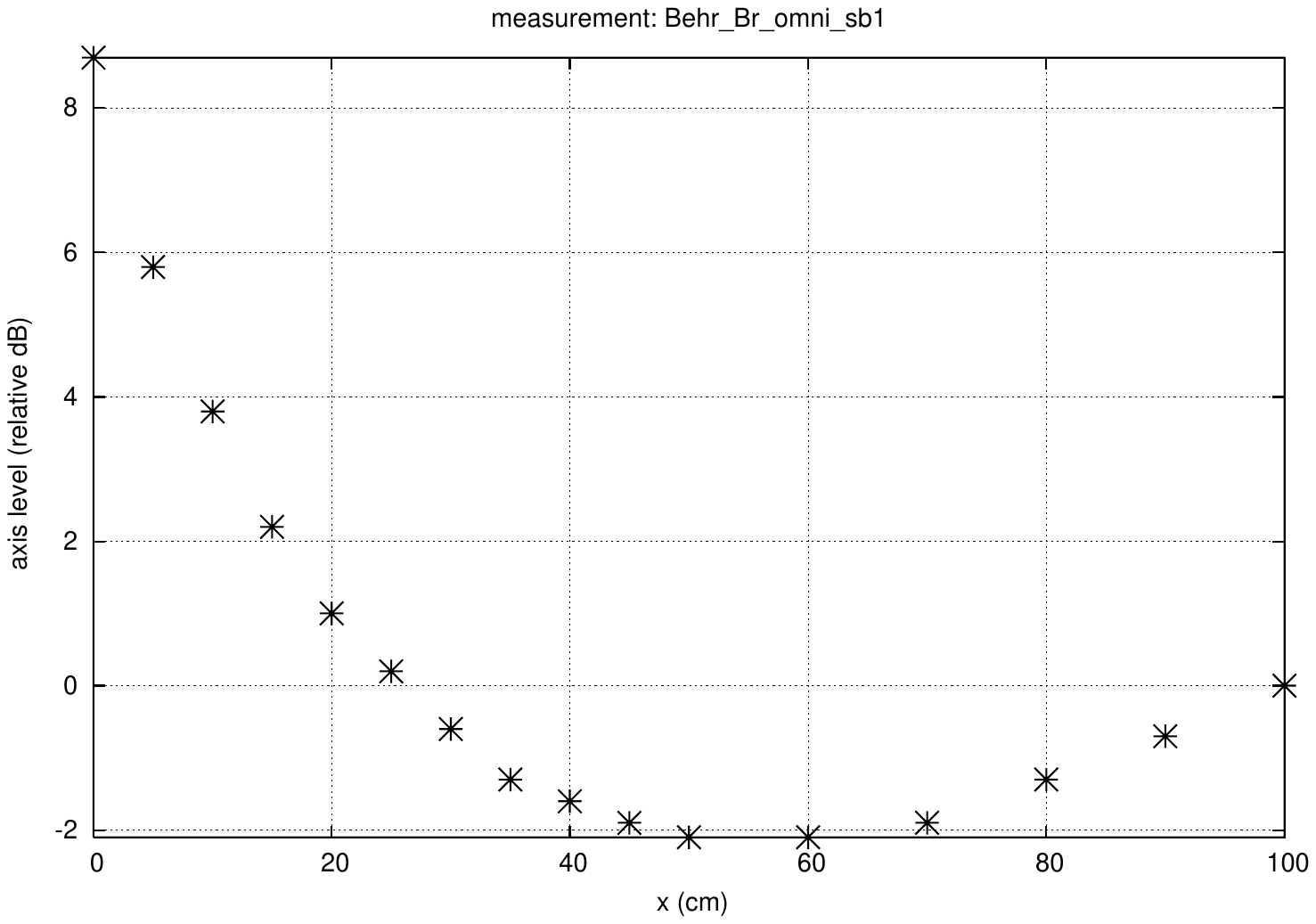}
\includegraphics[width=7.5cm]{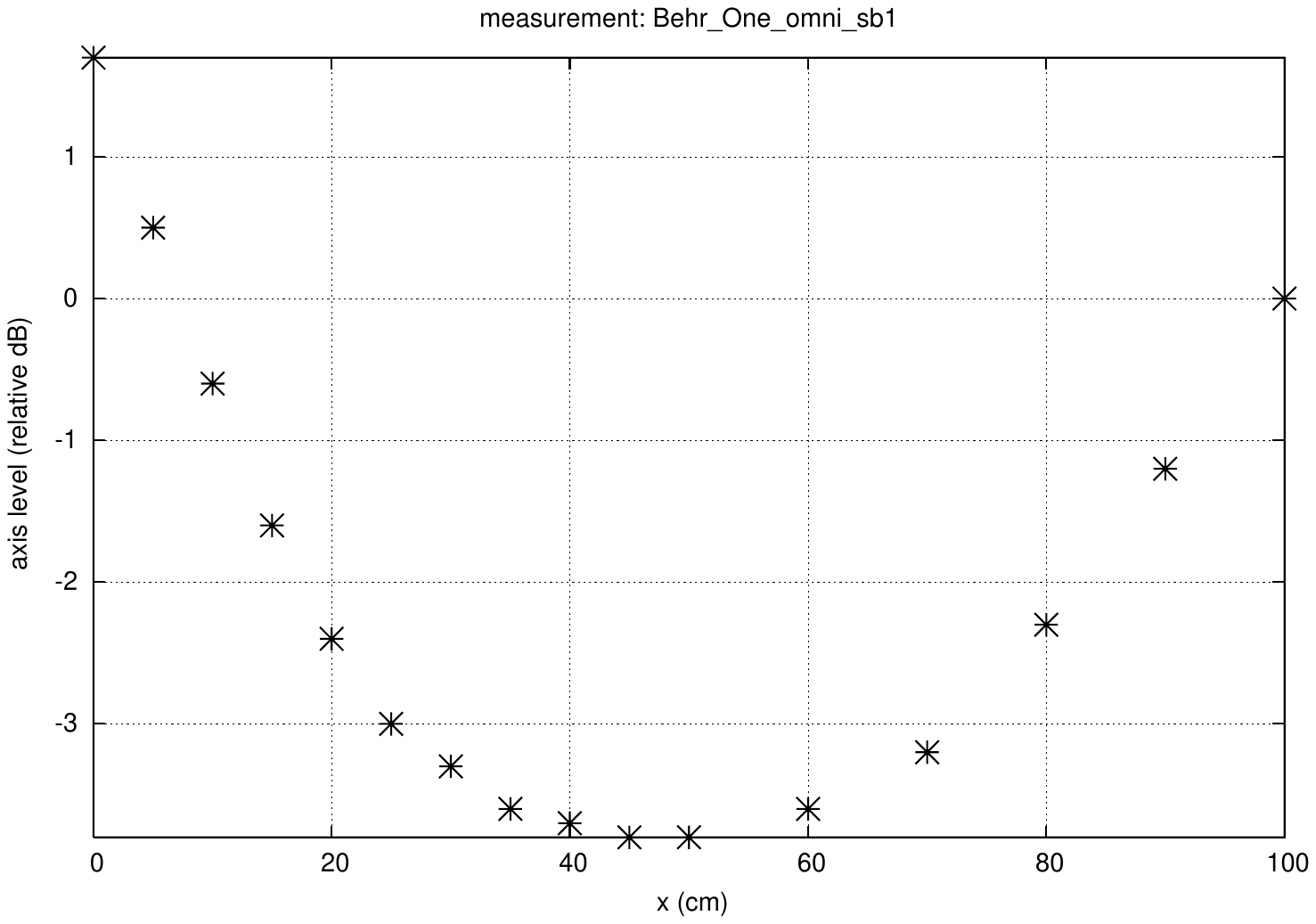}
\end{center}
\caption[9]{Evolution of the weight in the spectral balance of subband  1 as a function of $x$; pink noise  (left) and music (right) for a Behringer ECM8000 microphone with omnidirectional directivity: experimental points (crosses).}
\end{figure}

\begin{figure}[h!t]
\begin{center}
\includegraphics[width=7.5cm]{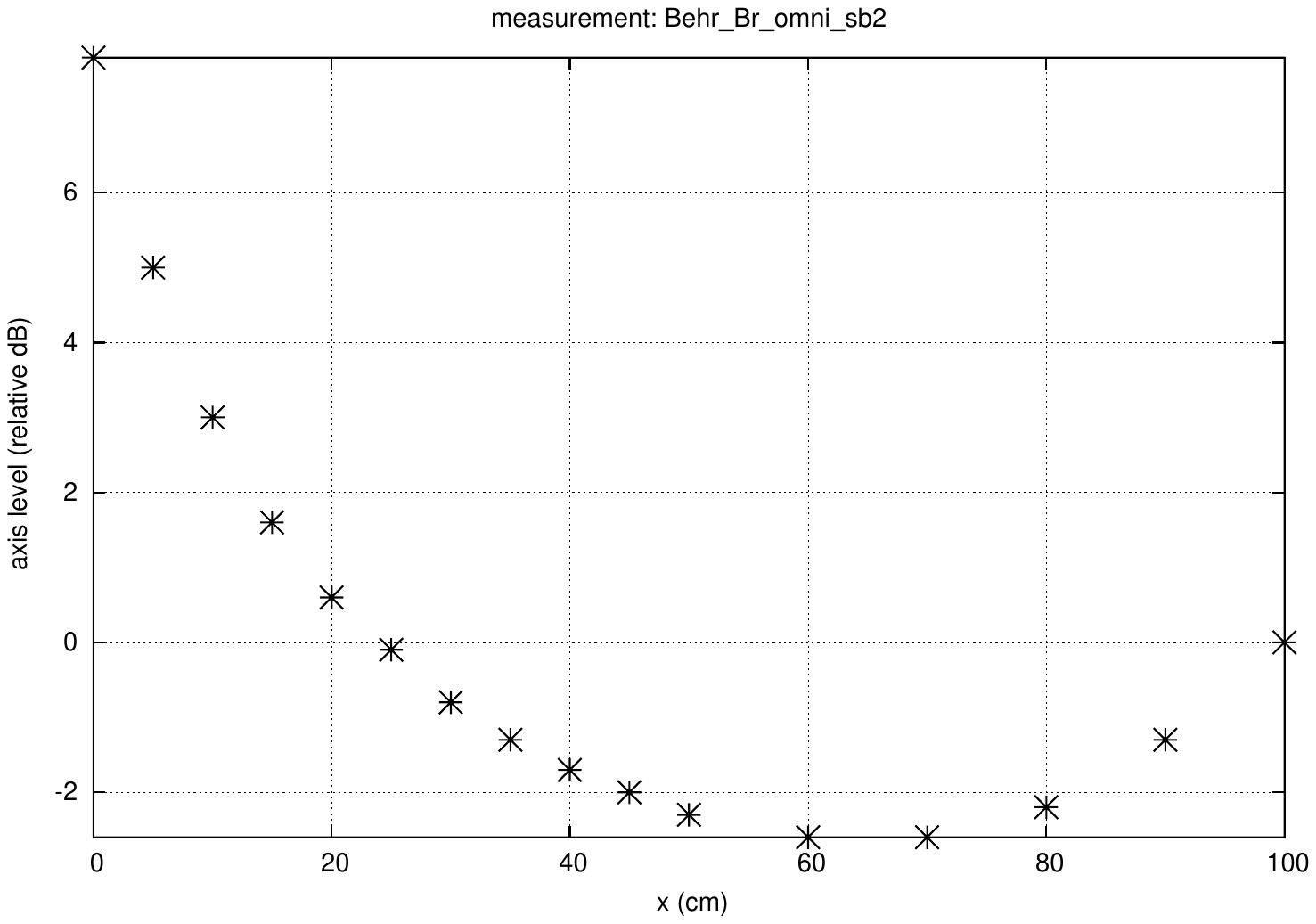}
\includegraphics[width=7.5cm]{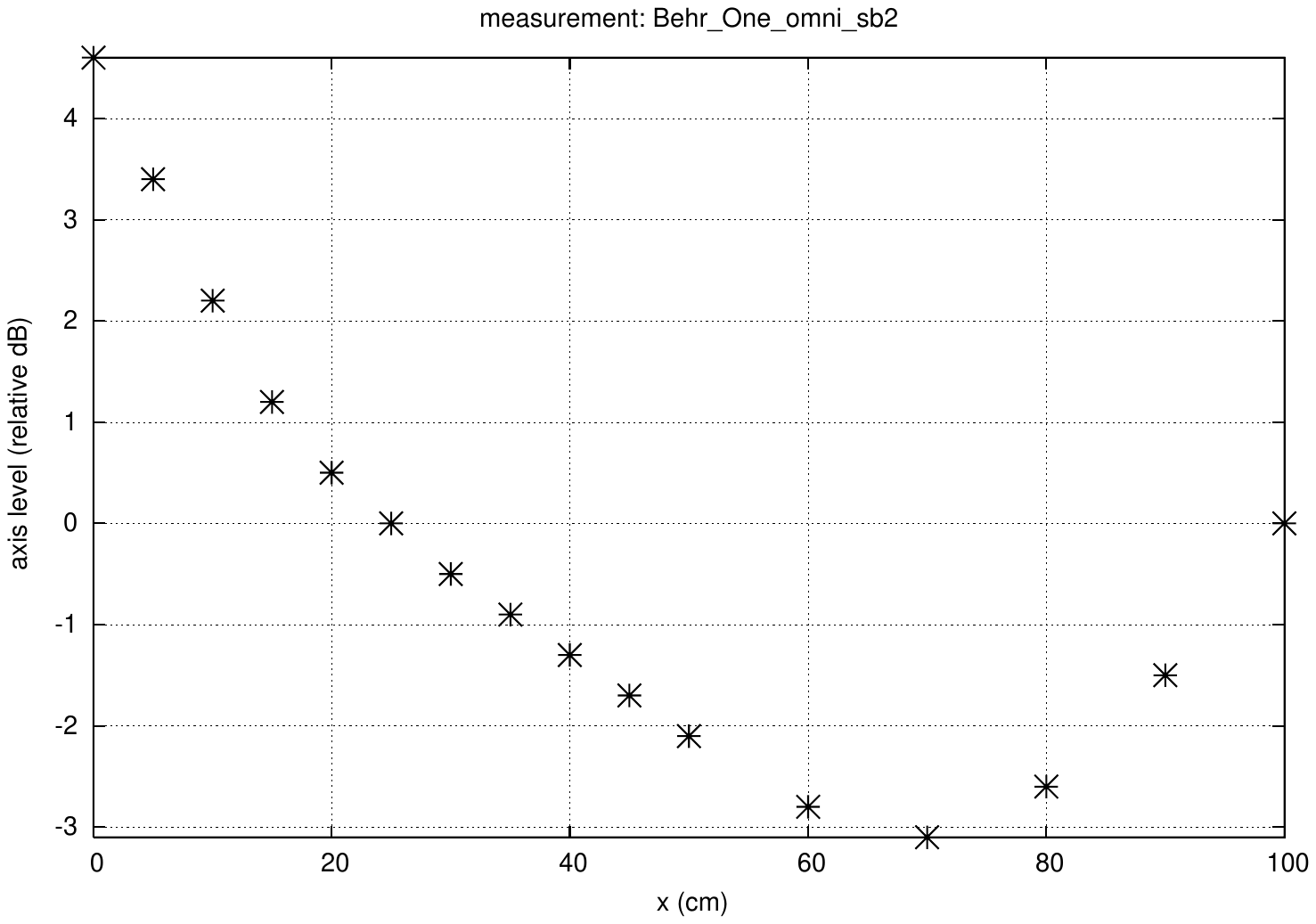}
\end{center}
\caption[10]{Evolution of the weight in the spectral balance of subband  2 as a function of $x$; pink noise  (left) and music (right) for a Behringer ECM8000 microphone with omnidirectional directivity: experimental points (crosses).}
\end{figure}

\pa Study of Fig. 9 to 18 underlines the fact that the curves in the pink noise case and in the music one differ in values  but, by the more, present some simi\-larities. We refind again the question of the most informative case and we still think that measurements made using musical stimuli are more informative. Furthermore, with musical stimuli, one can perform subjective and objective studies of an acoustical device using the same \textit{corpus}: musical recordings. As said before, we consider the pink noise measurements only as a bridge with classical information one may encounter in the literature.   

\pa In the following, we focus our attention on the global behavior for a given subband, as curves are different for both stimuli, but we consider that the curve associated with the music stimulus is more representative of the recording situation. So, if both curves are quite different, we assume that the music stimulus case should be taken as reference.

\pa In literature, proximity effect is associated with a boost of the low frequency when the distance from the source decreases.  Fig. 9 and 10, associated res\-pectively with 0-50~Hz and 50-200~Hz subbands, give the information about the behavior of low frequency register. 
 
\pa As said usually in literature, we find a decrease of the weight of subbands 1 and 2 when the distance from the source $x$ increases from $0$ to 50~cm, which is related to a boost effect for low frequencies when we get closer to the source.

\pa But, from 50~cm to 1~m, we have also an increase of the weight of subbands 1 and 2 which may be significant: at least 3~dB for music stimulus and at least 2~dB for pink noise. This effect seems more pronounced in the case of the music stimulus, for which we have silent moments, which may seem quite surprising. And, we have global magnitude for the variation of the weight which are rather similar for both stimulus.    

\pa So, for these both low-frequency subbands, we do not have only a diminution of their weight in the global balance but after a minimum weight range of $x$, something which may be a "re-equalization" phenomenon.  This makes complex the estimation of a distance associated with the validity of wave model as we still observe \textit{a priori} significant variations!

\pa Fig. 11 and 12 illustrate what happens for subbands 3 and 4, respectively associated to 200-400~Hz and 400-800~Hz frequency ranges, which correspond to the range where we may find most of the fundamentals, in case of music stimulus.

\pa The first comment we can make concerns the fact that the behavior of both subbands differ greatly according to the nature of the stimulus and that it seems again quite difficult to define a validity distance for waves assumption. 

\pa We have short variations of the weight for both subbands in the case of the pink noise stimulus (less than 2~dB) but with a diminution of the weight, fast for subband 3 and gradual for subband 4, associated with an increase for $x$ greater than 40~cm and then a second decrease of the weight: for $x>80$~cm for subband 3 and $x>70$~cm for subband 4. We also have a small increase between 90 and 100~cm, for subband 4, but which may not be significant.

\begin{figure}[h!t]
\begin{center}
\includegraphics[width=7.5cm]{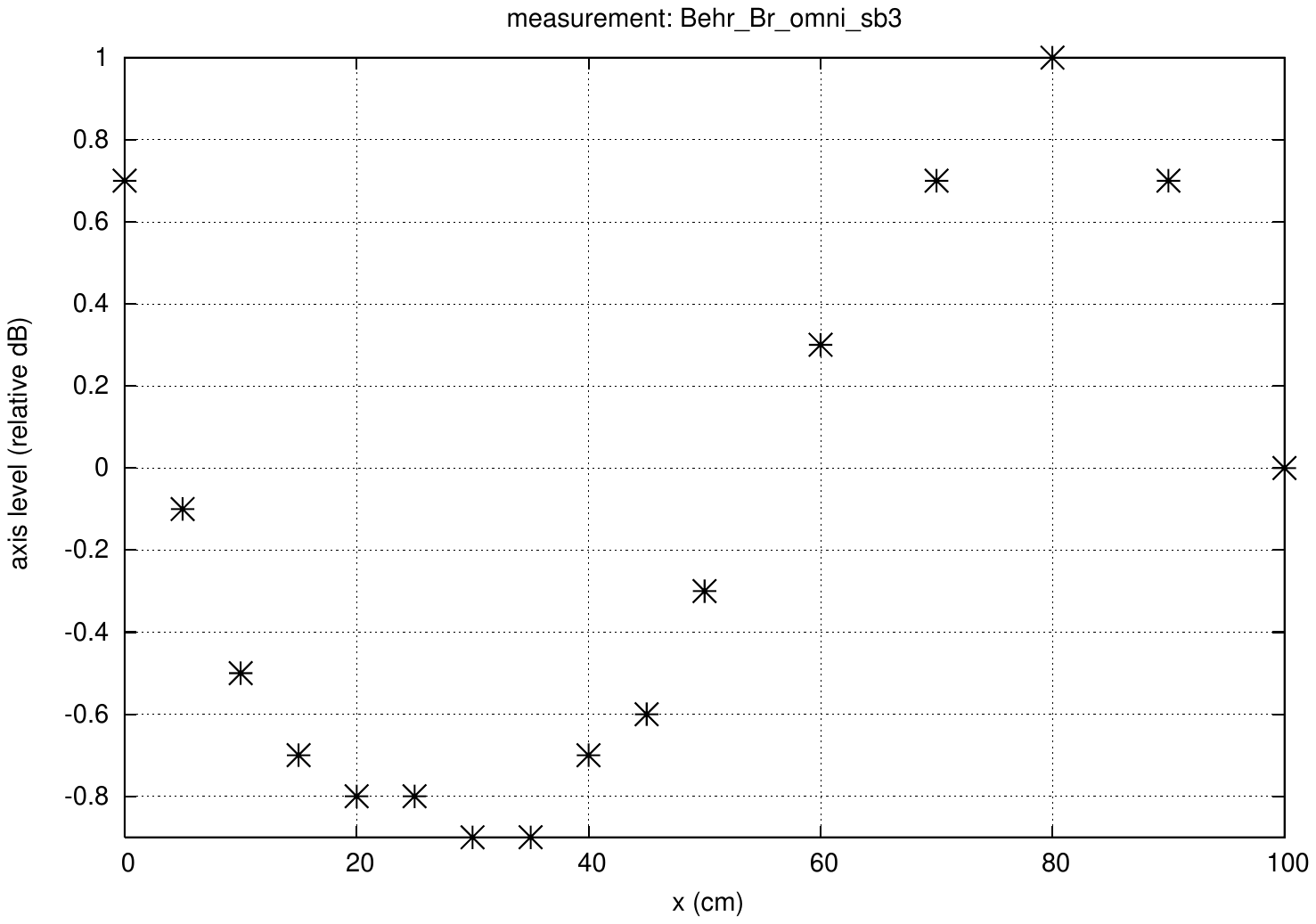}
\includegraphics[width=7.5cm]{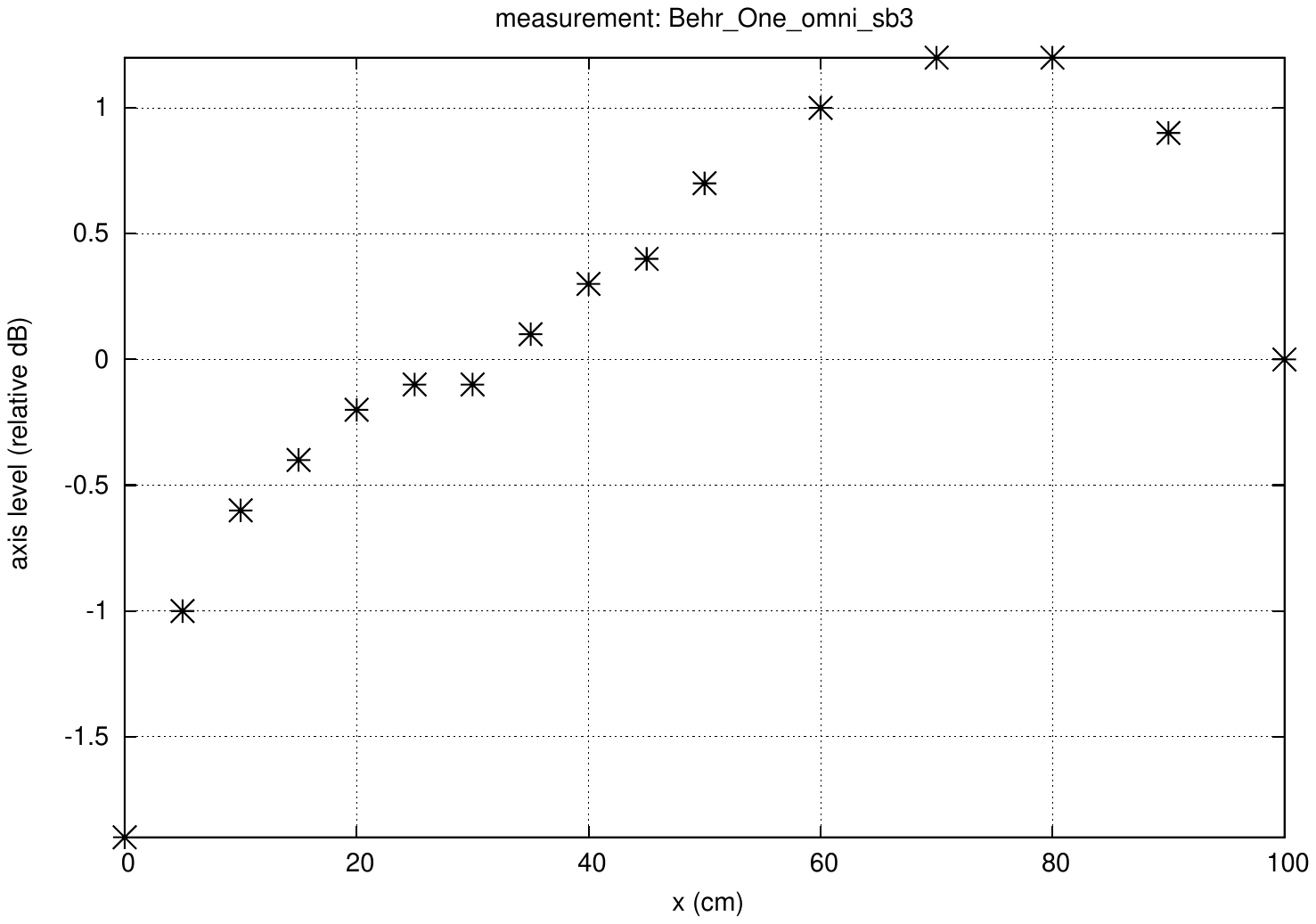}
\end{center}
\caption[11]{Evolution of the weight in the spectral balance of subband  3 as a function of $x$; pink noise  (left) and music (right) for a Behringer ECM8000 microphone with omnidirectional directivity: experimental points (crosses).}
\end{figure}

\pa But, the behavior with the music stimulus is rather different for both subbands. Indeed, we first have an increase of the weight for both subbands: from $x=0$ to $x=80$~cm for subband 3, from $x=0$ to $x=30$~cm for subband 4 followed with small "spatial modulation" from $x$ varying from 30 to 60~cm. After, we have a decrease of more than 1~dB for the weight for subband 3 and a decrease of the weight of subband 4 of almost 2~dB with a final increase of almost 1.25~dB between $x=90$ and $x=100$~cm.

\pa So, if we consider the music stimulus case as reference, we find first an increase of weight for subband 3 and 4 compatible with the literature description but followed by again a so-called "re-equalization" process.

\begin{figure}[h!t]
\begin{center}
\includegraphics[width=7.5cm]{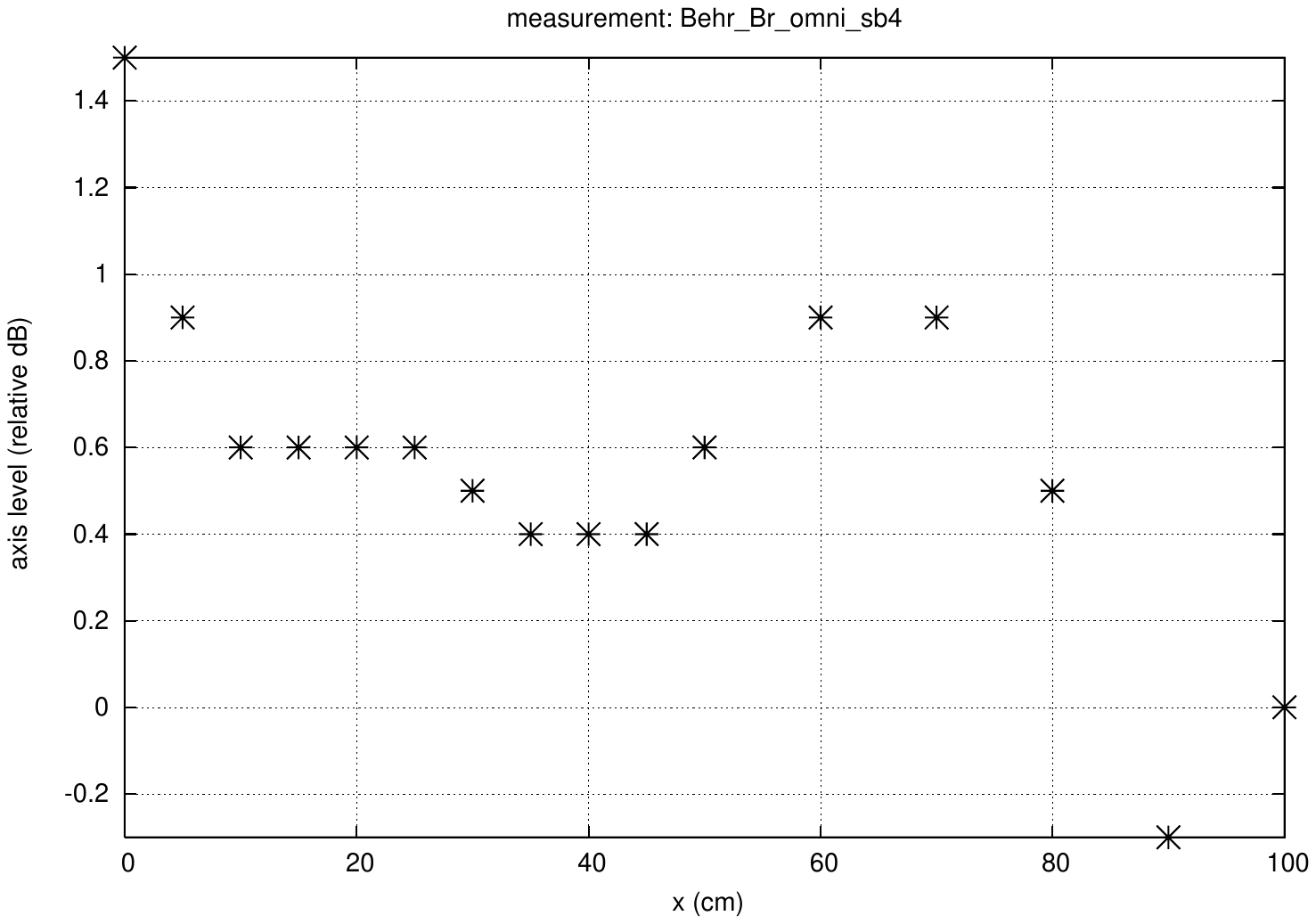}
\includegraphics[width=7.5cm]{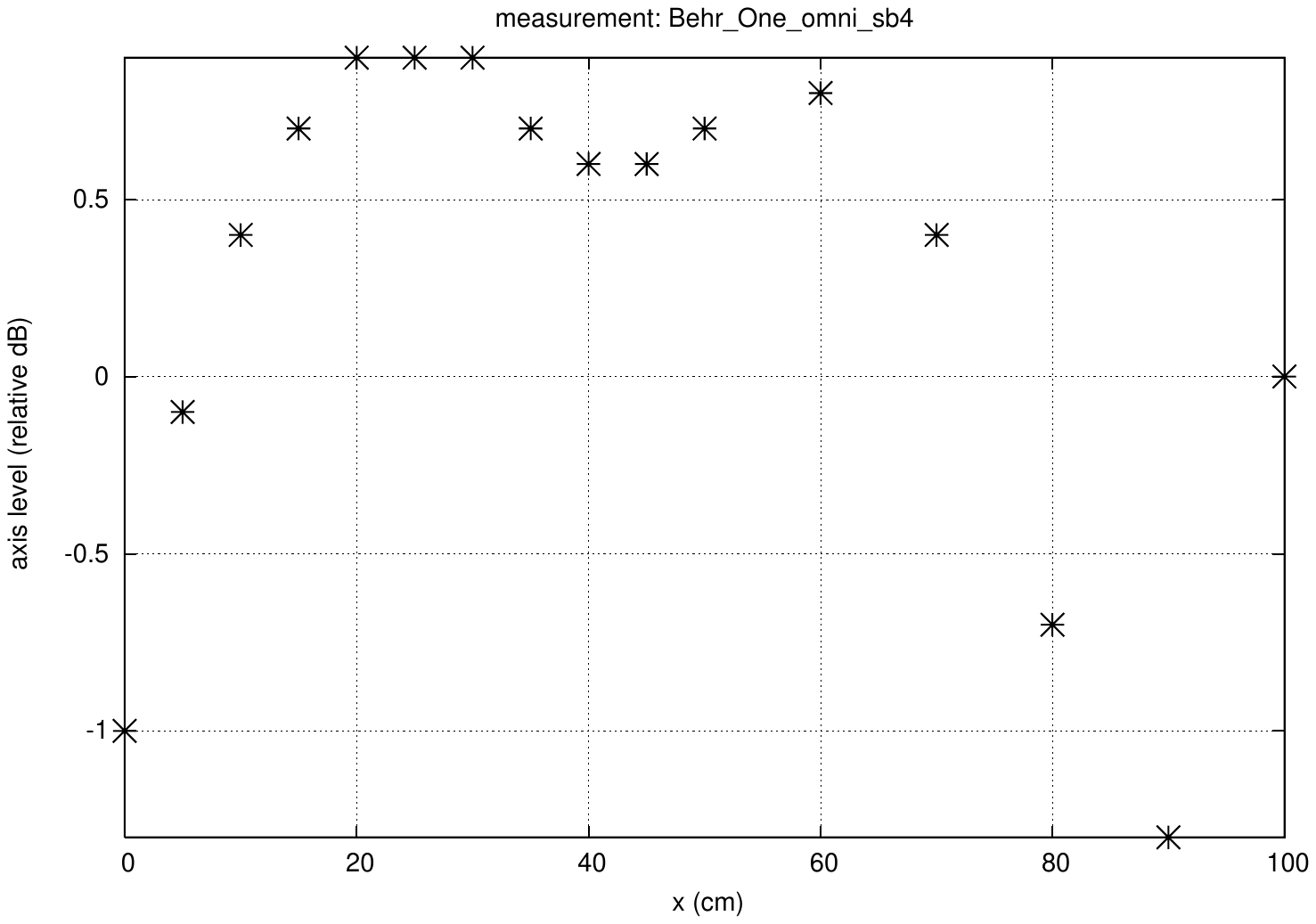}
\end{center}
\caption[12]{Evolution of the weight in the spectral balance of subband  4 as a function of $x$; pink noise  (left) and music (right) for a Behringer ECM8000 microphone with omnidirectional directivity: experimental points (crosses).}
\end{figure}

\pa Fig. 13 and 14, associated to subband 5 (400-800~Hz) and subband 6 (800-1200~Hz) present similar behavior whatever is the stimulus but differ a little from one subband to the other. 

\pa For subband 5 (see Fig. 13), we have an increase of the weight up to 35~cm followed by a "spatial modulation" with a magnitude of approximatively  1.5~dB before a final increase for $x=1$~m. As every weights are mean or global ones, we consider that a 1.5~dB spatial variation should not be compatible with the validity area of wave approximation: this variation magnitude seems too much important.

\pa For subband 6 (see Fig. 14), we first encounter a decrease of the weight from 0 to 10~cm followed by an increase of the weight from 10 to around 40~cm for music stimulus, gradual from 40~cm to 1~m for the pink noise stimulus. For the music stimulus, we have a range from around 40 to 90~cm where there are small variations of the weight before an increase between 90 and 100~cm (more than 1~dB). For subband 6, the limit validity for wave is rather difficult to define because of the final increase, concerning the music stimulus case.

\pa It is interesting to note that for last points we have an increase of the weight in all four situation (both stimuli and both subband) while we have first a decrease of the weight for subband 4 which may be perceived as a boost effect going from around 10~cm to 0~cm !

\begin{figure}[h!t]
\begin{center}
\includegraphics[width=7.5cm]{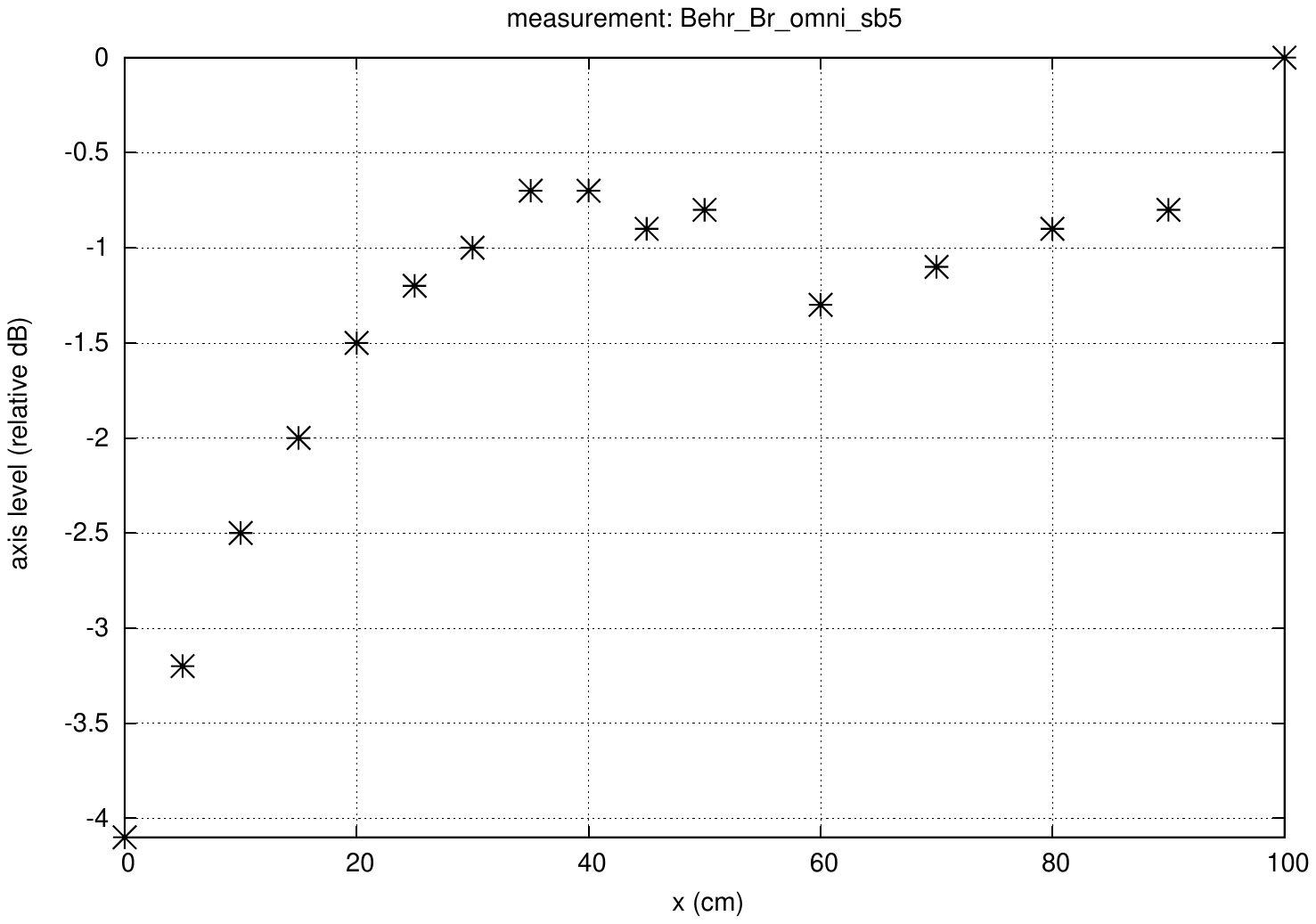}
\includegraphics[width=7.5cm]{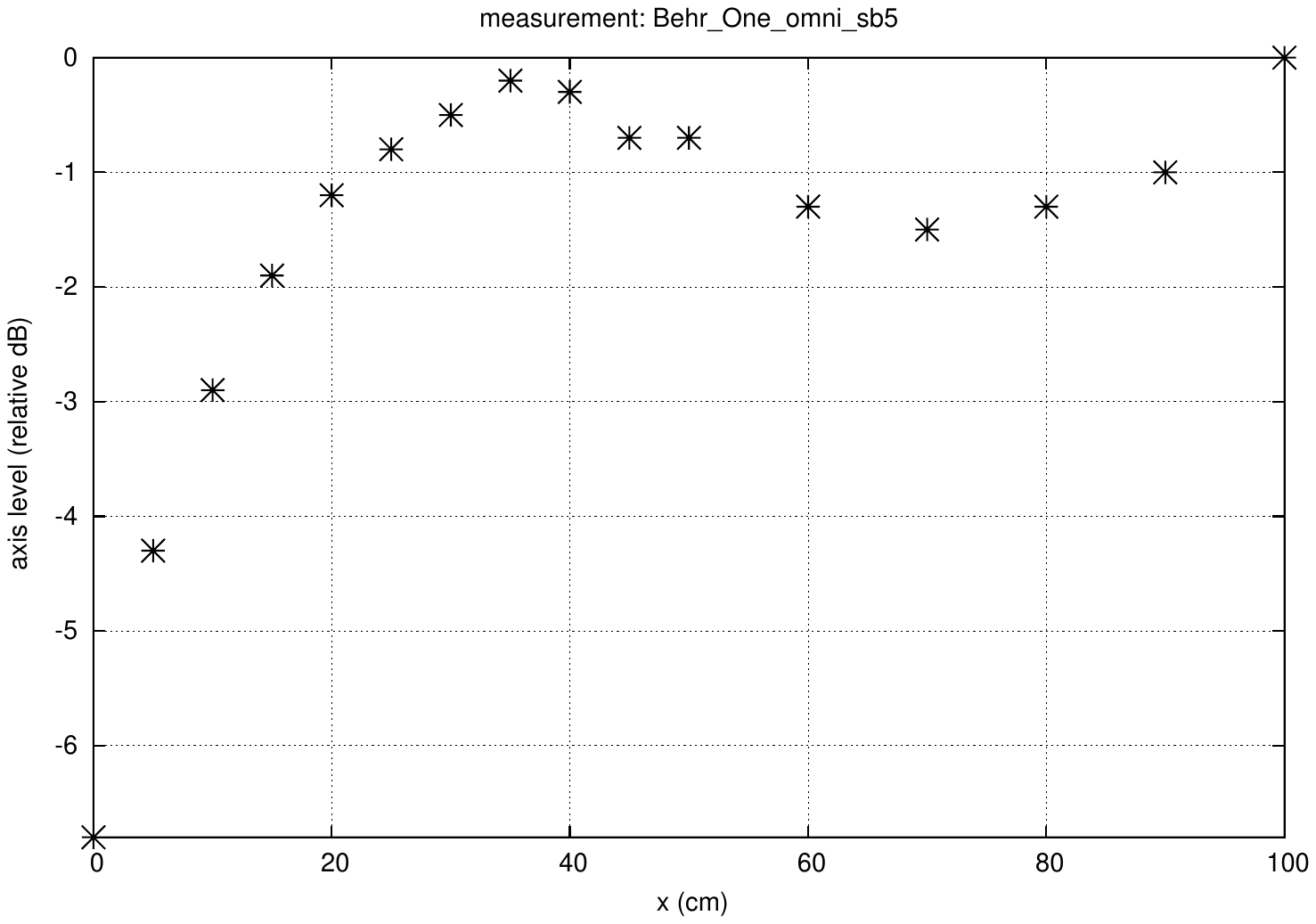}
\end{center}
\caption[13]{Evolution of the weight in the spectral balance of subband  5 as a function of $x$; pink noise  (left) and music (right) for a Behringer ECM8000 microphone with omnidirectional directivity: experimental points (crosses).}
\end{figure}

\begin{figure}[h!t]
\begin{center}
\includegraphics[width=7.5cm]{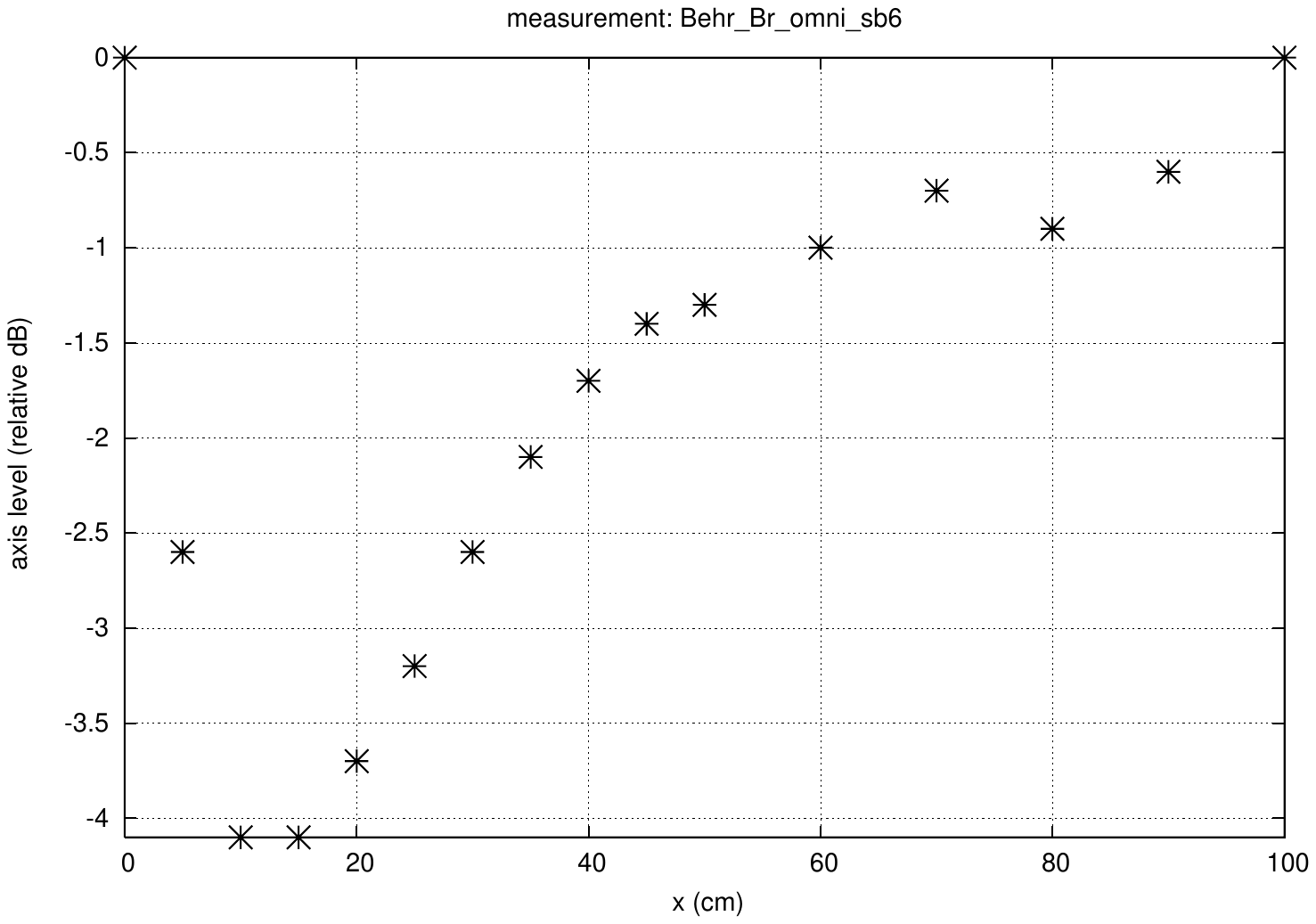}
\includegraphics[width=7.5cm]{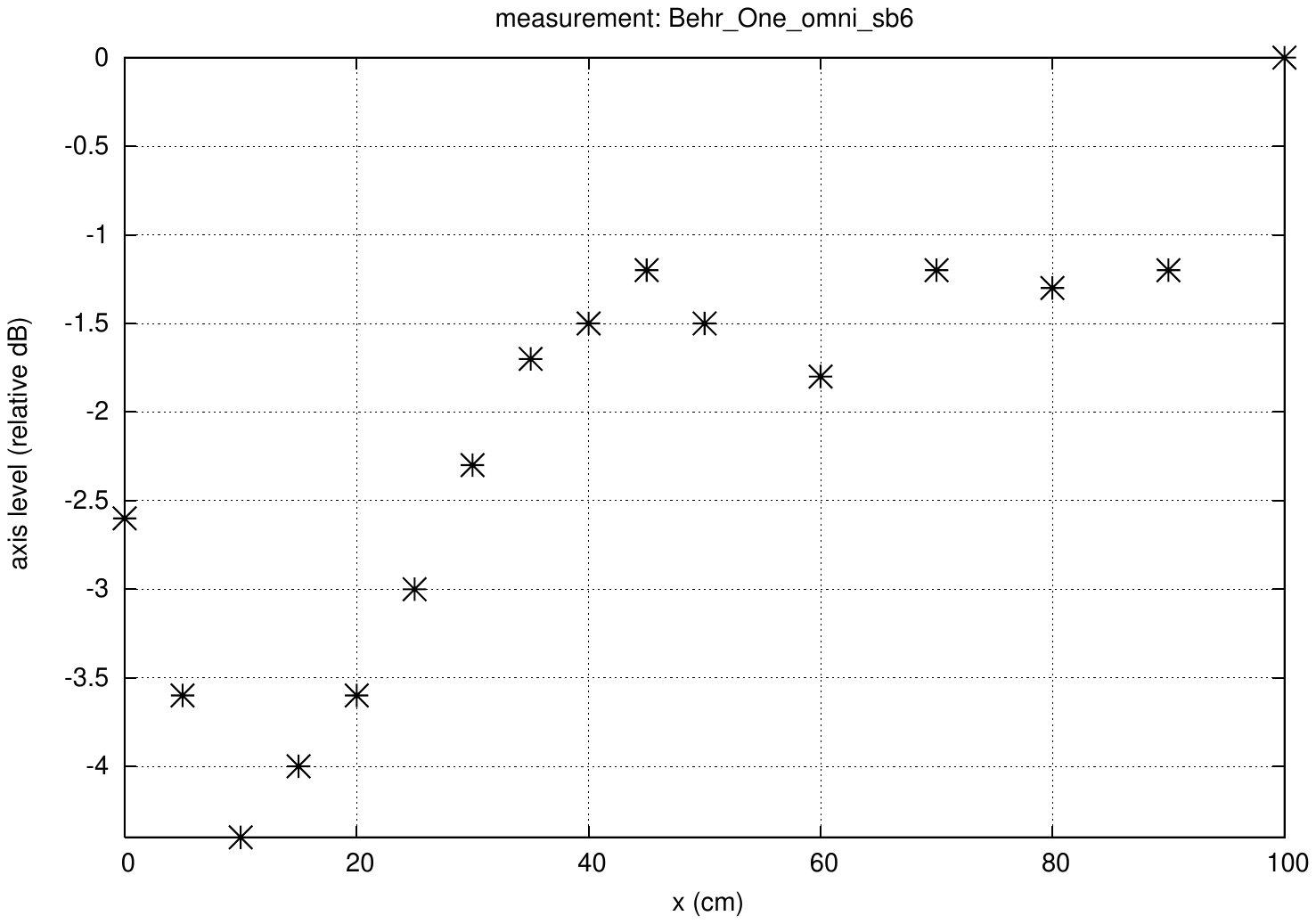}
\end{center}
\caption[14]{Evolution of the weight in the spectral balance of subband  6 as a function of $x$; pink noise  (left) and music (right) for a Behringer ECM8000 microphone with omnidirectional directivity: experimental points (crosses).}
\end{figure}

\pa With Fig. 15 we have the information for subband 7 (1.2-3~kHz), associated with the greatest sensibility frequency range for hearing. The curves for pink noise stimulus and music one are rather different but both present a first increase of the weight (between 0 and 5 or 10 ~cm), a decrease of the weight (quasi linear for pink noise stimulus up to 60~cm; after a constant weight range from 10 to 25~cm for music stimulus followed by a decrease of the weight from 25 to 60~cm) and finally a quasi constant range from 60 to 100~cm for both stimuli.

\pa We find a violent increase and a gradual decrease in both cases, just before a quasi constant weight range so the "validity limit" for wave approximation could be chosen equal to 60~cm for this subband.   

\begin{figure}[h!t]
\begin{center}
\includegraphics[width=7.5cm]{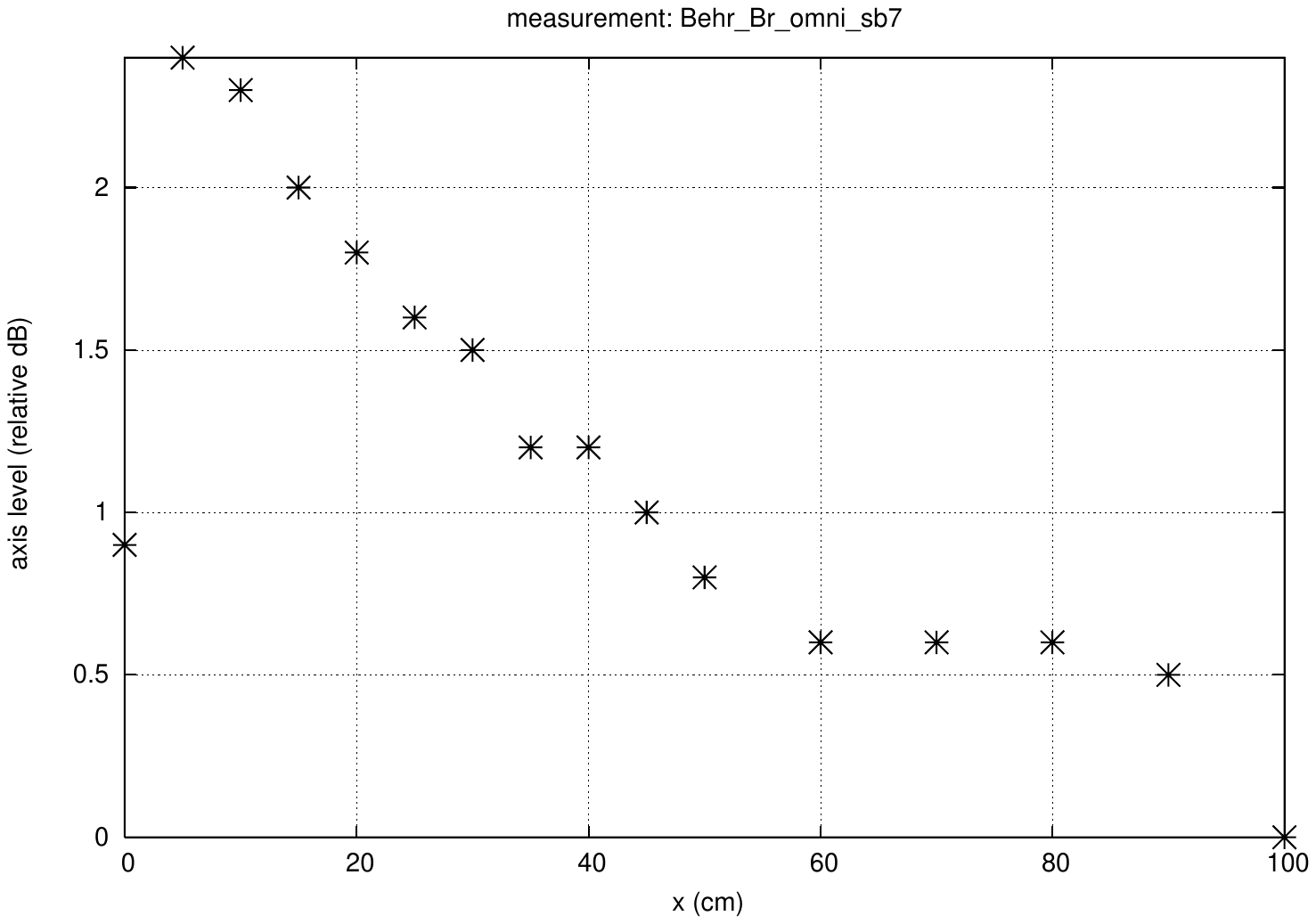}
\includegraphics[width=7.5cm]{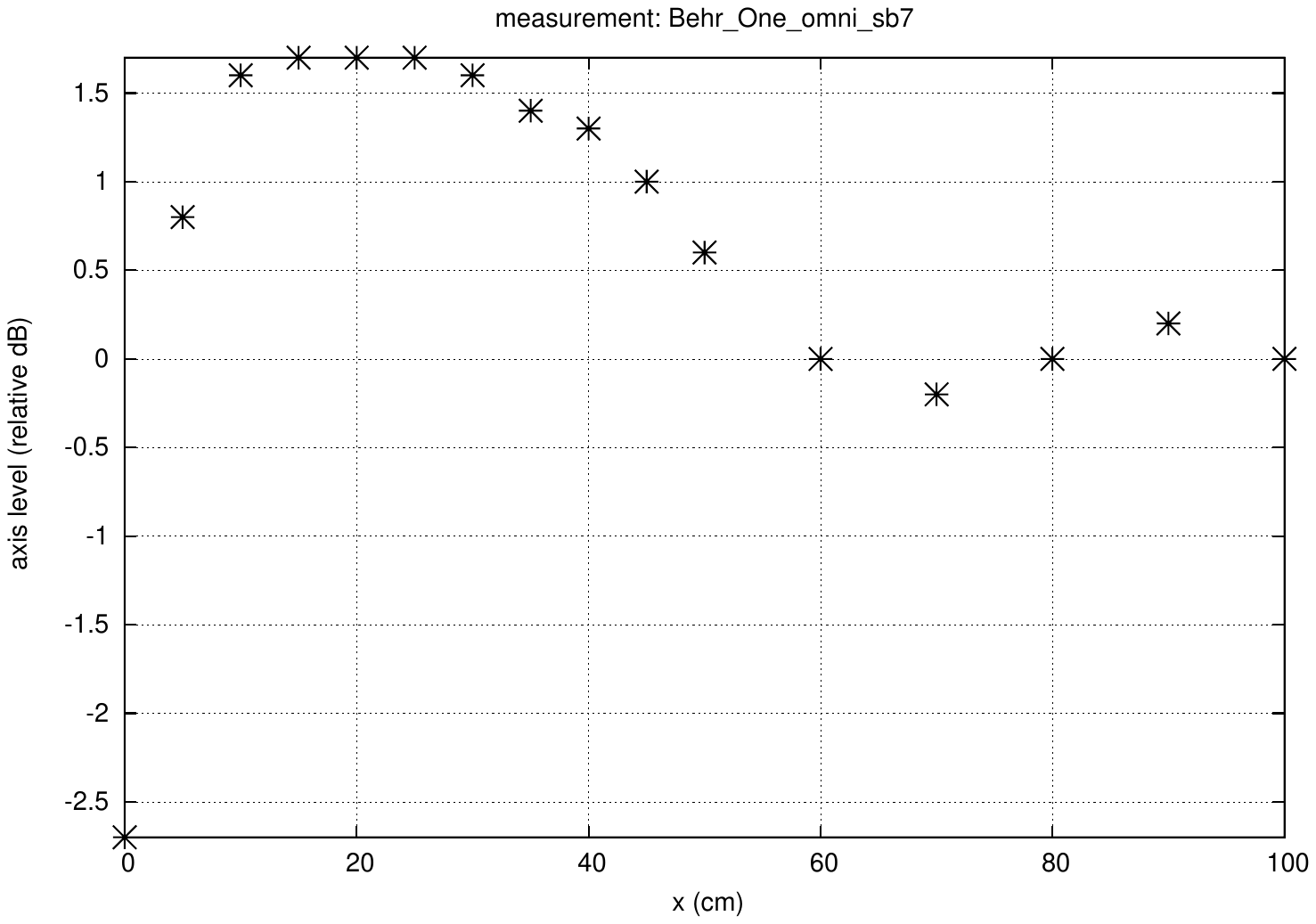}
\end{center}
\caption[15]{Evolution of the weight in the spectral balance of subband  7 as a function of $x$; pink noise  (left) and music (right) for a Behringer ECM8000 microphone with omnidirectional directivity: experimental points (crosses).}
\end{figure}

\pa Fig. 16 to 18 present the evolution of relative weight for subband 8 (3-6~kHz), 9 (6-15~kHz) and 10 (15-22.05~kHz). 

\pa For music stimulus, these three subbands have a simi\-lar behavior that is to say an increase of the weight from 0 to 20/25~cm followed by such a "spatial modu\-lation" with a global magnitude around 1~dB, which perhaps may not be significant or rather significant even if we are studying mean or global weight for each subband. 

\pa If we consider non significant such a spatial modulation, the "validity limit" could then be chosen around 20/25~cm for subband 8 to 10 and, in the case of the music stimulus.

\begin{figure}[h!t]
\begin{center}
\includegraphics[width=7.5cm]{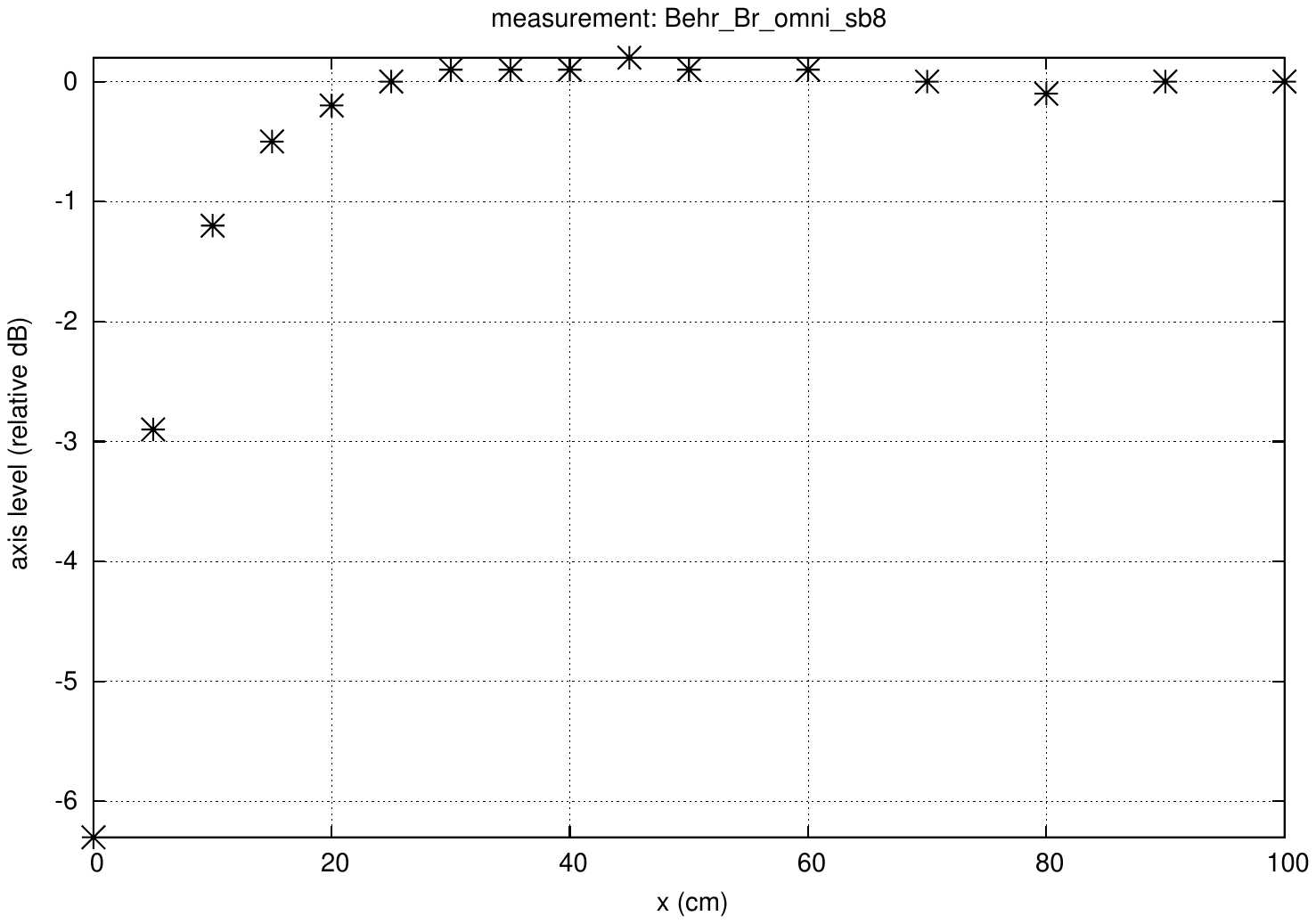}
\includegraphics[width=7.5cm]{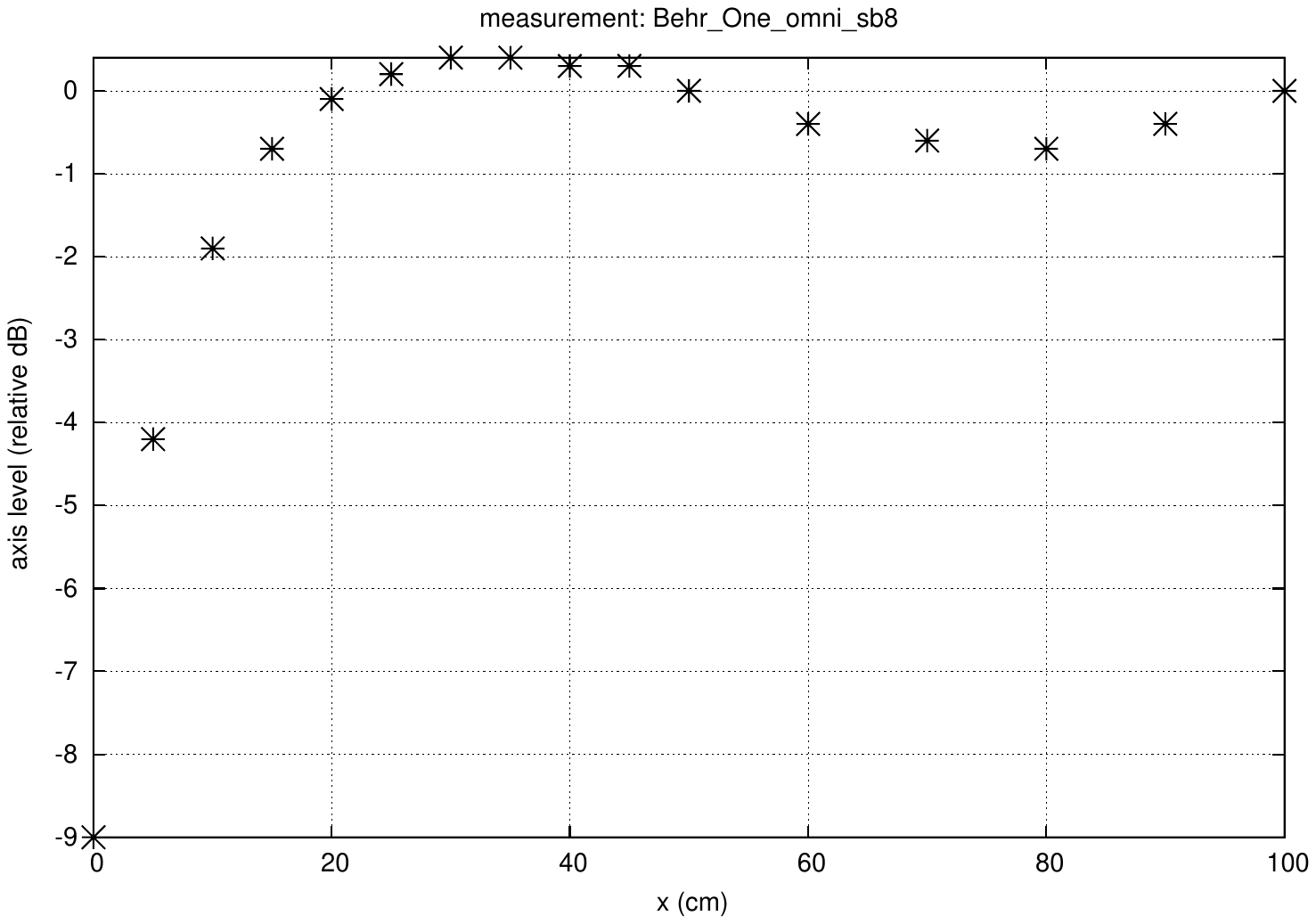}
\end{center}
\caption[16]{Evolution of the weight in the spectral balance of subband  8 as a function of $x$; pink noise  (left) and music (right) for a Behringer ECM8000 microphone with omnidirectional directivity: experimental points (crosses).}
\end{figure}

\pa Considering the case of the pink noise stimulus for subbands 8 to 10 we can notice that  we have similar behavior for subbands 8 and 9 and that this behavior is also similar to the one found for the music stimulus : a fast increase from 0 to 20/25~cm followed by some spatial modulation of magnitude lower than 0.5~dB, which may be considered as a quasi constant weight situation.

\pa On the contrary, the evolution of subband 10 weight is rather different. Indeed, if we refind a fast increase of the weight for first two points (0 and 5~cm), we then encounter a fast decrease from 5 to around 20/25~cm followed by small variations from 20/25 to 100~cm. 
 
\pa One more time, we have a very different behavior for pink noise and music stimuli. 

\begin{figure}[h!t]
\begin{center}
\includegraphics[width=7.5cm]{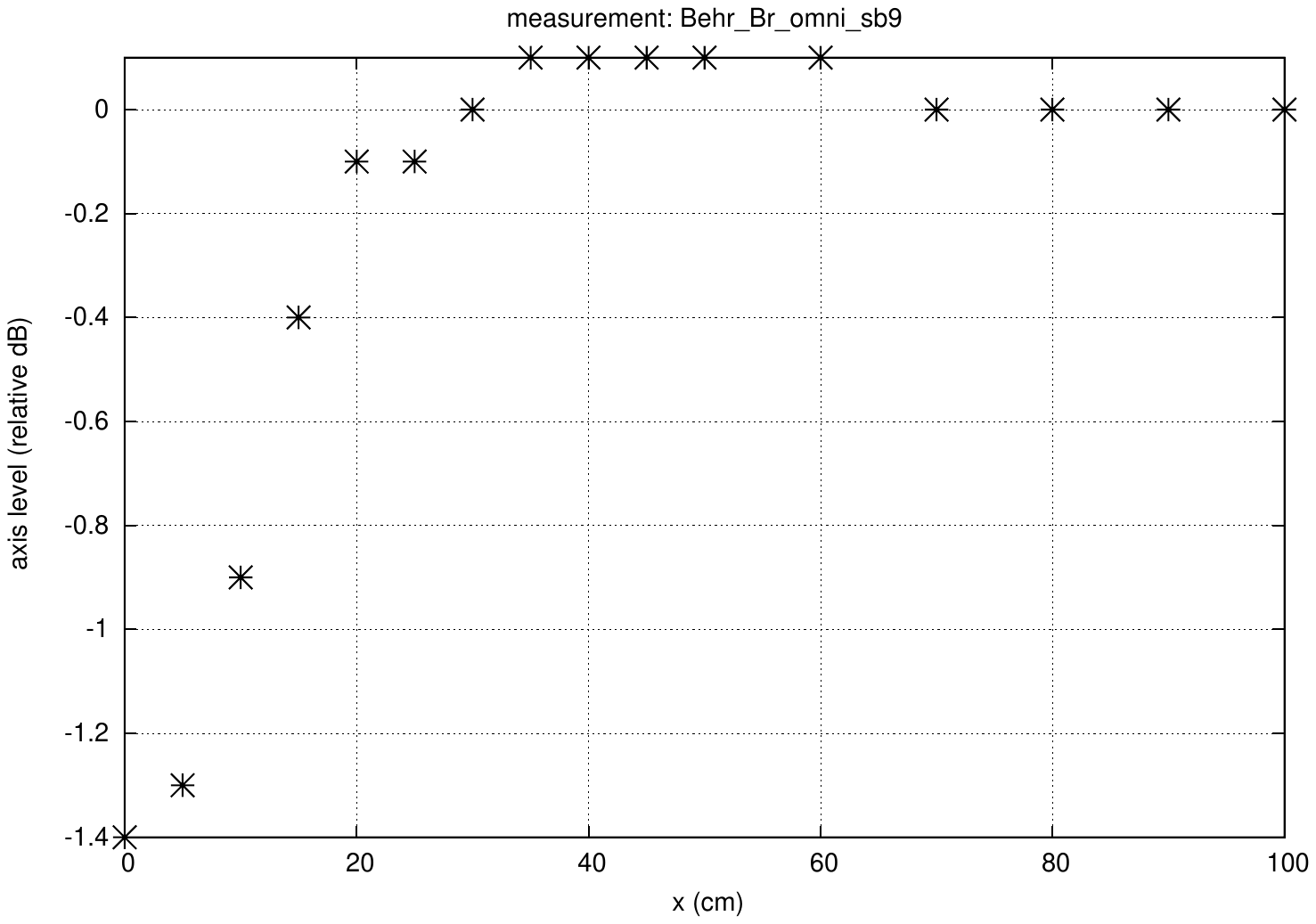}
\includegraphics[width=7.5cm]{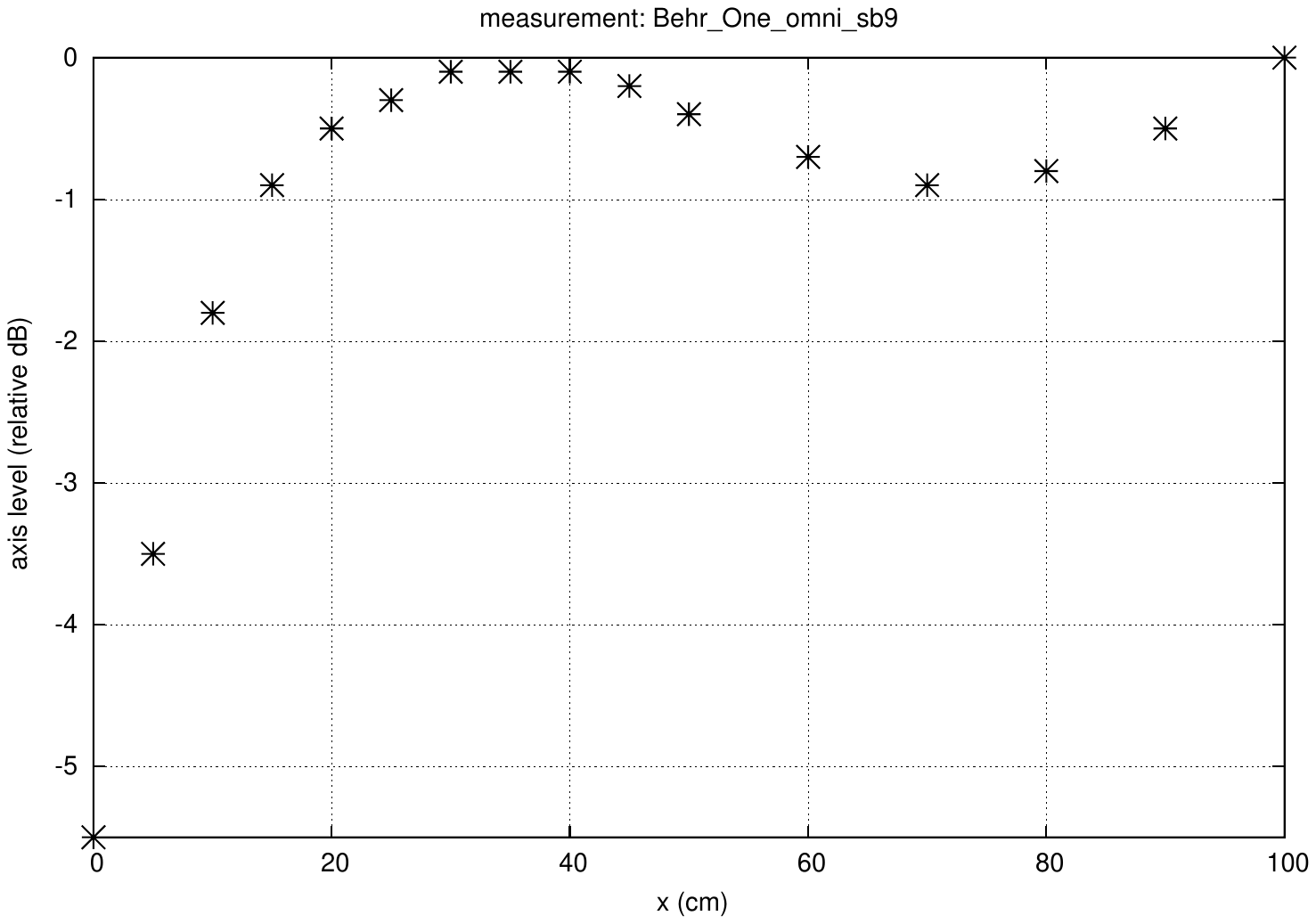}
\end{center}
\caption[17]{Evolution of the weight in the spectral balance of subband  9 as a function of $x$; pink noise  (left) and music (right) for a Behringer ECM8000 microphone with omnidirectional directivity: experimental points (crosses).}
\end{figure}

\pa We should consider again the idea that the music stimulus case is rather more representative of a recording situation and not forget that with music we have a signal with temporal coherence which is not the case for pink noise. So, it should be interesting to do again the same experiment using another music stimuli and also a bigger anechoic chamber to get some more information about the evolution of the weight which mimics a "spatial modulation" but would not be considered as an eigenmode phenomenon, according to us. 

\pa Changing the size of the anechoic room would permit to determine if this so-called "spatial modulation" depends of the room size and characteristics or if it is an inner characteristic of the physical phenomenon.

\begin{figure}[h!t]
\begin{center}
\includegraphics[width=7.5cm]{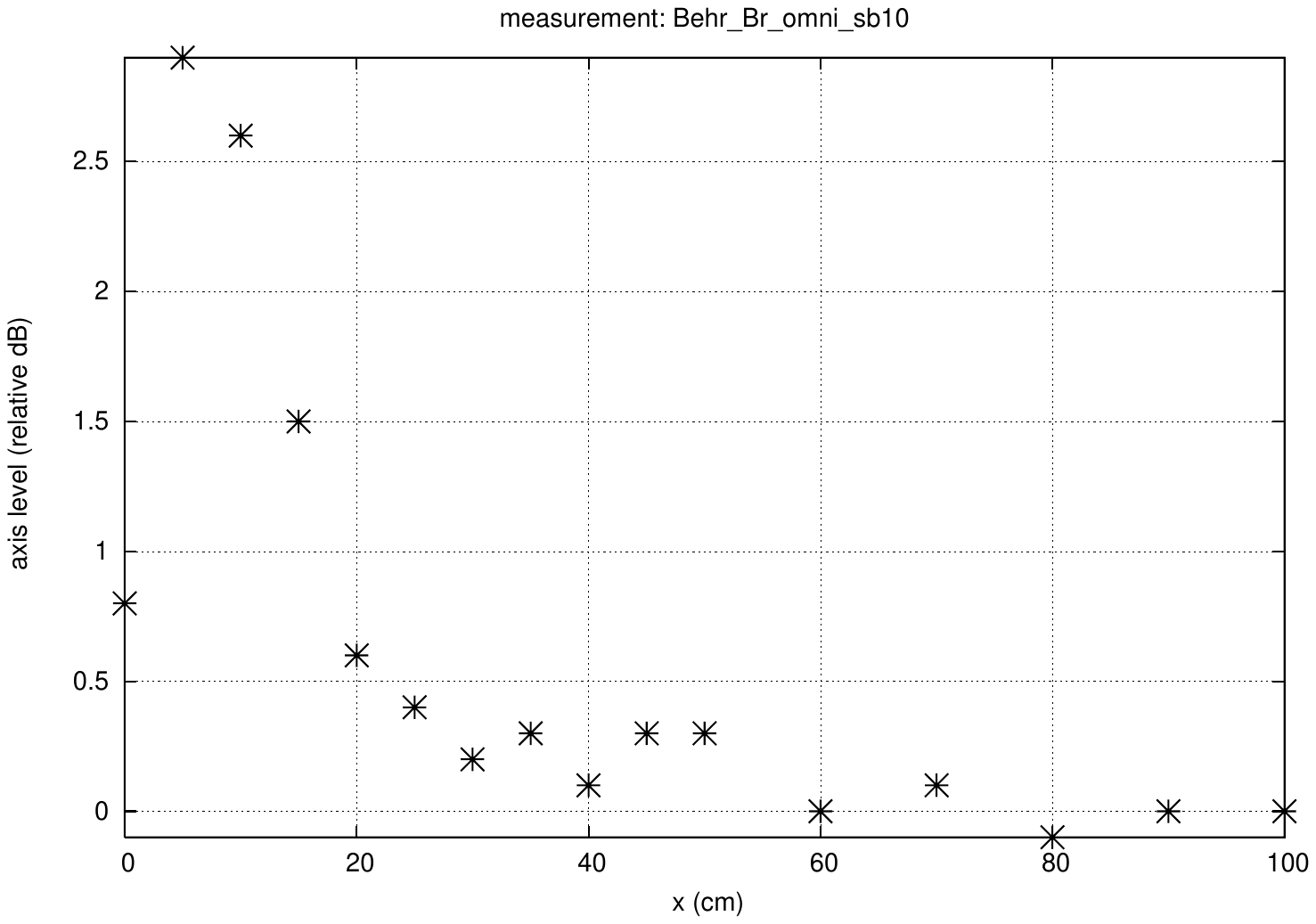}
\includegraphics[width=7.5cm]{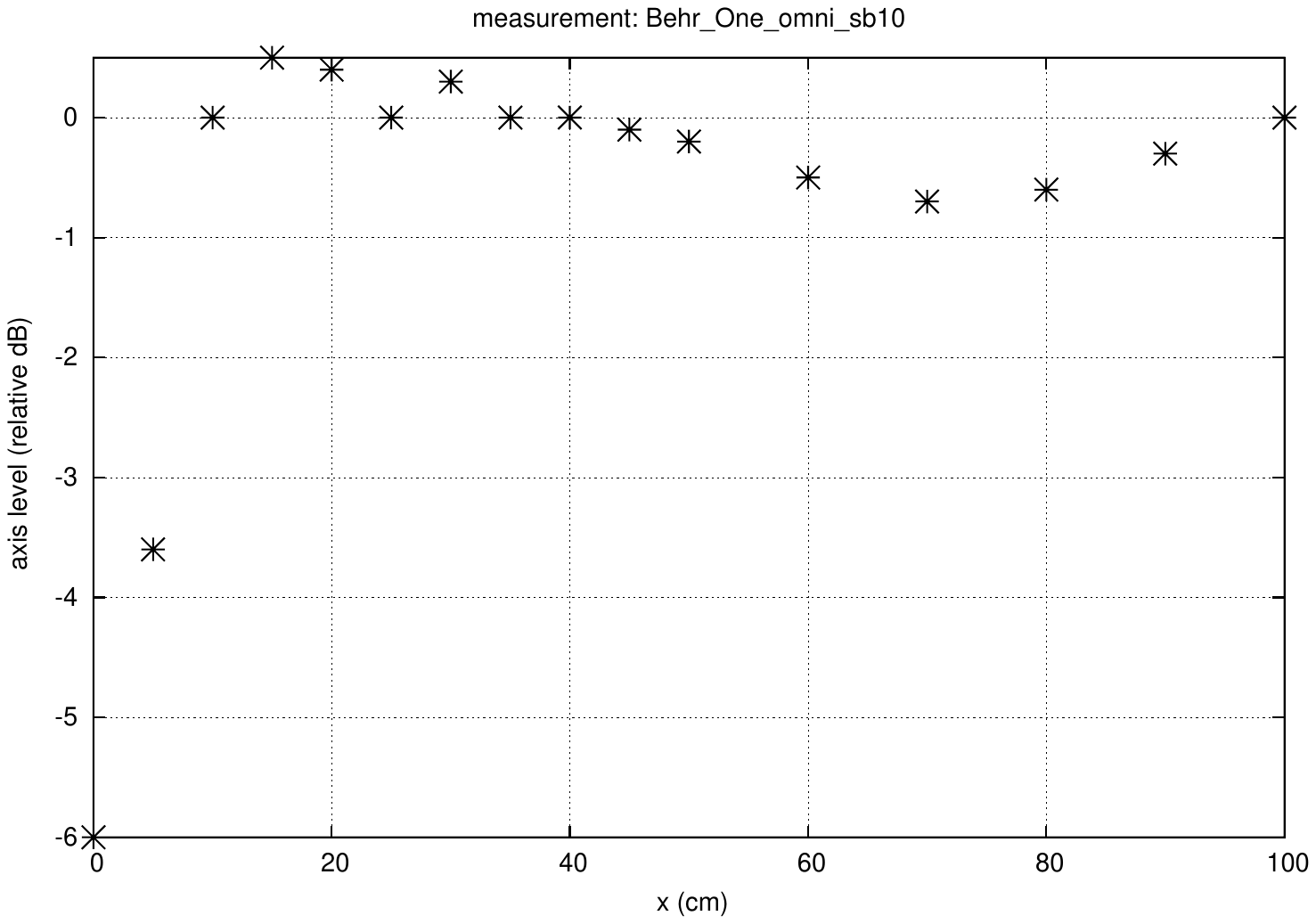}
\end{center}
\caption[18]{Evolution of the weight in the spectral balance of subband  10 as a function of $x$; pink noise  (left) and music (right) for a Behringer ECM8000 microphone with omnidirectional directivity: experimental points (crosses).}
\end{figure}

\pa So, reviewing the experimental results for subbands 1 to 10, above all for music stimulus which seems more pertinent, we can underline the following conclusions:
\begin{itemize}
\item we still assume that the measurements made for the ECM8000 omnidirectional microphone give access to the "real" pressure field or physical phenomena so, in the following we keep on consi\-dering these measurements as reference when studying what happens for the other microphones;

\item as discussed previously, we propose to consider the music stimulus as a good indicator giving access to informative knowledge about the behavior of the microphone in the perspective of real recordings;

\item although the ECM8000 gives the information about the "real" pressure field, we have found a significant proximity effect with a bass boost for the 0-200~Hz frequency range (subbands 1 and 2) and a so-called "re-equalization"  for the other subbands (perhaps except for subband 4) which is associated with an increasing weight at last over a short range of distance with some diminution of the weight (subbands 3, 4 and 7) or at least some variations (sometimes significant) of the weight when the distance from the source increases;

\item for subbands 1 to 6 (0-1800~Hz frequency range) we think we cannot define a distance over which the wave approximation could be a fair approximation (the so-called "validity limit") because the weight variations are too big over the whole tested distance range from the source;

\item for other subbands we can propose a so-called "validity limit" but it seems to vary according to the considered subband:  60~cm for subband 7 (1.8-3~kHz) or 20/25~cm for subbands 8 to 10 (3-22.05~kHz);

\item we must recall that studying the mean level amplifications curves for the same microphone (see Fig. 3), we have found a so-called "validity limit" around 50~cm which means that to determine the nature of the physical phenomenon, and thus the validity of the wave assumption, we should consider both mean level and all subband weight curves, but, which may also be an indicator of the weight of the subband 7 corresponding to the most sensitive frequency range for human perception;

\item we have found a proximity effect for the "real" pressure field and this effect is rather different from the one described for directional microphone as we can note an increase of the weight of low frequencies after a minimum and often a decrease for higher frequencies which all vary with the distance from the source according to the subband;

\item such phenomenon is not compatible with dissipative wave description because viscous losses need several meters to have a significant effect, but this phenomenon corresponds to a "re-equalization" of the spectral balance for longer tested distances, which needs another physical explanation. 
\end{itemize}

\pa We have now some information, which are taken as reference in the following, to study the behavior of the other tested microphones. 

\pa We focus our attention on the most representative subbands and only consider the case of the music stimulus as we assume it to be more informative in a recording prospect.

\pa Readers interested to get the information about the other cases (subbands not given here and expe\-rimental results for the pink noise stimulus) may contact the main author by email at following email address: contacts.lmillot@free.fr.

\subsubsection{Comparisons of omnidirectional microphones}
\pa In this paragraph, we compare both omnidirectional microphones:  ECM8000, considered as a pressure sensor, and U89i used with an omnidirectional directivity. We only give the information about the most representative subbands: subbands 1, 2, 5 and 8. It is important to keep in mind the fact that the minimum tested distance from the source is different for both microphones:  5~cm for the U89i microphone, 0~cm for the ECM8000.

\pa Indeed, differences of behavior for subband 3, 4 and 6 are rather similar to the ones found for subband 5 while differences of behavior for subband 8 are quite representative of the ones found for subbands 7, 9 and 10.

\pa Fig. 19 and 20 show what happens respectively for subbands 1 and 2 for both omnidirectional microphones.

\pa We can note that, in both cases, the curves have a similar look: a fast decrease followed by a quasi mini\-mum range (subband 1) or location  
(subband 2). And the minimum location varies with the subband and, for subband 1, with the microphone: 35~cm for subband 1 and U89i microphone, 50~cm for subband 1 and ECM8000 microphone; 70~cm for both microphones and for subband 2. After this minimum distances range, we have a significant weight increase for $x$ varying from the minimum distance to 1~m.  

\begin{figure}[h!t]
\begin{center}
\includegraphics[width=7.5cm]{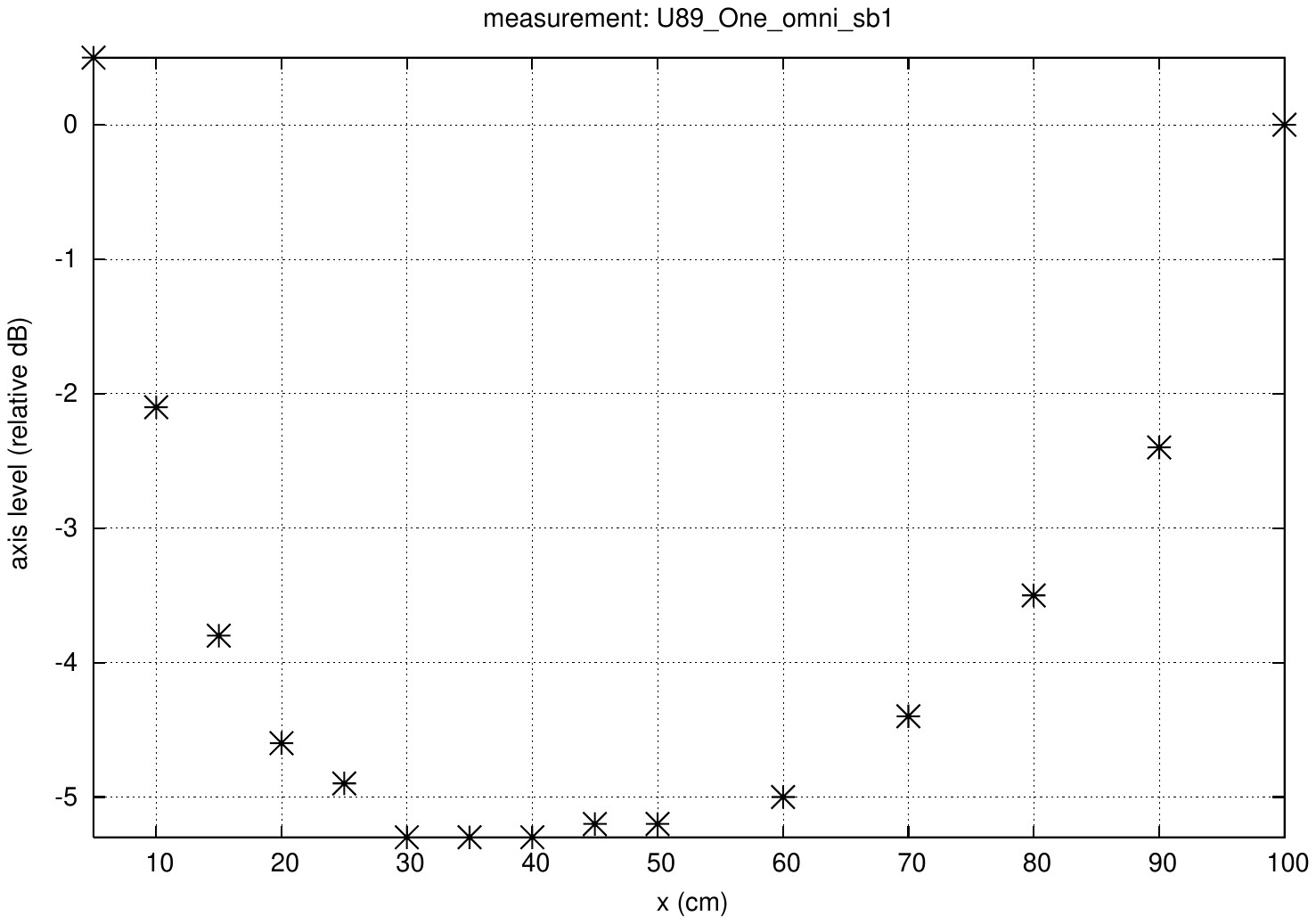}
\includegraphics[width=7.5cm]{Behr_One_omni_sb1.pdf}
\end{center}
\caption[19]{Evolution of the weight in the spectral balance of subband 1 as a function of  $x$; music and omnidirectional microphones: (left) Neumann U89i and Behringer ECM8000 (right).}
\end{figure}

\pa Global magnitudes of the  weight variations seem rather similar for both subbands but the weight at the 5~cm distance from the source depends on the microphone and one can note that this weight is bigger for the reference, the ECM8000 microphone, for both subbands (with a small difference for subband 1). This may mean that the design of the U89i microphone could have been made intending to attenuate a little the bass boost or proximity effect for low frequency compared to the weight found for a distance from the source equal to 1~m. But, if there is almost not increase in level at 5~cm location, for following distances from the source, we have greater decreases of the weight compared to the 1~m one: so increasing the distance from the source would have a great influence on the proximity effect for the 30 first centimeters (attenuation of almost 6~dB for the global weight) for subband 1 and the same attenuation but for the the first 70 centimeters for subband 2.

\begin{figure}[h!t]
\begin{center}
\includegraphics[width=7.5cm]{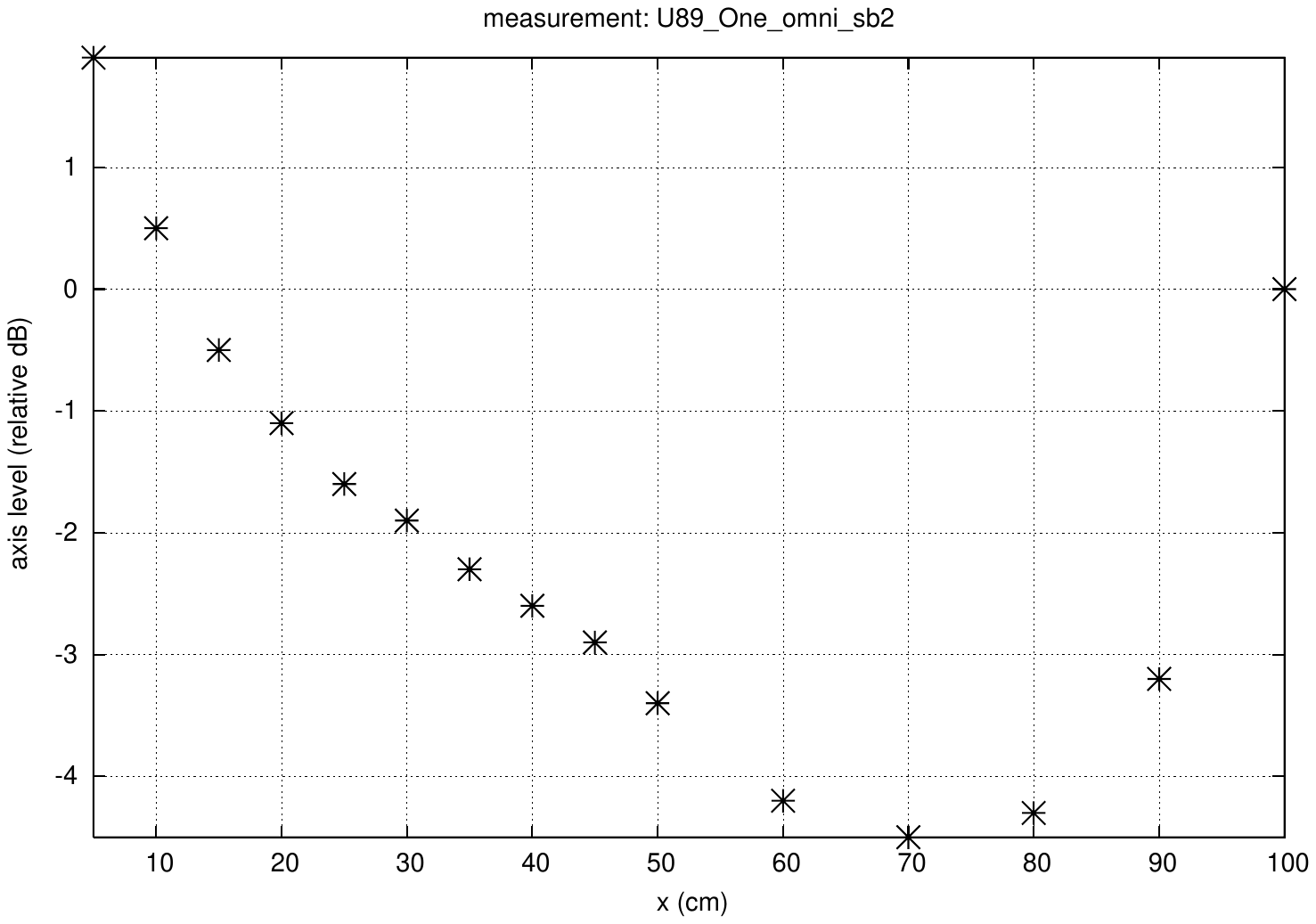}
\includegraphics[width=7.5cm]{Behr_One_omni_sb2.pdf}
\end{center}
\caption[20]{Evolution of the weight in the spectral balance of subband 2 as a function of  $x$; music and omnidirectional microphones: (left) Neumann U89i and Behringer ECM8000 (right).}
\end{figure}

\pa The small difference between both curves for subband 1 may be understood by the fact that this frequency range is often not used (U89i microphone frequency range is given to be 20~Hz-20~kHz) and would, by the way, be quite attenuated using a popscreeen or a windscreen.

\pa We refind that the phenomenon depends at least of the frequency range and that we cannot consider a spatial range (under 1~m)
where the wave assumption can represent a fair approximation for both microphones and both subbands.

\pa Fig.  21 gives the result for subband 5, representative of what happens also for subbands 3 and 4, and almost representative of weight evolutions for subband 6. 

\begin{figure}[h!t]
\begin{center}
\includegraphics[width=7.5cm]{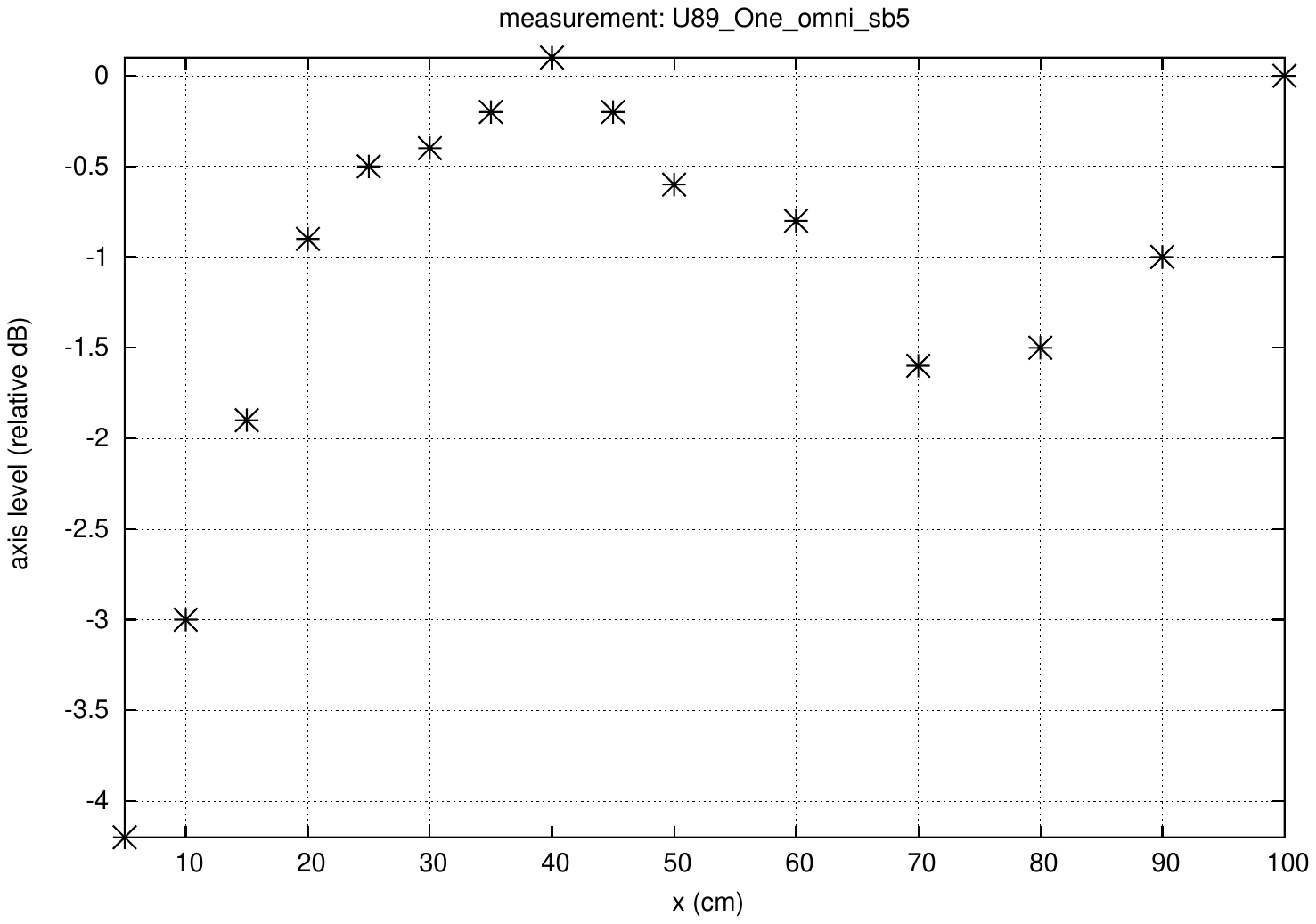}
\includegraphics[width=7.5cm]{Behr_One_omni_sb5.pdf}
\end{center}
\caption[21]{Evolution of the weight in the spectral balance of subband 5 as a function of  $x$; music and omnidirectional microphones: (left) Neumann U89i and Behringer ECM8000 (right).}
\end{figure}

\pa We find a rather fast increase of the weight for both microphone with a maximum located at 40~cm for the U89i microphone and at 35~cm for the ECM8000 microphone. After, we refind the so-called "spatial modulation" with a magnitude of around 1.5~dB for both microphones. For subband 3 and 4 the "spatial modulation" is bigger, and similar for subband 6, so we think that we cannot assume that these "spatial modulations" are negligible. 

\pa With this assumption, we refind the fact that the wave approximation may not be valid under 1~m for at least the frequency range of 0-1800~Hz. 

\pa Moreover, the whole magnitude of the weight evolution is bigger for U89i microphone for subbands 3, 4 and 6 while it is lower for subband 5, with a maxi\-mal difference of magnitude around to 2~dB. So except for subband 6, the U89i microphone design seems to amplifies the variation of the spectral balance for frequency ranges of 200-800~Hz and 1.2-1.8~kHz, when the distance from the source varies, and to reduce this variation for frequency range of 800-1200~Hz. 

\pa Fig. 22 gives the evolutions of weight for subband 8 which is also representative of what happens for subbands 7, 9 and 10.

\begin{figure}[h!t]
\begin{center}
\includegraphics[width=7.5cm]{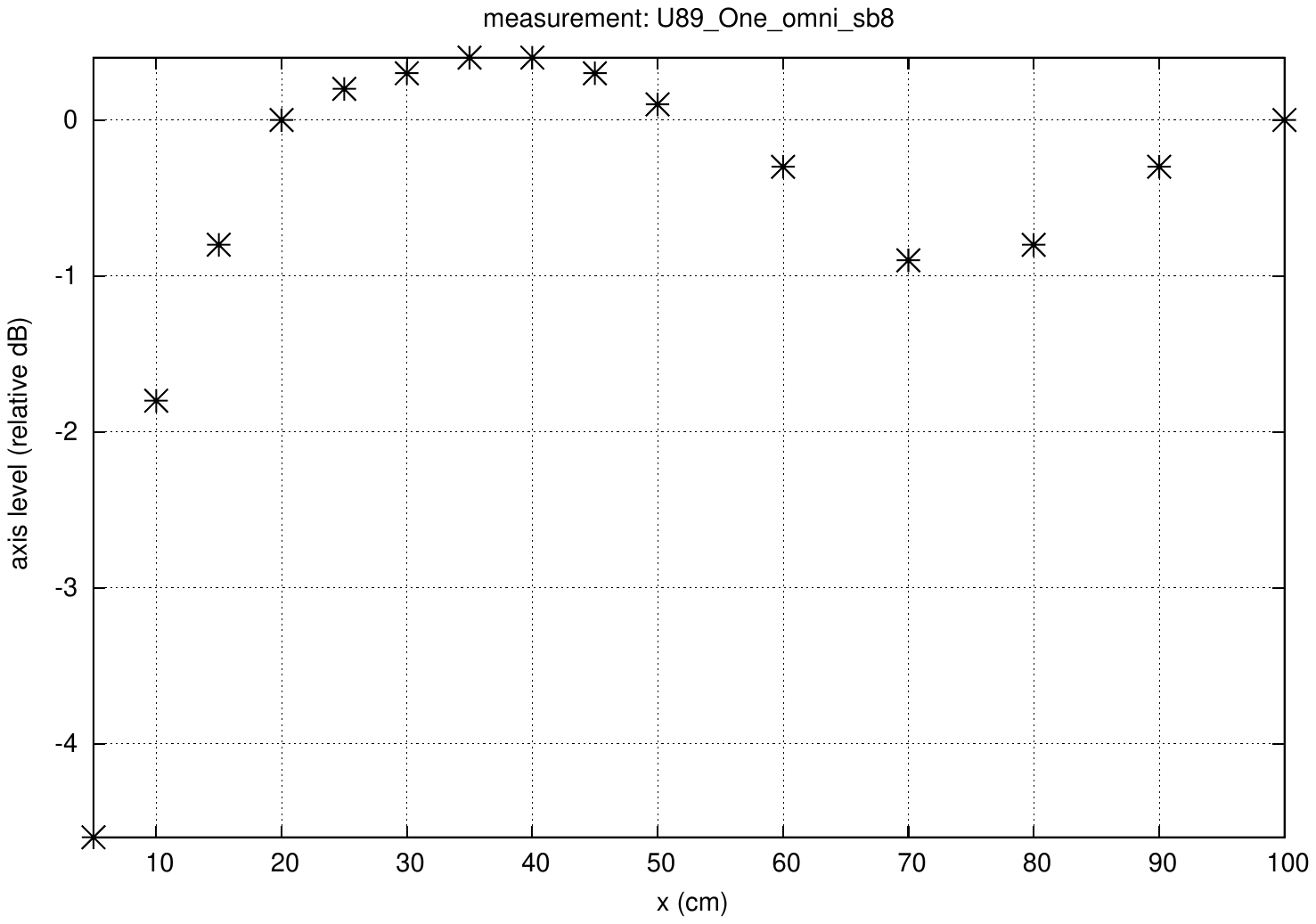}
\includegraphics[width=7.5cm]{Behr_One_omni_sb8.pdf}
\end{center}
\caption[22]{Evolution of the weight in the spectral balance of subband 8 as a function of  $x$; music and omnidirectional microphones: (left) Neumann U89i and Behringer ECM8000 (right).}
\end{figure}

\pa We can underline the fact that the curves have a similar look for both microphones.  We first find a rather fast increase of the weight up to a maximum located at 25~cm from the source and followed by a so-called "spatial modulation" up to the 1~m distance but with a modulation magnitude lower than 1~dB, which is also the case for subbands 7 (but with a bigger modulation magnitude, less than 1.5~dB, just after the maximum which diminishes just before the minimum to the 1~m point), 9 and 10.

\pa We can note that the whole magnitude is greater for ECM8000 microphone compared to U89i one, a conclusion which is also valid for subbands 7, 9 and 10 (for which the U89i microphone presents a significant decrease after the maximum and before a quasi constant weight distance range). This may mean that the design of the U89i, for the omnidirectional directivity, could have be made intending to reduce a little the influence of the proximity effect (associated with the first rise of the weight) and in a sense the effect of the whole variation of the weight for the tested distance range.

\pa Due to the spatial magnitude lower than 1~dB  after the minimum for subbands 8, 9 and 10 or for distances $x\geq 50$~cm for subband 7 (for the U89i microphone), we refind the idea that for these distance ranges, the spherical wave model may be a fair assumption. 

\pa But, for these measurements made with these two microphones, the spherical wave model is only valid over at least a 20/25~cm distance (50~cm for subband 7 and 60~cm for subband 10 for U89i microphone) and only for frequencies upper than 1.8~kHz, which means that the spherical model cannot be used for the whole audio range, even for significant distances from the source one would not consider as its proximity.

\pa This conclusion may be considered as a very bad news for classical acoustical models and assumptions, and for microphone modeling!

\subsubsection{Comparisons for three directivities of Neumann U89i microphone}
\pa We have chosen to present the curves for the three tested directivities (bidirectional, cardioid and omnidirectional) for subbands 1,2, 5 and 8 and only for music stimulus.

 \begin{figure}[h!t]
\begin{center}
\includegraphics[width=5cm]{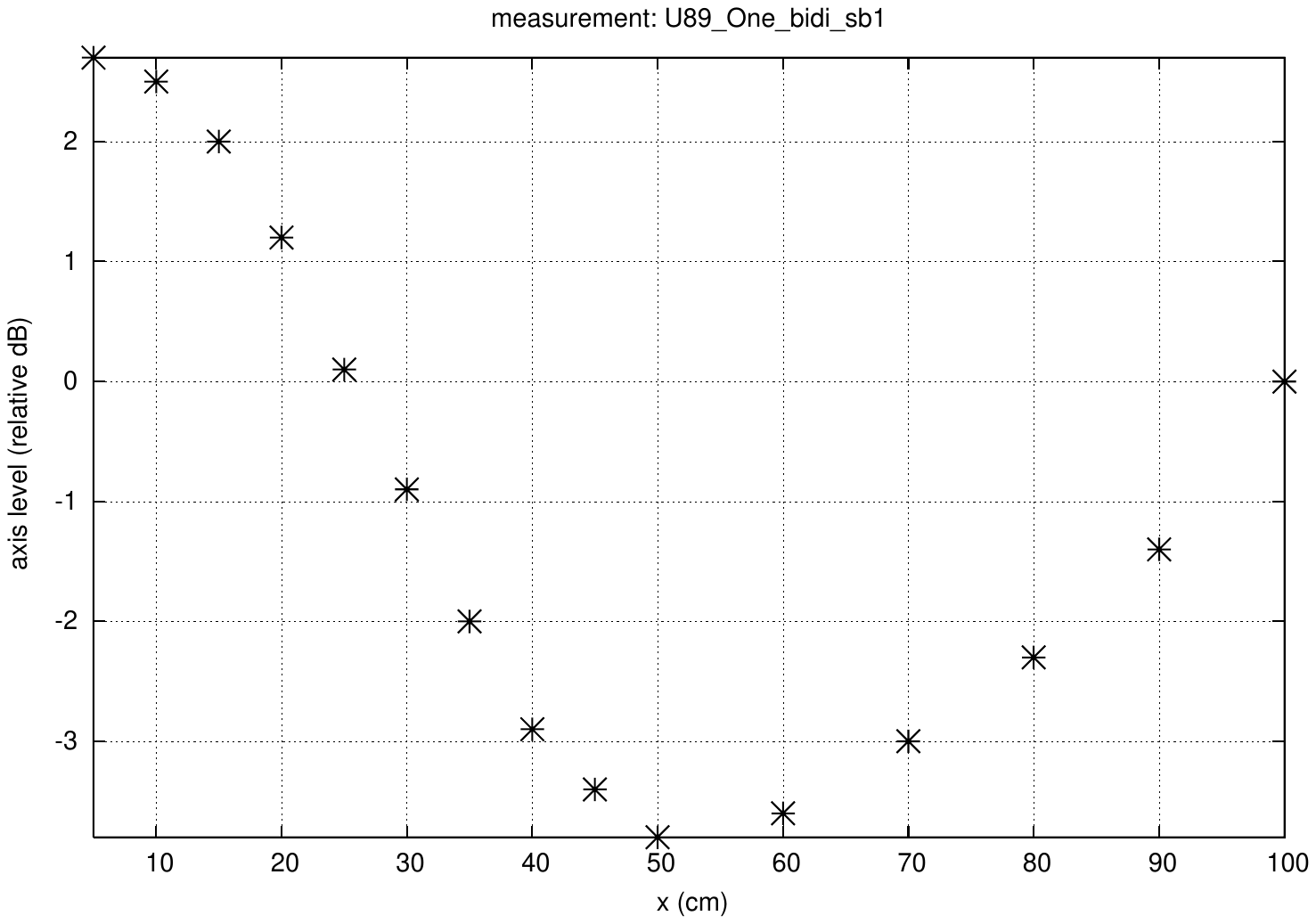}
\includegraphics[width=5cm]{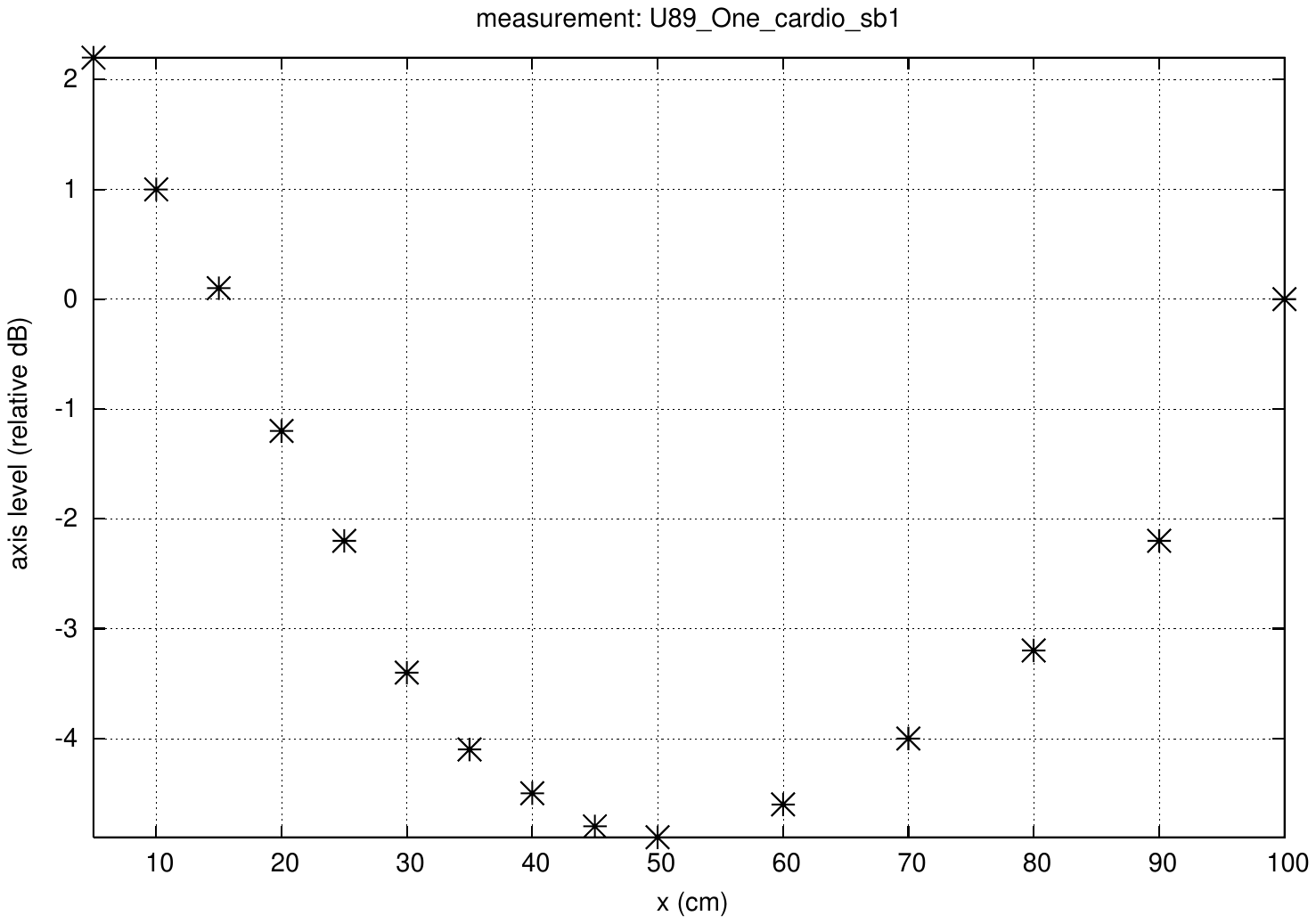}
\includegraphics[width=5cm]{U89_One_omni_sb1.pdf}
\end{center}
\caption[23]{Evolution of the weight in the spectral balance of subband  1 as a function of $x$; music and a Neumann U89i microphone with bidirectional (left), cardioid (center) and omnidirectional (right) directivities: experimental points.}
\end{figure}

\pa Fig. 23 gives the weight evolutions curves for subband 1 (0-50~Hz) with the curve for the cardioid directivity in order to underline the fact that this curve is quite close to an average of both other curves associated with bidirectional and omnidirectional directivities.

\pa This order of presentation of the three curves is kept in the following to  demonstrate that this remark is valid for all subbands. 

\pa Studying the curves of Fig. 23, we can notice that their look is rather similar: a progressive decrease of the weight down to a minimum, whose location varies according to the directivity (50~cm for bidirectional and cardioid directivities, around 35~cm for the omnidirectional one) with similar global weight variation according to the distance from the source. 

\pa We can note that the 5~cm weight is around 3~dB for bidirectional directivity and around 0.5~dB for the omnidirectional directivity. We can also note that the minimum is equal to almost -4~dB for the bidirectional while it is around -5.2~dB for the omnidirectional directivity. And, when considering the curve for the cardioid directivity we refind a weight bigger than 0~dB for $x=5$~cm (a little more than 2~dB) and a low value for the minimum, around -5~dB, which corresponds to the mix, with some attenuation, of the behaviors found for both bidirectional and omnidirectional directivity.

\pa Due to the great variation over the whole distance range for the three directivities, the spherical wave assumption cannot be considered as  valid for the subband 1, which is coherent with the previous results for this subband using notably the ECM8000. 

\pa But, contrary to the omnidirectional case, the weight for $x=5$~cm is a little bigger for U89i microphone with cardioid and bidirectional directivities and the weight for the minimum are bigger for bidirectional and cardioid directivities compared to the case of the ECM8000 microphone. So, using the bidirectional directivity, the proximity effect will be more sensitive in subband 1 for the U89i for a short variation of the distance from the source from the 5~cm location and also when going from the minimum location to the 1~m one. 

\pa So, the proximity effect will be faster for omnidirectional directivity, slower but deeper for cardioid directivity, more progressive for bidirectional directivity but with a upper minimum.  For the distance effect (distance range from 50~cm to 1~m), the weight rise (the so-called "re-equalization") is rather progressive but with stronger rise for omnidirectional, intermediate rise for cardioid directivity and smaller rise for bidirectional microphone.

\begin{figure}[h!t]
\begin{center}
\includegraphics[width=5cm]{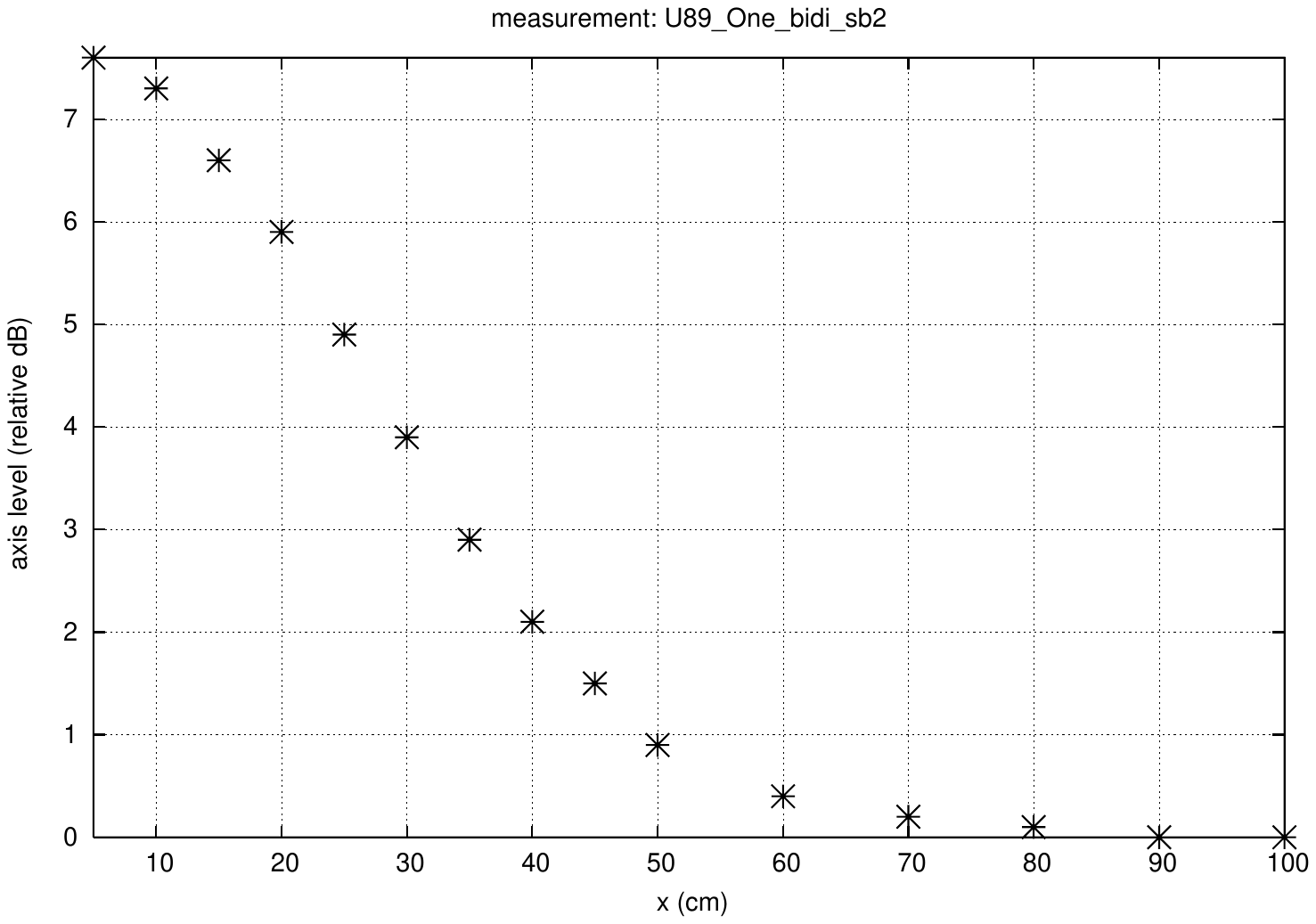}
\includegraphics[width=5cm]{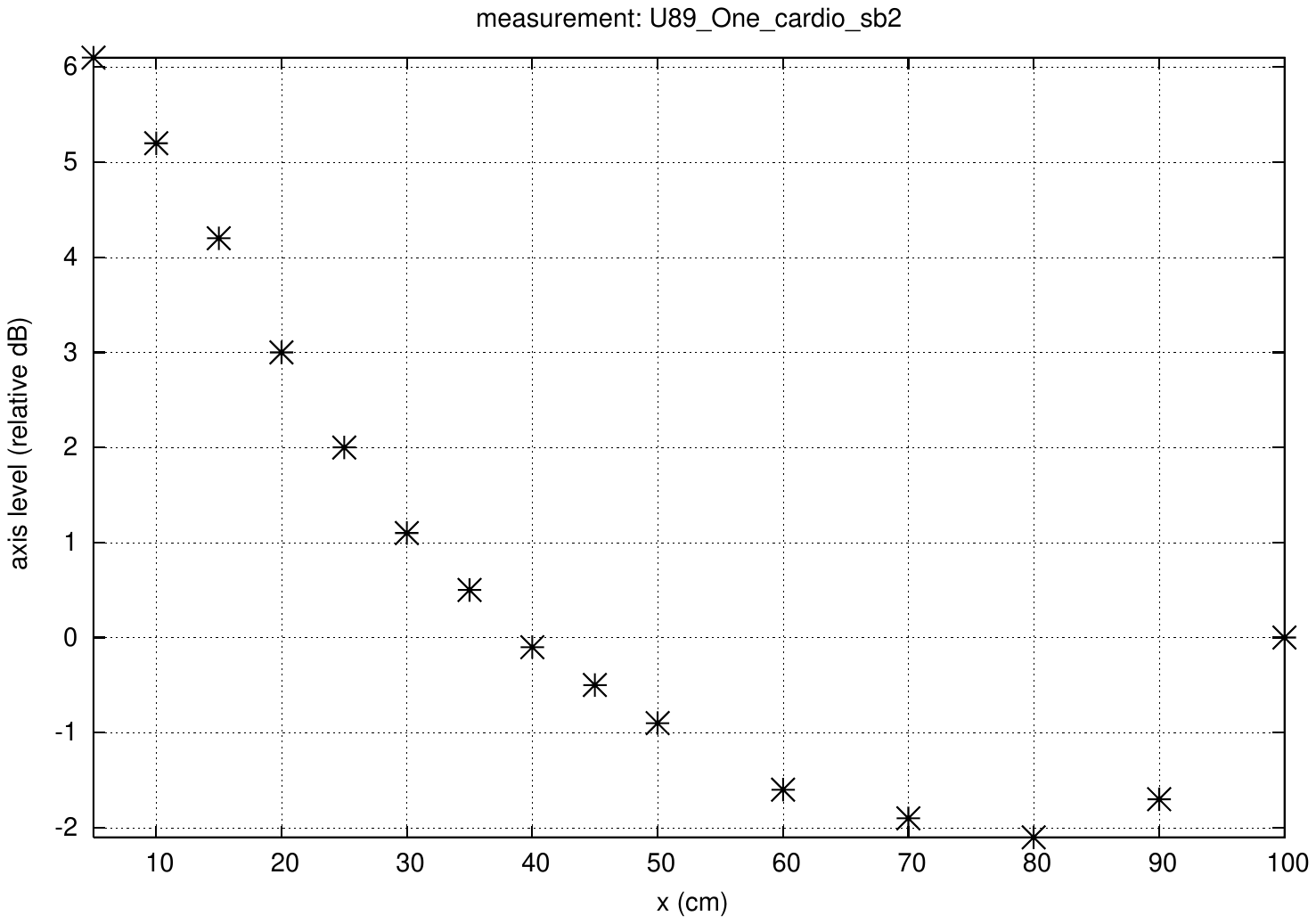}
\includegraphics[width=5cm]{U89_One_omni_sb2.pdf}
\end{center}
\caption[24]{Evolution of the weight in the spectral balance of subband  2 as a function of $x$; music and a Neumann U89i microphone with bidirectional (left), cardioid (center) and omnidirectional (right) directivities: experimental points.}
\end{figure}

\pa  Thus this means that, contrary to what one may think or find in the literature, for the U89i microphone, the directivity which presents the greater weight variations, for proximity and so-called "re-equalization" effects,  is the omnidirectional one not the bidirectional one...

\pa Fig. 24 presents the results for subband 2 (50-200~Hz) still in the low register of the audio range. 

\pa Studying the three curves associated with the three directivities we can refind the principle of a mixing of the curves for omnidirectional and bidirectional directivities to get the one for the cardioid directivity.

\pa For all three directivities, we find a decrease of the weight from $x=5$~cm to the minimum location: $x\approx 90$~cm for bidirectional directivity, $x\approx 70$~cm for omnidirectional directivity and $x\approx 80$~cm for cardioid one. In all three cases, the 5~cm weight is bigger than 0~dB with a weight lower than the one found for the ECM8000 microphone (around 4.5~dB) for the bidirectional directivity.

\pa The minimum for the ECM8000 is -3~dB so the minimum weight is lower for bidirectional and cardioid directivity, bigger for omnidirectional one. 

\pa Just before the minimum weight we find a quasi cons\-tant weight for the bidirectional directivity but refind an increase after the minimum as in the case of the ECM8000 microphone. As said before, the behavior of the cardioid directivity can be explained assuming that this case
corresponds to the mix of the bidirectional and omnidirectional ones.

\pa Except for the case of the bidirectional directivity for which we can postulate a "validity limit" around 60~cm, the two other cases cannot permit to consider the wave assumption as valid, due to the rough rise of the weight after the minimum. 

\pa The bidirectional directivity may corresponds to the prototype of a classical description of the proximity effect while the two other directivities exhibit a non classical rise of the weight, or "re-equalization" effect, after the minimum and a faster decrease of the level down to the minimum weight. 

\pa Considering the weight variation over the whole distance range, the omnidirectional directivity would be associated with the stronger distance effects (proxi\-mity and "re-equalization") which corresponds 
to the proposal made for subband 1.

\begin{figure}[h!t]
\begin{center}
\includegraphics[width=5cm]{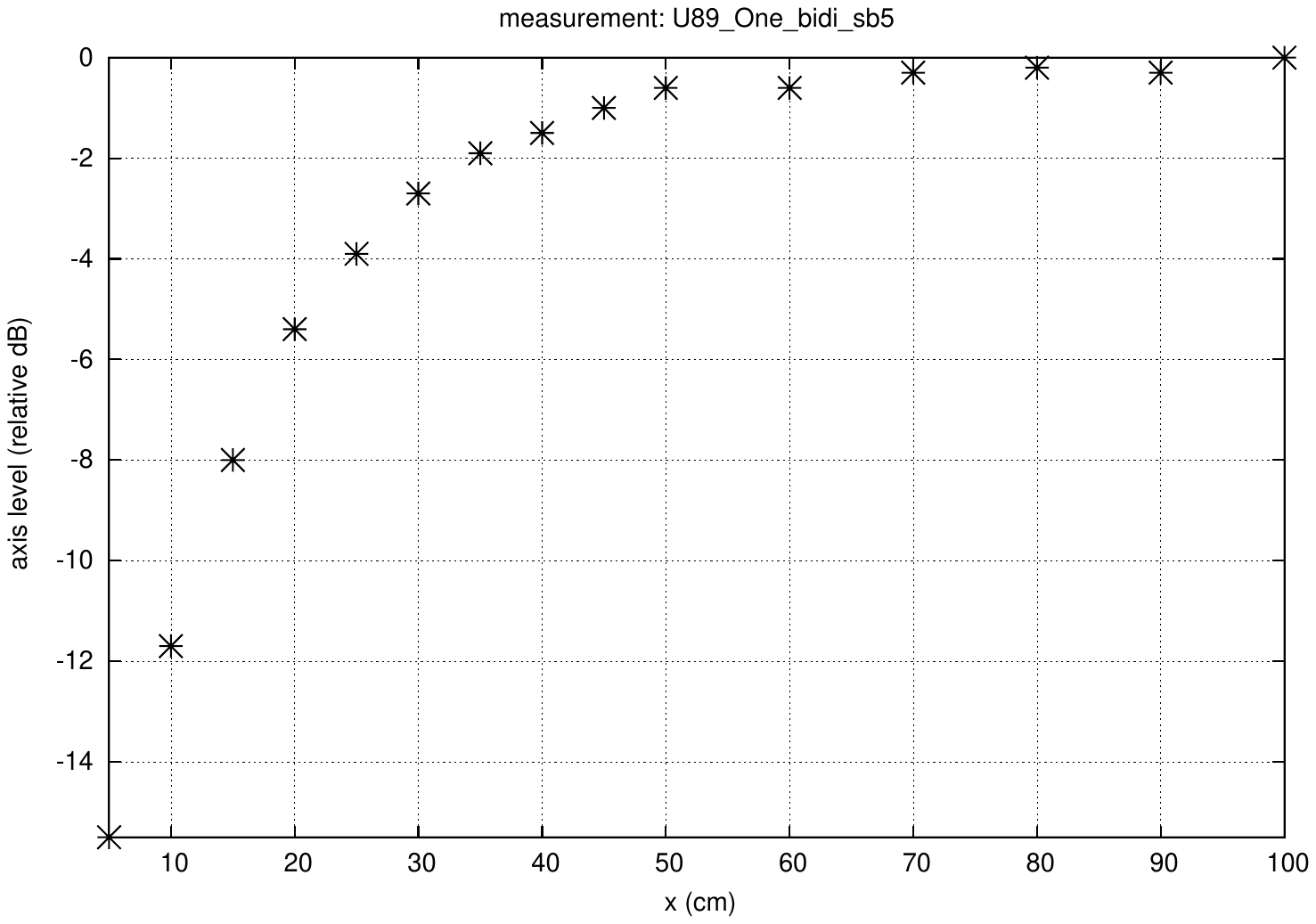}
\includegraphics[width=5cm]{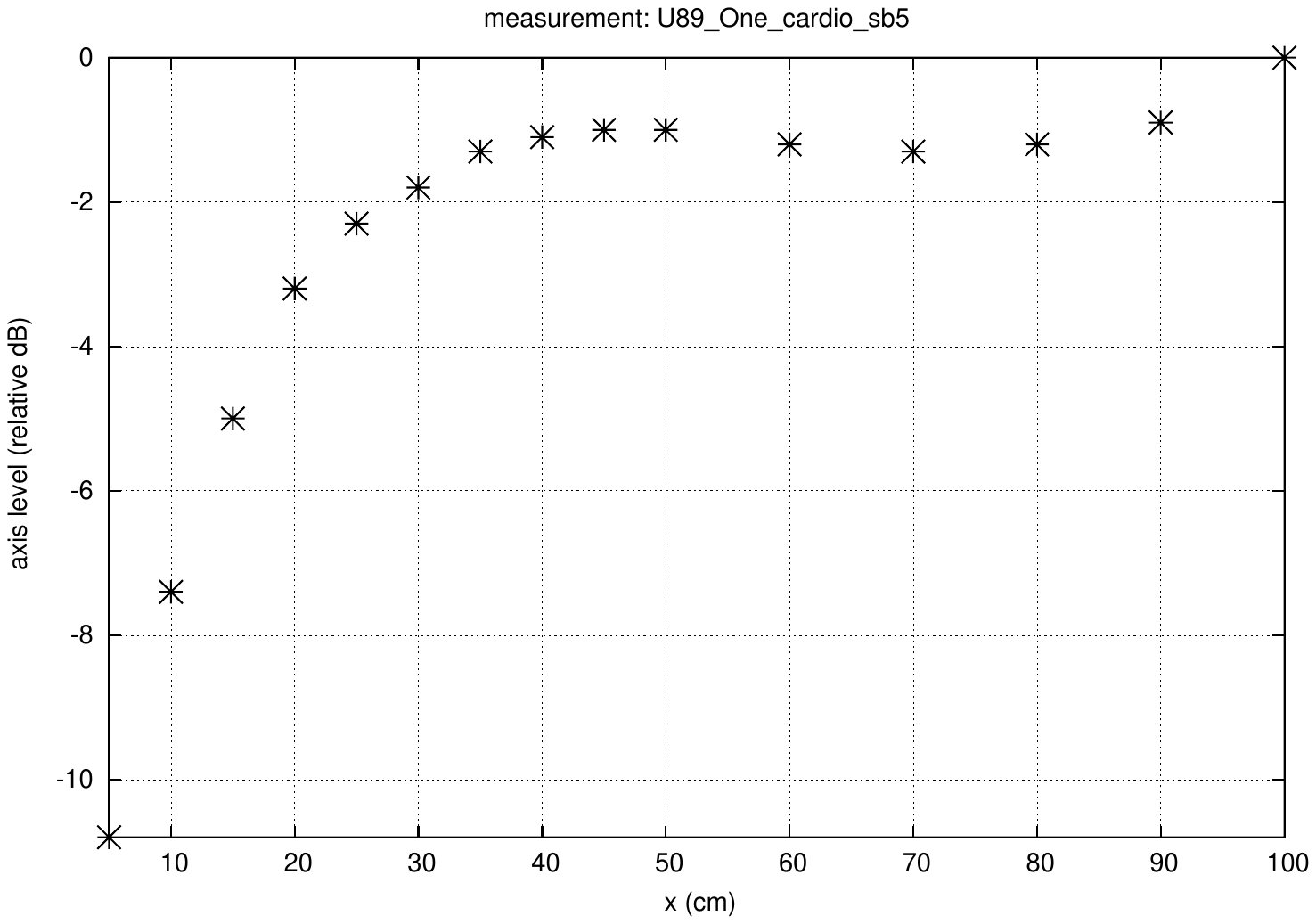}
\includegraphics[width=5cm]{U89_One_omni_sb5.pdf}
\end{center}
\caption[25]{Evolution of the weight in the spectral balance of subband  5 as a function of $x$; music and a Neumann U89i microphone with bidirectional (left), cardioid (center) and omnidirectional (right) directivities: experimental points.}
\end{figure}

\pa Fig. 25 gives the weight curves for subband 5 (800-1200~Hz) for three tested directivities. But the weight curves for subbands 3,4  and 6 are similar to the ones of subband 5.

\pa We can again underline the fact that the cardioid curve is a mix of the bidirectional and omnidirectional ones and that it exhibits the main characteristics of both other directivities, with some attenuation.

\pa We can also note that we are only able to define a "validity limit" for wave assumption in the case of bidirectional directivity, for subbands 3 to 6, as found for subband  2. But, for subbands 3 to 6, it is not possible to find a "validity limit" for wave assumption for the two other directivities because weight variations are bigger than 1~dB for the so-called "spatial modulation" effect.

\pa Compared to the ECM8000, the curve is rather simi\-lar for omnidirectional directivity, presents a quite lower weight for 5~cm distance from the source for the cardioid and a much lower weight   or 5~cm distance associated with a quasi constant weight after the rise for the bidirectional directivity. 

\pa Thus, the classical description of the proximity effect is refound for the bidirectional directivity but, on the contrary, we have quite different weight curves for the two other directivities.  

\pa As in the cases of subbands 1 to 4, it seems that the microphone with the bigger global effect is again the omnidirectional microphone while the bidirectional one corresponds to the classical representation of a proximity effect for subbands 3 to 6: a quite low weight of these subbands in the spectral balance with a progressive but big rise of this weight when $x$ increases.

\pa  So, except for the bidirectional case, we refind the observation made for  the ECM8000 microphone: we cannot consider the wave model as a valid assumption for these subbands and need another explanation for the physical phenomenon.

\pa Fig. 26 gives the information for the three directivities for subband 8 which is also representative of the curves associated with subbands 9 and 10.

\begin{figure}[h!t]
\begin{center}
\includegraphics[width=5cm]{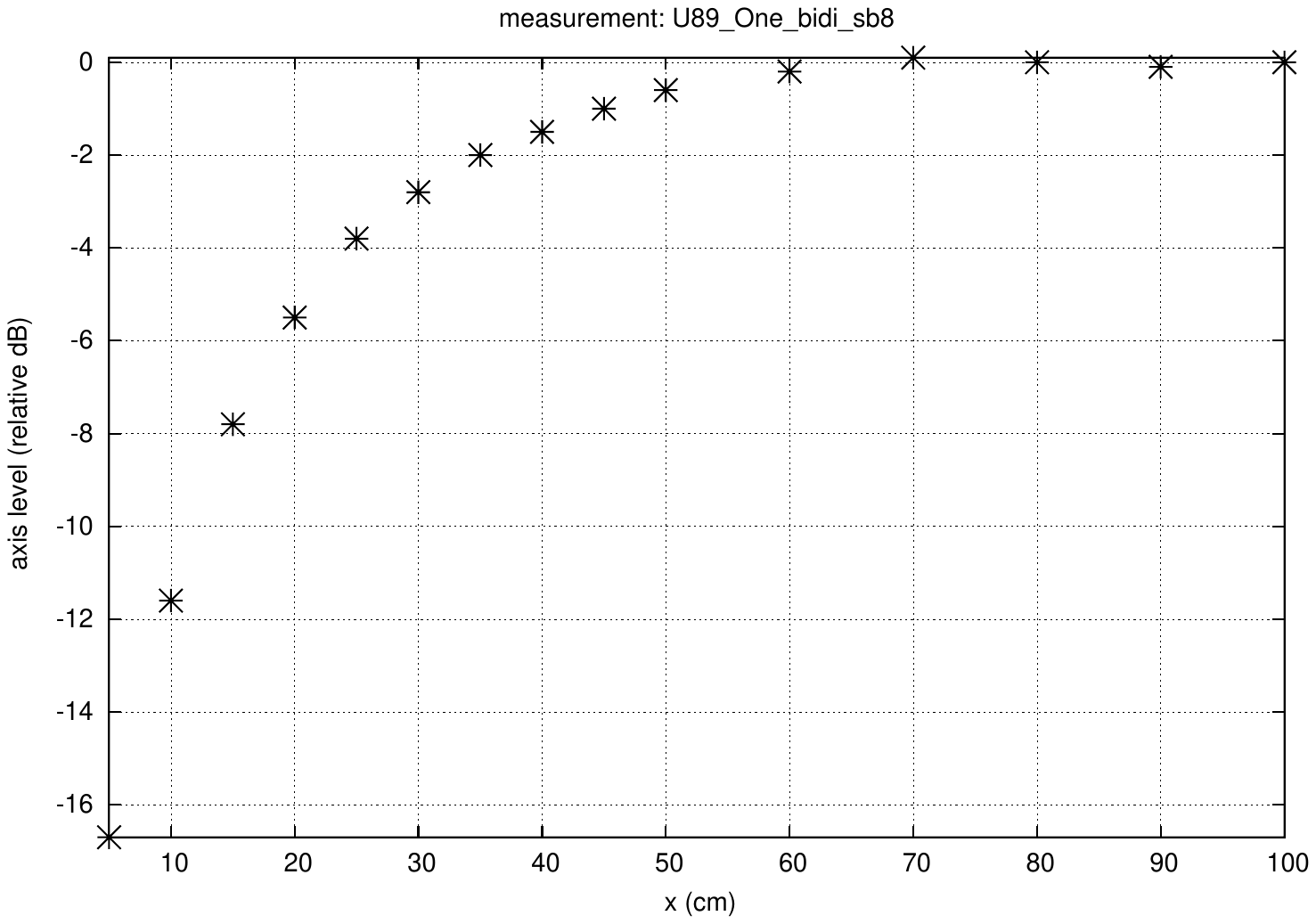}
\includegraphics[width=5cm]{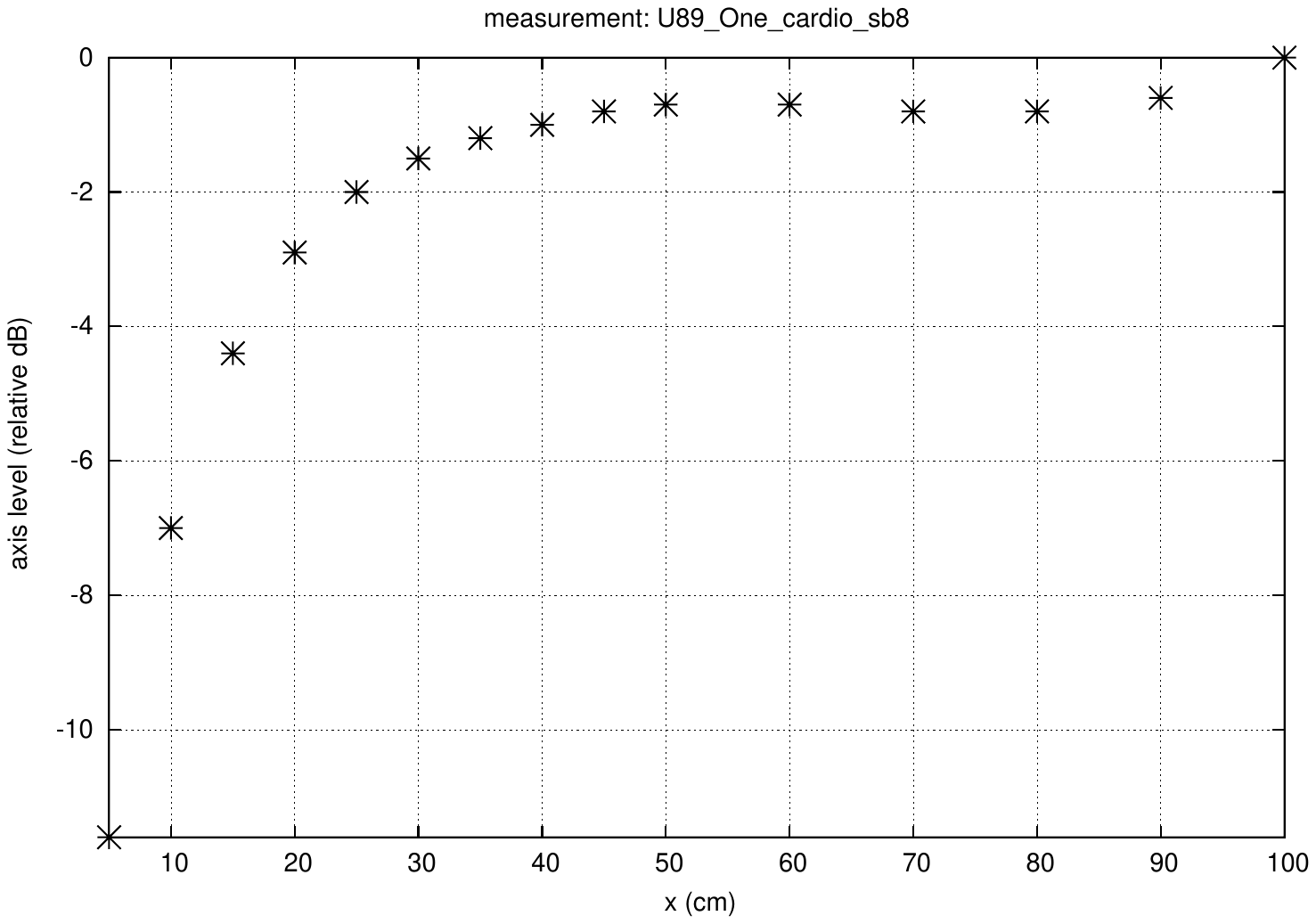}
\includegraphics[width=5cm]{U89_One_omni_sb8.pdf}
\end{center}
\caption[26]{Evolution of the weight in the spectral balance of subband  8 as a function of $x$; music and a Neumann U89i microphone with bidirectional (left), cardioid (center) and omnidirectional (right) directivities: experimental points.}
\end{figure}

\pa As for the other subbands, the cardioid curve corres\-ponds to a mix of the omnidirectional and bidirectional ones, which means that the cardioid directi\-vity case exhibits the main characteristics, with some attenuation or averaging, of the two other directivities.

\pa As found for the ECM8000, we have a fast (omnidirectional directivity) or progressive (bidirectional and cardioid directivities) rise of the weight  with either so-called "spatial modulation" with magnitudes lower than 1~dB (omnidirectional and cardioid directivity) or an asymptote with almost constant weight (bidirectional directivity), which corresponds to the classical description of the proximity effect in lite\-rature. 

\pa With small modulation magnitude or quasi zero one, we refind the idea of a "validity limit" for wave assumption for sufficient distances from the source (varying according to the directivity) as found for subbands 8 to 10 for ECM8000 microphone.

\pa But compared to the ECM8000 weight curve of Fig. 16 (or Fig. 22), the minimum weight at 5~cm distance from the source is bigger for all three directivities which means that the effect for these subbands is stronger for the U89i microphone whatever is  the chosen directivity compared to the ECM8000.

\pa For the subbands 8 to 10, we  would have almost a proximity effect and this effect would be more signi\-ficant for the bidirectional directivity in term of magnitude of the global variation of the weight, but, faster for the omnidirectional directivity. So the microphone with the stronger effect will depend on the chosen parameter, but we think one may find that the omnidirectional directivity has the main effect due to the faster variation of the weight when varying the microphone quite closely to the sound source.

\pa So, we have found a proximity effect and also the "re-equalization" effect for all three directivities. As for the ECM8000; we have found that using omnidirectional and cardioid directivities it would not be possible to consider a wave model for the frequency range 0-1800~Hz, except for the bidirectional directivity for which the frequency limit is lower: wave assumption could be valid for frequencies upper than 200~Hz.

\pa But this result is somehow a paradox as we find that a spherical wave model would constitute a fair approximation for the case of a bidirectional microphone, for a sufficient distance from the source for almost the whole audio range (this approximation is not valid for frequency range 0-200~Hz). Indeed, even if we are concerned with in-axis measurements (so zero incidence excitation), we find the larger validity of the spherical wave model for the bidirectional directivity, classically defined as a mix of two close capsules with opposite phases. Normally, one would \textit{a priori} think that it would be the omnidirectional microphone which would be associated with the best validity of the wave assumption and it is clearly not the case at this validity cannot be found for frequencies under 1.8~kHz!

\pa These experimental results clearly ask the question of the model to use to explain and synthesize the microphones behavior, for "low" frequencies and above from all small distances to significant distances from the source. It also clearly indicates that, it would be quite interesting to find a physical model where pressure field description would coincide with spherical waves as a limit case, where spherical waves constitute a fair approximation. 

\pa In the next section, we keep on studying the beha\-vior of microphones, making the comparison between our three cardioid microphones, which would exhibit a mean or a kind of a compromise between the omnidirectional and bidirectional behaviors.

\subsubsection{Comparisons of cardioid microphones}
\pa With Fig. 27 to 30, we present the experimental results for subbands 1, 2, 5 and 8 in the case of the music stimulus. As in the former paragraph these four subbands are representative of the other subbands (3, 4 and 6 for subband 5; subbands 9 and 10 for subband 8).

\pa In this subsection, we study the case of the three tested cardioid microphones: Neumann U89i with cardioid directivity, Audio-Technica AT2020 and Behringer C-2.

\pa Fig. 27 illustrates what happens for subband 1 (0-50~Hz) for these three microphones.

\pa We can underline the fact that all three curves exhibits a decrease of the weight down to a minimum and after a rise up to the 1~m location. 

\begin{figure}[h!t]
\begin{center}
\includegraphics[width=5cm]{U89_One_cardio_sb1.pdf}
\includegraphics[width=5cm]{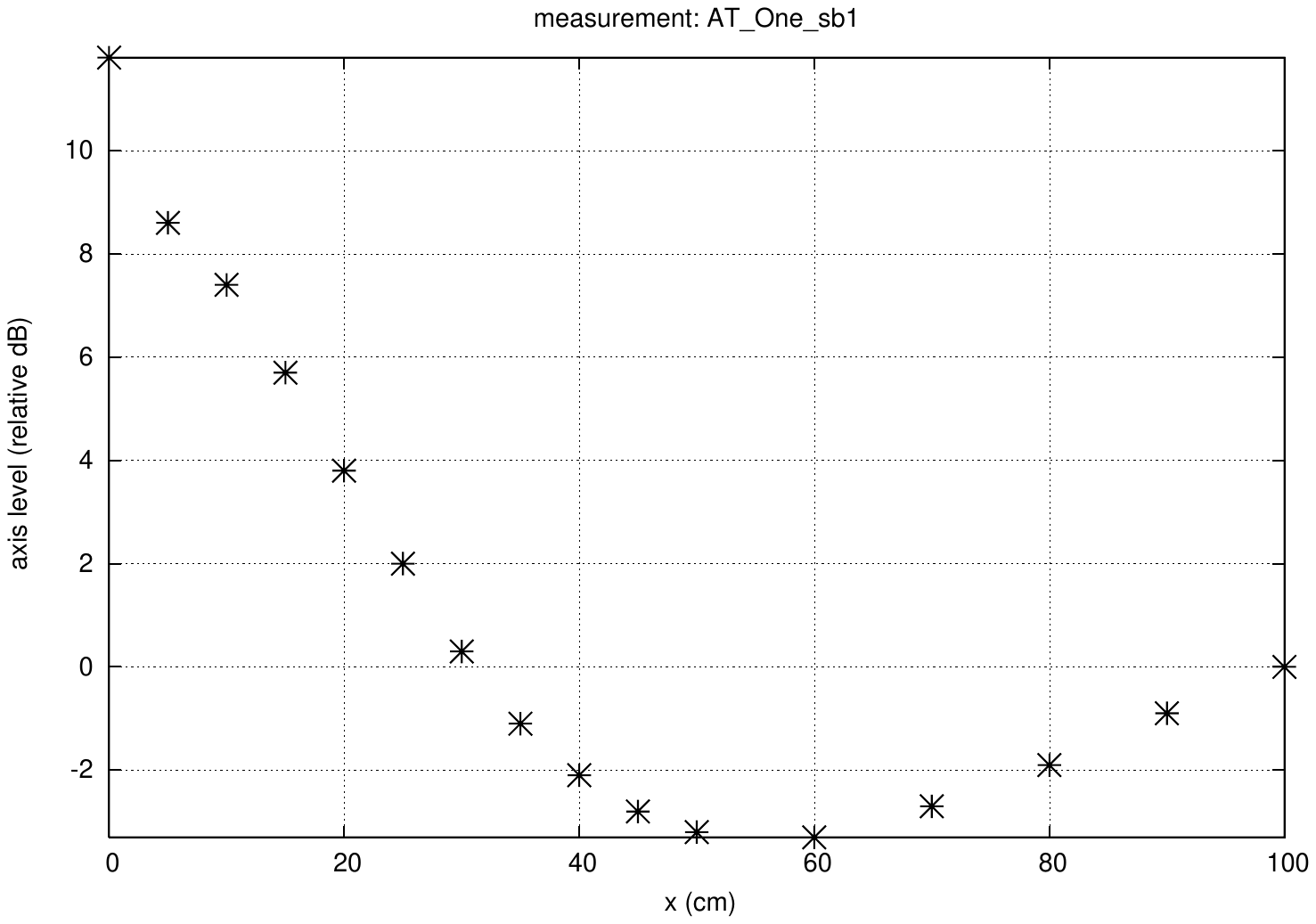}
\includegraphics[width=5cm]{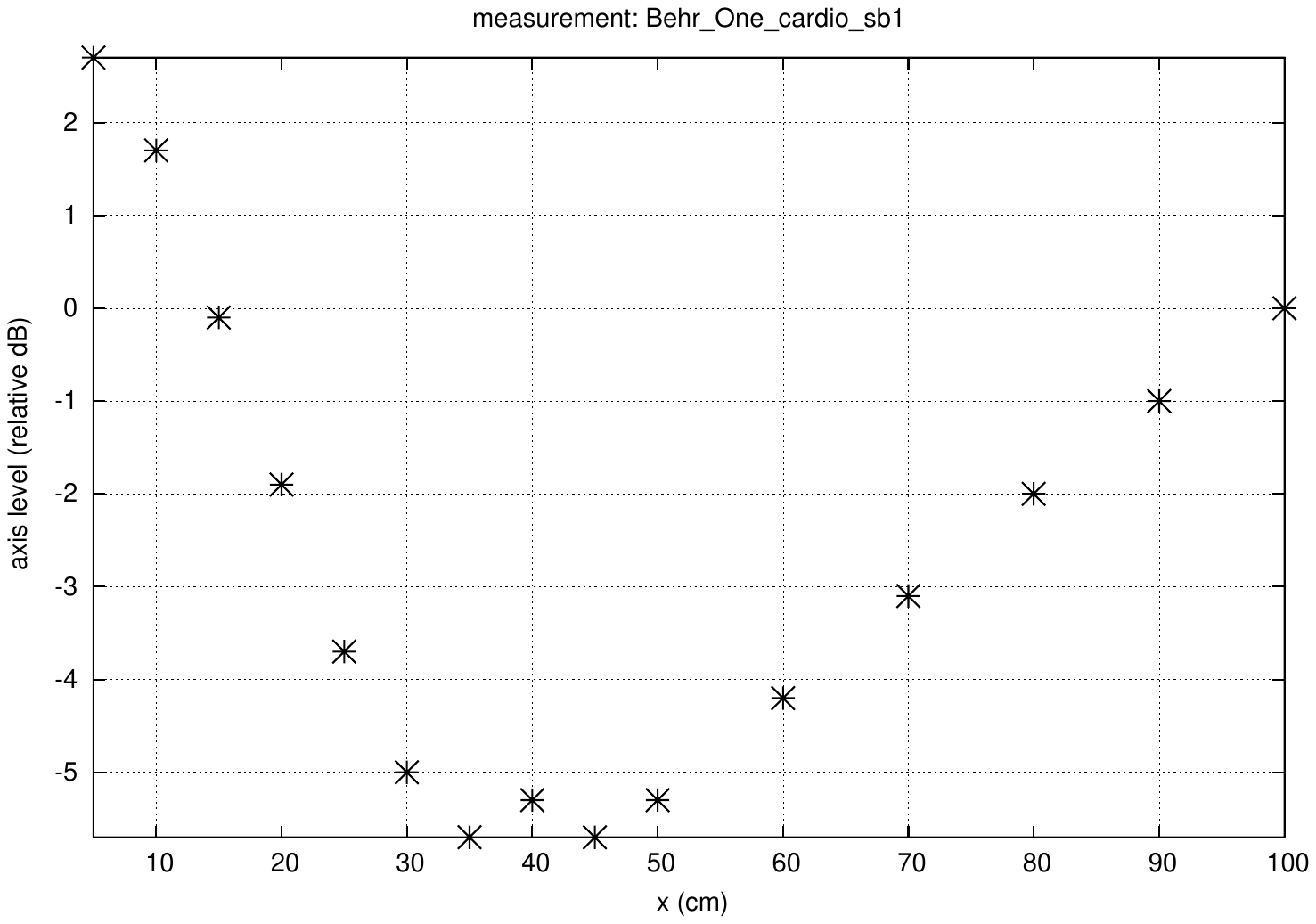}
\end{center}
\caption[27]{Evolution of the weight in the spectral balance of subband  1 as a function of $x$; music and cardioid microphones: (left) Neumann U89i, Audio-Technica AT2020 (center) and Behringer C-2 (right).}
\end{figure}

\pa Considering the behavior during the rising part of the curve we can point out that the weight variations seem to important to consider that the wave assumption is valid for any of the three microphone.

\pa But, we can underline the fact that if the U89i and C-2 microphones have weight curves with similar look, excep\-ting the location of the minimum point (35/45~cm for the C-2 microphone and 50~cm for the U89i), the boost for distance $x=5$~cm is much bigger for the AT2020 microphone (around 8.5~dB instead of around 2.1/2.3~dB for the two other microphones).

\pa It is also interesting to note that the AT2020 is the microphone for which considering a wave appro\-ximation after the minimum distance would be the less wrong case as we have a "spatial modulation" of 2~dB  magnitude.

\pa Fig. 28 illustrates what happens for subband 2 (50-200~Hz) for the three microphones.

\pa The three weight curves are similar: a decrease down to a minimum and then a rise up to 1~m distance from the source. But the global magnitudes of the weight evolutions are again rather different as there are more than 4.5~dB with the following decreasing weight order for microphones:  C-2, AT2020 and finally U89i microphones.

\pa But, if we now consider the rising part of the curves we can note that the magnitude for the C-2 and AT2020 microphones are lower than 1~dB while the magnitude for the U89i microphone is equal to 2~dB. 

\pa This may indicate that we can consider the wave model as a fair model for two of the three microphones after their minimum weight location: the C-2 and AT2020 microphones. But, as we have found it in previous subsection, we cannot consider the wave assumption as valid for the U89i microphone.

\begin{figure}[h!t]
\begin{center}
\includegraphics[width=5cm]{U89_One_cardio_sb2.pdf}
\includegraphics[width=5cm]{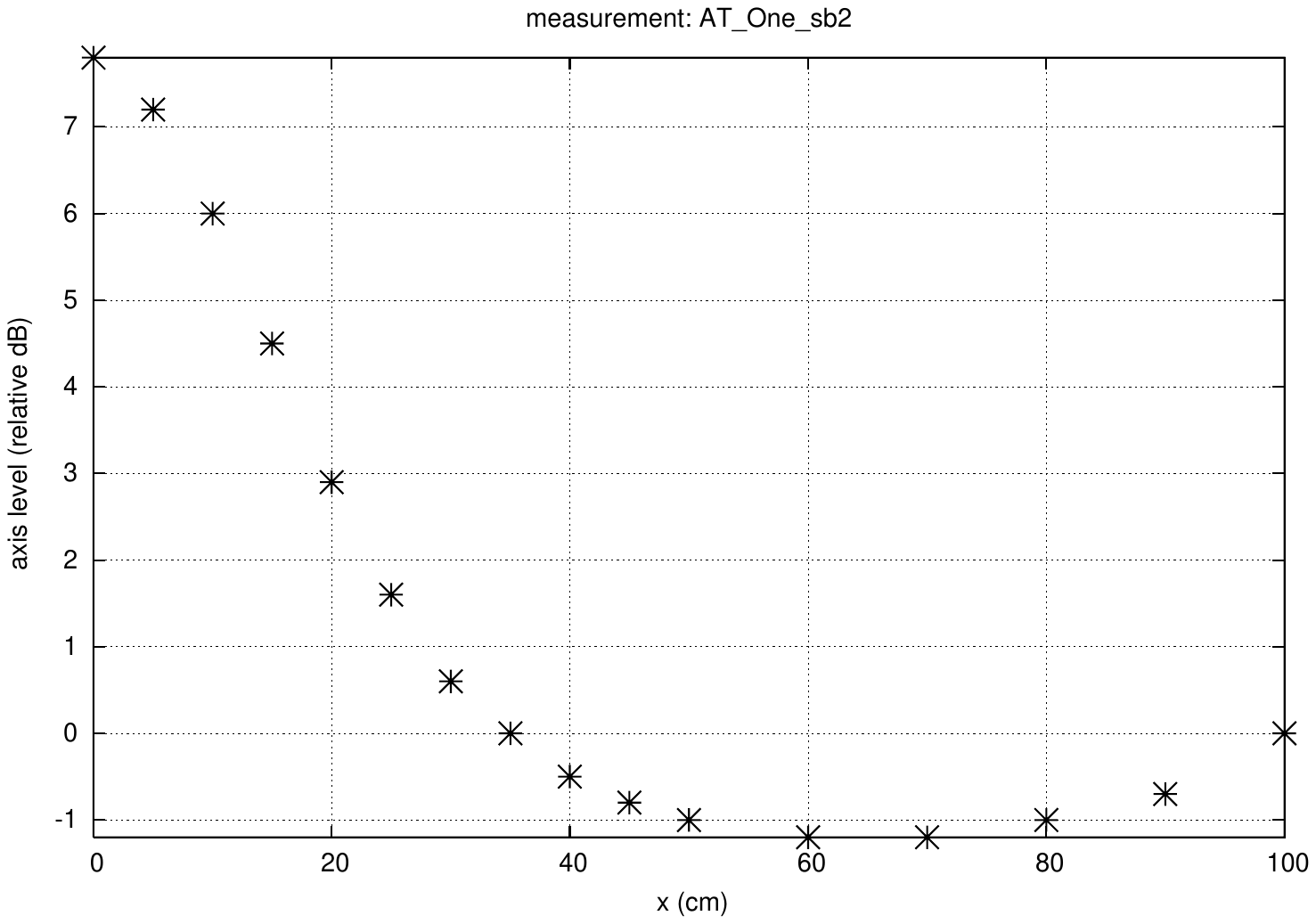}
\includegraphics[width=5cm]{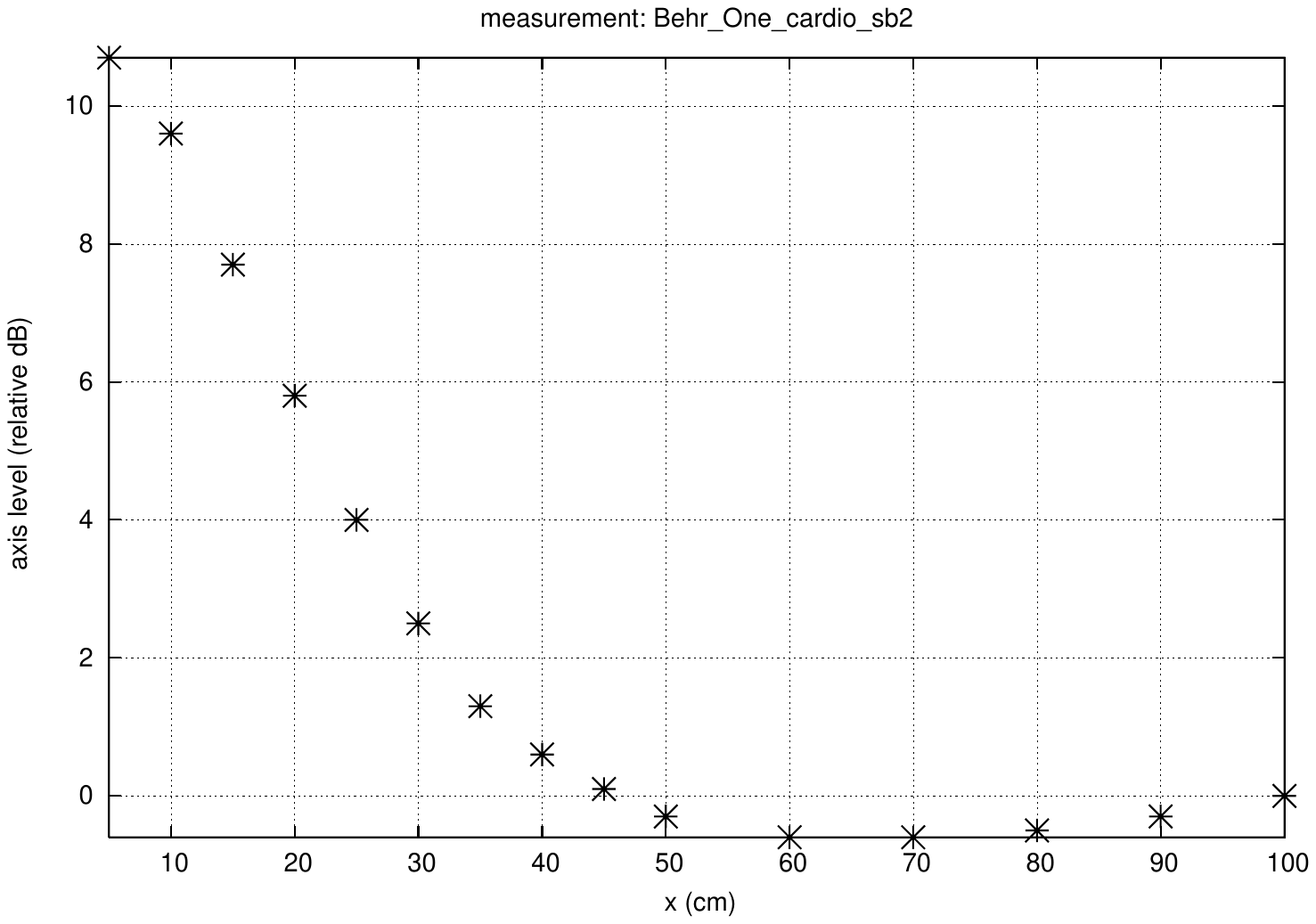}
\end{center}
\caption[28]{Evolution of the weight in the spectral balance of subband  2 as a function of $x$; music and cardioid microphones: (left) Neumann U89i, Audio-Technica AT2020 (center) and Behringer C-2 (right).}
\end{figure}

\pa Fig. 29 gives the weight evolutions according to the distance from the source for subband 5, which are representative of what happens for subbands 3, 4 and 6.

\pa We can observe that the curves are similar for the three microphones which is also rather the case for subbands 3, 4 and 6.

\begin{figure}[h!t]
\begin{center}
\includegraphics[width=5cm]{U89_One_cardio_sb5.pdf}
\includegraphics[width=5cm]{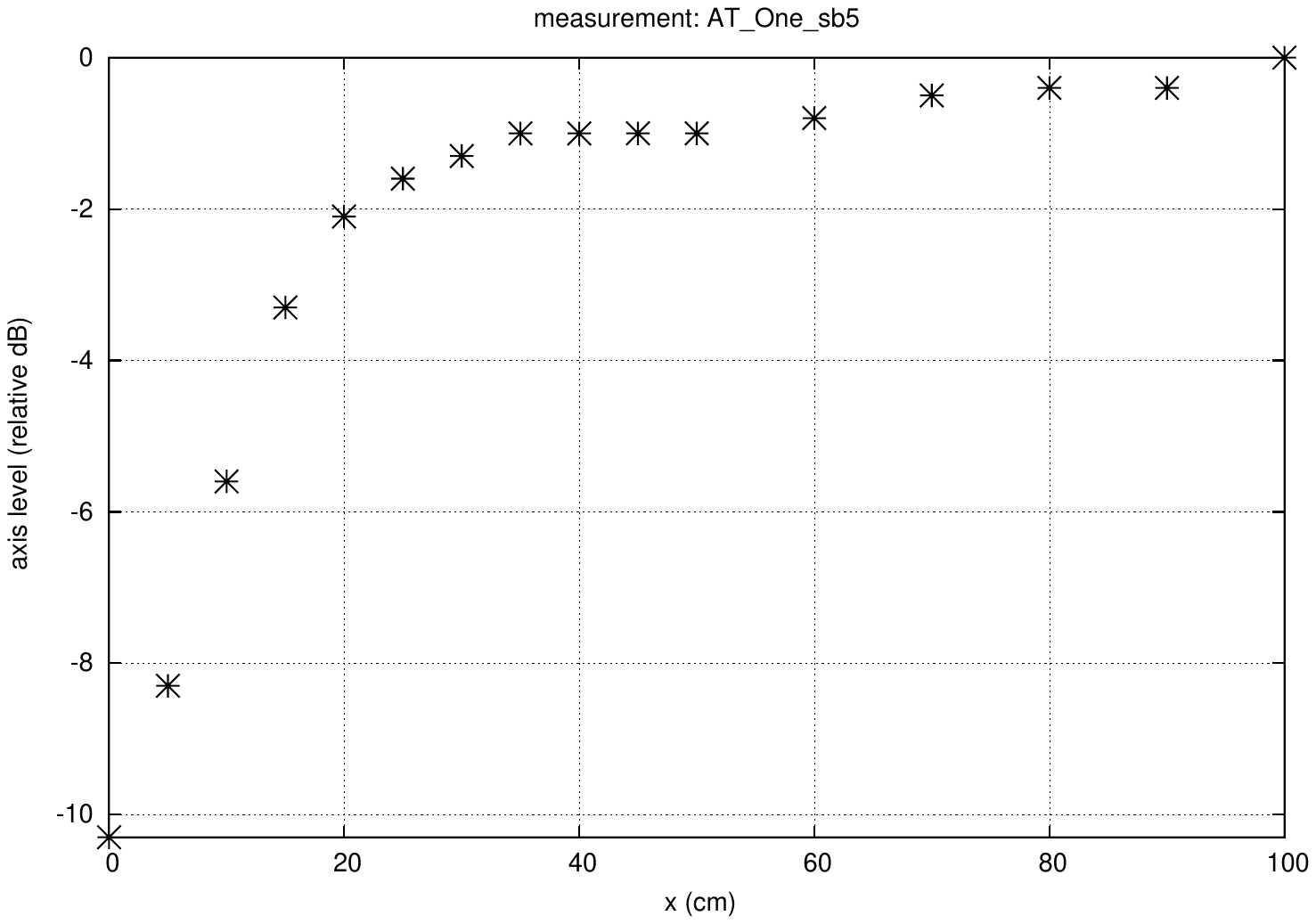}
\includegraphics[width=5cm]{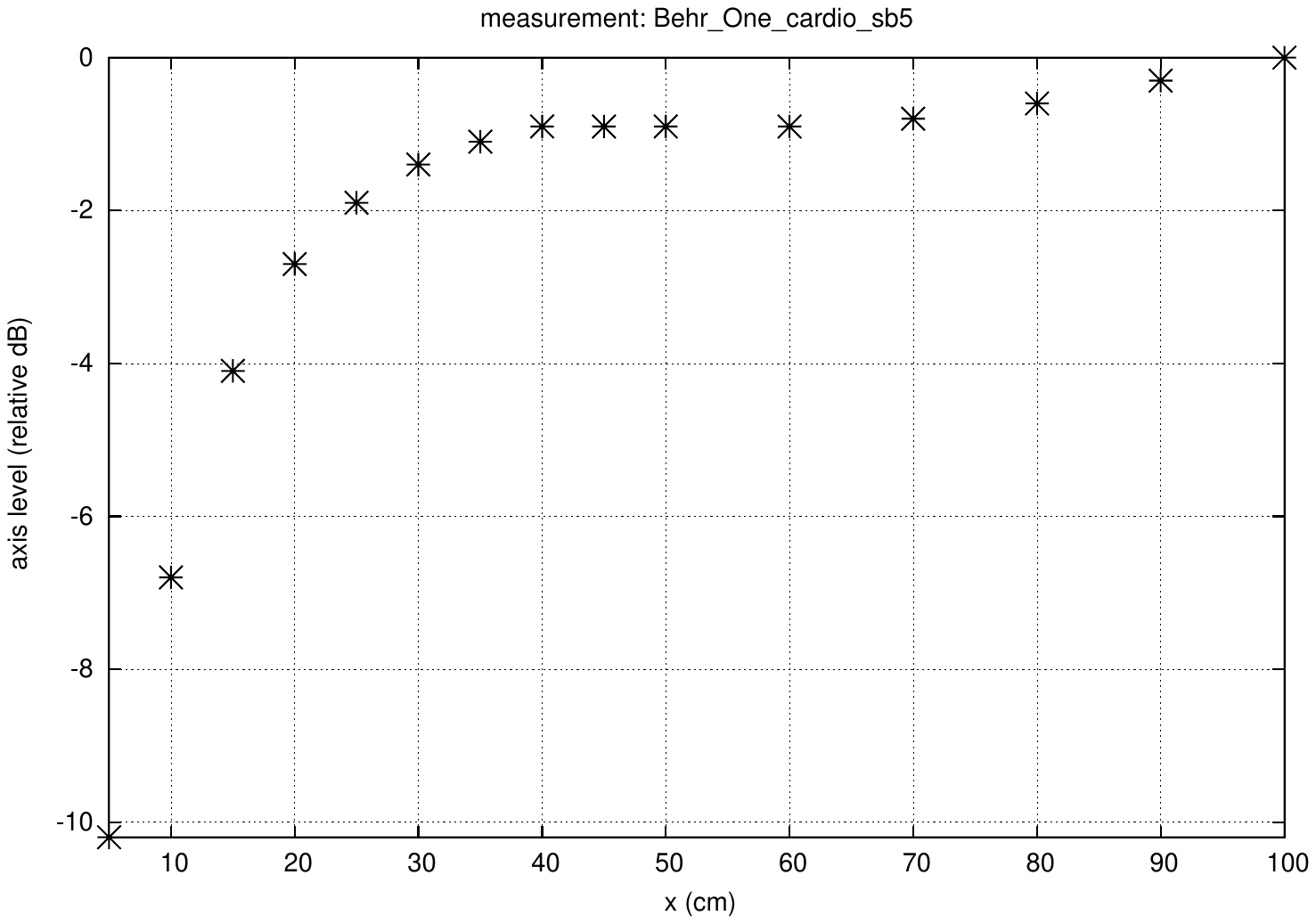}
\end{center}
\caption[29]{Evolution of the weight in the spectral balance of subband  5 as a function of $x$; music and cardioid microphones: (left) Neumann U89i, Audio-Technica AT2020 (center) and Behringer C-2 (right).}
\end{figure}

\pa We find a rather fast rise of the weight followed by a slowly rise up to the 0~dB point at 1~m. We can consider that the spatial modulation is too important for the U89i microphone which may indicates that the wave assumption is not really valid for subbands 3 to 6, as said before. But, we do not have the same situation for the two other microphones because, when it exists, the spatial modulation is lower than 1~dB which may permit to consider the wave assumption as a fair approximation for a sufficient distance from the source (approximatively 40~cm in the case of the subband 5 but this "validity limit" which varies with the subband). 

\pa Even if for the U89i microphone, we do not have a validity range for wave model, we have yet a curve look which corresponds to the classical description of the proximity effect, similar to the ones found for the two other microphones.

\pa So for the subbands 3 to 6, except for the U89i microphone, we can consider a wave model as a fair approximation as soon as the weight has raised the $\pm$1~dB, which happens at around 40~cm for subband 5 for both microphones, but varies with the subband. 

\pa Fig. 30 give the weight curves for subband 8, but similar to the ones found for subband 9 and 10, for the three microphones.

\pa We find similar curves with a fast rise and then a slowly rise of the weight up to 1~m. For this case, and subbands 9 and 10, we clearly can assume that the wave model is a fair assumption as the weight evolution becomes lower than 1~dB quite quickly. But the "validity limit" for spherical wave varies from one microphone to the other and also according to the subband. 

\pa We can also notice that the behavior of the weight for these microphones and these subbands matches the classical theoretical description found in the lite\-rature.

\pa So, if for the U89i microphone the wave assumption becomes valid for frequencies upper than 1.8~kHz, we have seen that, for the AT2020 or the C-2 microphones, this approximation becomes valid, for sufficient distances, for  frequencies over 50~Hz.

\pa This result demonstrates that the design of a microphone may permit to mimic the theoretical wave model for low frequencies but with a "validity limit" distance which may vary with the frequency range and with the microphone. 

\begin{figure}[h!t]
\begin{center}
\includegraphics[width=5cm]{U89_One_cardio_sb8.pdf}
\includegraphics[width=5cm]{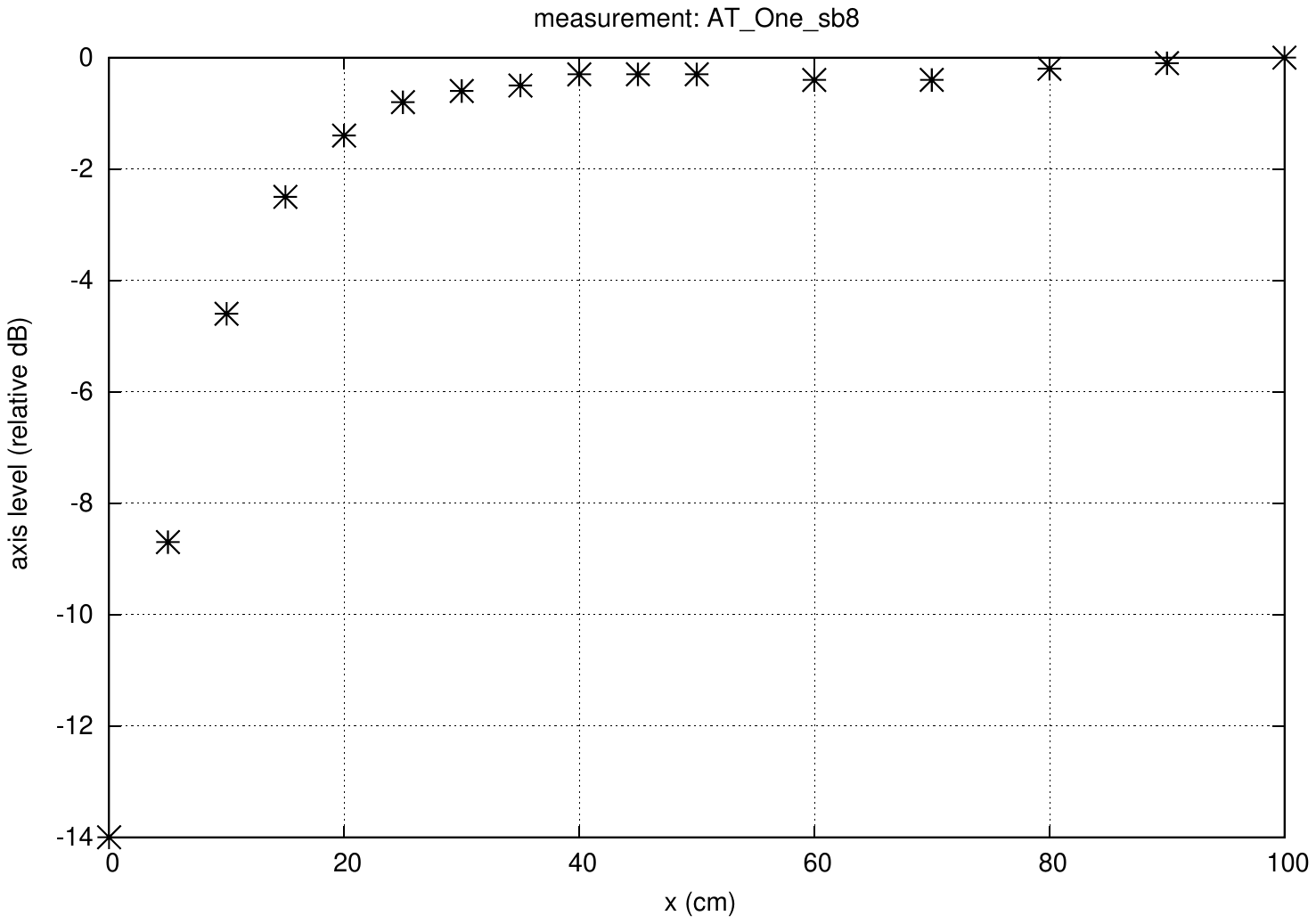}
\includegraphics[width=5cm]{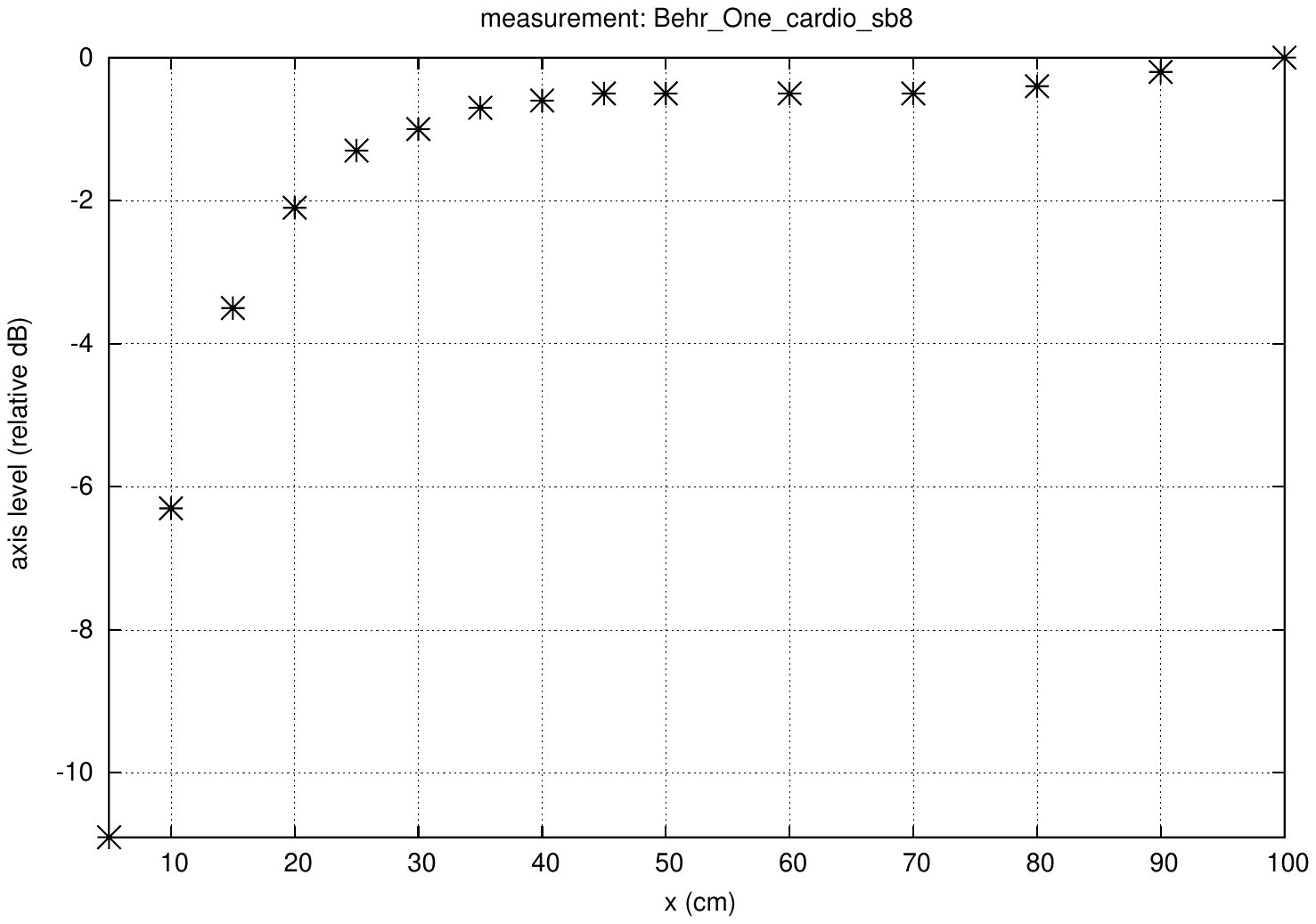}
\end{center}
\caption[30]{Evolution of the weight in the spectral balance of subband  8 as a function of $x$; music and cardioid microphones: (left) Neumann U89i, Audio-Technica AT2020 (center) and Behringer C-2 (right).}
\end{figure}

\pa This fact may constitute an argument to prefer measurements using music stimuli rather than pink noise, or worse, sinusoids. But, according to the destination of the microphone, the music stimuli may change.

\pa We can also note that to get some informative know\-ledge about the physical phenomena we should consider both the mean level and subbands weight curves. 

\pa It is also interesting to underline the fact that we have only global attributes of the measurements, mean level and spectral balance evolutions accor\-ding to the distance from the source, but we may also try to see if the subbands signals can also give some useful knowledge about the physical phenomena, in the proximity and at significant distances to the source.

\pa In \cite{Millot:06}, we have used for the oral presentation some subband directivity diagrams derived from the IDS subband analysis of simulations of recordings using an extension of the classical directivity-based model for microphone. Drawing directivity diagram using music stimuli and subband analysis seem also a fair method. 

\pa But, even with the information we have collected and partially presented here, we do not have enough information to derive a better simple physical model at this time. We just know that the model we propose in \cite{Millot:06} may constitute a fair approximation, when spherical wave assumption is valid, if we use a very much over-sampled version of the calculus core for the model, which may be tested in future.

\pa We still have to review some more experimental results for a naked loudspeaker in order to get more information to be able to propose another way to model the acoustical phenomena and discuss some future measurements campaigns.

\section{Naked loudspeaker study}
\subsection{Main information about the experimental setup}
\pa We have used the second experimental setup with notably an Endveco pressure sensor model 8507C-2 (with its associated signal conditioner), a National Instruments DAQ card model PCI-6036E using BNC inputs and LabView 7.1 to control the signals acquisition. The loudspeaker is a Celestion G12M Greenback with a diameter of 305~mm, a 75-5000~Hz frequency range  and a measured cut-off frequency of 75~Hz.

\pa The Endevco pressure sensor has tiny dimensions (a diameter of 2.34~mm and a length of 12.7~mm)  and a sensitivity of 17.8~mv/kPa 
which permits a maximum level of 176,8~dB SPL. But this pressure sensor is quite more adapted for the very high level, due to its quite small sensitivity, so as soon as we increase the distance from the source, the noise level becomes higher which was very annoying when we used sinusoids as excitations.

\pa To limit its drawback, we diffuse sinusoids with very high levels in order to get signal emerging from the noise as far as possible from the loudspeaker: 120~dB SPL for the 65~Hz sinusoid under the cut-off frequency of the loudspeaker (75~Hz) and 130~dB SPL for the 100~Hz sinusoid (over the cut-off frequency).

\pa But, to limit the noise in the signals, we filtered them using the IDS analyzer using adapted frequency mapping with zero-phase and quite stiff (16383 samples impulse responses) FIR filters:  0-50~Hz, 50-80~Hz and 80~Hz-22.05~kHz subbands for the 65~Hz sinusoid; 0-85~Hz, 85-115~Hz and 115~Hz-22.05~kHz subbands for the 100~Hz sinusoid. So the experimental curves for the mean level amplification correspond to the filter signals with a bandpass of 30~Hz for both sinusoids: 50-80~Hz for 65~Hz sinusoid, 85-115~Hz for 100~Hz sinusoid.

\pa We also recorded the signal for the pink noise and music stimuli and consider the unfiltered signal to plot the mean level amplification curves. But, we diffused the pink noise with a level of 120~dB and the piece of music with 112 to 119~dB level. 

\pa All these level indications are given for the zero distance from the source. 
 
\pa In this section, we focus our attention only on the question of the in-axis mean level curves and just consider the distances from the sources for which the signal is enough strong to emerge, which explains that, for the sinusoids, the distance range is only equal to 0-60~cm. Doing again the same measurements in a bigger anechoic room would perhaps permit to get information for bigger distances from the source and perhaps to use lower diffusion levels.

\pa As for the other measurements, we have chosen the largest distance from the source as 0~dB reference and again assume that it may correspond to a location where we should have the biggest probability to conclude that spherical wave model is a fair approximation for local pressure field. So, we still plotted the curves for the mean level amplification when we go closer to the source.

\subsection{Mean level vs distance from the source for broadband stimuli}

\pa Fig. 31 gives the mean level amplification curves for pink noise and music stimuli.

\begin{figure}[h!t]
\begin{center}
\includegraphics[width=7.5cm]{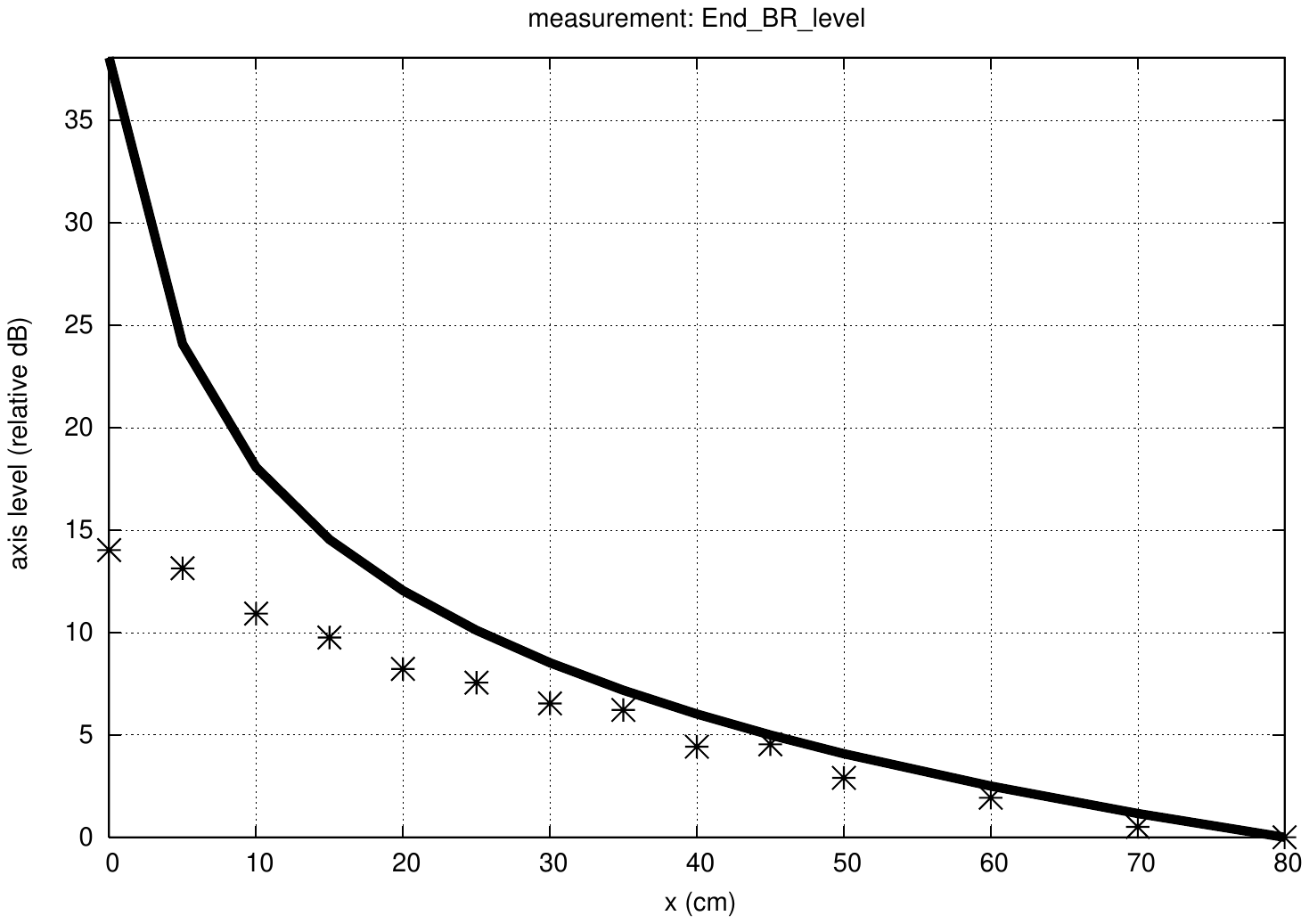}
\includegraphics[width=7.5cm]{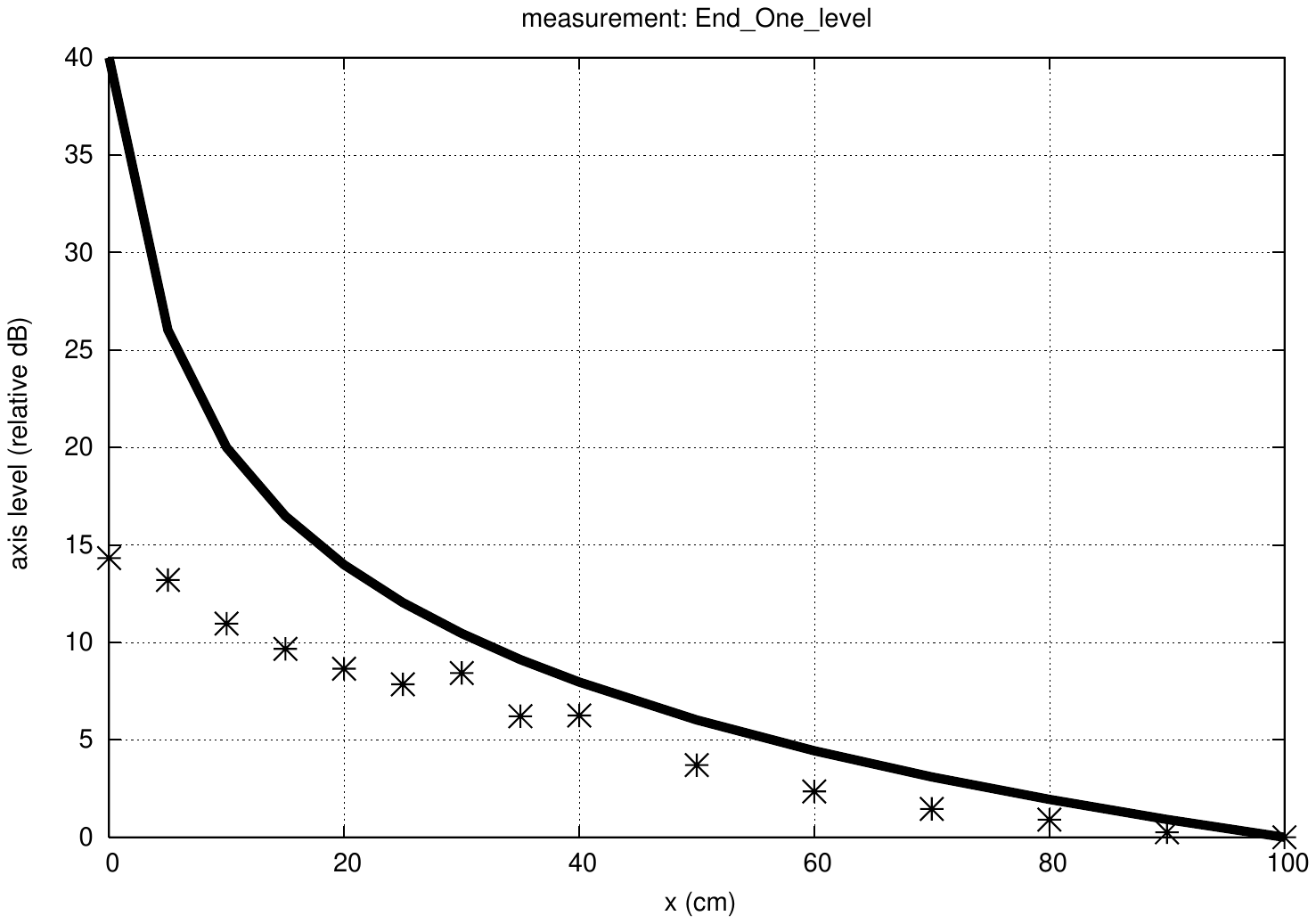}
\end{center}
\caption[31]{Mean level as a function of $x$: experimental points (crosses); theoretical amplification (solid line). (left) pink noise, (right) music.}
\end{figure}

\pa We must precise that with the DAQ system controlled by LabView, it seems to be not possible to record the measurements as wave files, even with calibration process permitting to get some measurements in floating points format with a value range of [-1, 1]. This fact has been confirmed by our contacts at National Instruments. So we had no choice except recording experimental signals as text files with limited precision, which means that we did not have the possibility to listen to the recordings during the measurements campaign. Indeed, the translation of the text files into wave files takes a lot of time and we were not able to do it, in parallel, during the measurements. 

\pa But the great advantage of the use of the Endevco sensor capture and the NI DAQ card was to be able to get measurements down to 0~Hz up to the 22050~Hz with a linear behavior of the acquisition chain. It would be better, in the future to try to control the NI DAQ with, for instance, Matlab, which may permit to record a wave file directly.

\pa Considering the curves of Fig. 31, for pink noise and music stimuli, we can note that both curves are a bit different. Indeed, in the case of the pink noise stimulus, the experimental points and the theoretical wave $1/x$ law mean level amplifications are much more closer than in the case of the music stimulus. In fact, we refind a trend already observed with the combo guitar amplifier and the other microphones: the gap between theory and experimental points is bigger for music stimulus. But, for the case of the naked loudspeaker, the gap is quite big compared to the measurements made for the combo guitar amplifier. 

\pa As both sources are quite different, it would not be careful to make further comparisons between both cases, thus, in the following, we just consider the case of the naked loudspeaker without forgetting that music stimuli should be rather more informative for people intending to get some useful information  to better understand what happens for usual recordings in a room.

\pa it is interesting to note that for the pink noise stimulus, we can propose a "validity limit" around 45/50~cm for wave approximation while such proposal is more  difficult to make for music stimulus as the distance between theory and experimental points seems significant for most of the distances from the source. One may propose to take the "validity limit" around to 80~cm but it is just to find such a limit. In other terms, the wave assumption seems to be not relevant for the music stimulus which may encourage us to discard the wave assumption and propose another explanation further. 

\pa But, we can underline the fact that for both stimuli, the gap between experimental points and theoretical curve increases significantly as we go closer to the source. And, even if the gap is small or negligible for distances over 45~cm in case of the pink stimulus, this gap follows the same trend for both cases for distances shorter than 25~cm. And, for $x=0$~cm, we have close differences for both cases: around 24/25~dB under the theoretical prediction which corresponds to a bigger difference than the one found for the ECM8000 microphone with the combo guitar amplifier.

\pa Thus, it appears that the physical phenomena are quite different from waves, for both stimulus, in a large distance range: at least 0-40~cm for pink noise case; whole distance range for music case! 

\pa These results make us think that an alternative explanation of physical phenomena should, eventually, have spherical wave model as a limit case for pressure field but should not assume the existence of waves as a an \textit{a priori} principle. In fact, we do think that waves may naturally emerge where they represent a fair approximation for pressure field.

\subsection{Mean level vs distance from the source for sinusoids}
\pa Fig. 32 gives the mean level amplification curves for two sinusoids: one (65~Hz) under the cut-off frequency (75~Hz) of the naked loudspeaker, the other (100~Hz) above this cut-off frequency.

\pa According to the classical explanation of the acoustical short-circuit for naked loudspeaker we should not have a signal under the cut-off frequency because front and back waves, having opposite phases, may interfere destructively in any point far away enough from the source. 

\pa Introducing a plane finite baffle, one may observe that for frequencies under the cut-off frequency the new arrangement permits to listen and to get some sound.

\pa We must precise that adding a plane finite baffle, for instance, realizes a decoupling between back and front phenomena, or back and front waves according to classical explanation.

\pa But, Fig. 32 demonstrates that we have quite similar evolutions of the mean level for both sinusoids at least for the  0-40~cm distance range. And, listening the signals for all tested distances, we can note that both sinusoids are audible even for the 1~m distance but with modulation amplitude and noise for distances $x\geq 25$~cm, which is confirmed if we listen to normalized versions of the recordings, using a normalization at the same mean level for all distances. 
 
\begin{figure}[h!t]
\begin{center}
\includegraphics[width=7.5cm]{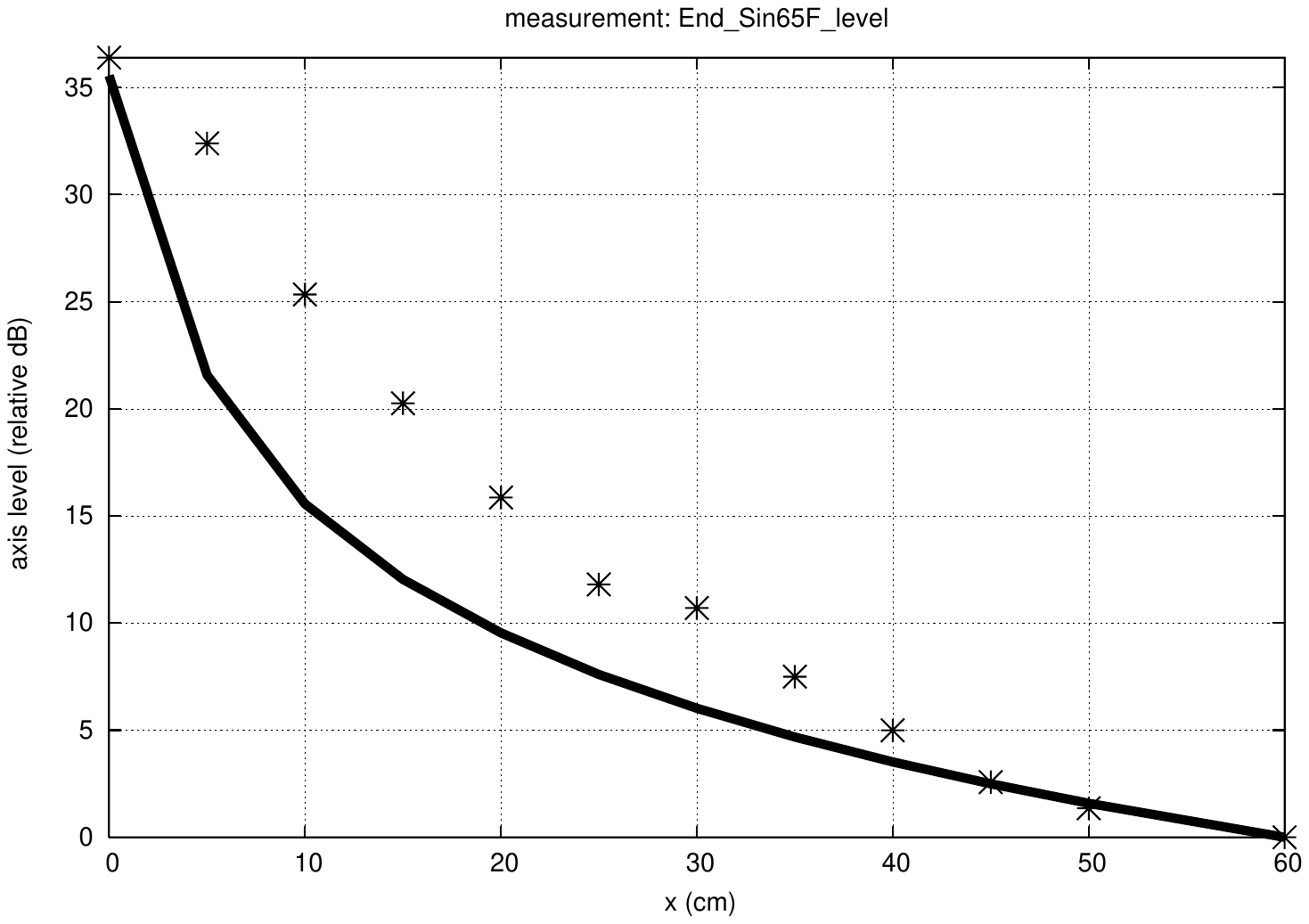}
\includegraphics[width=7.5cm]{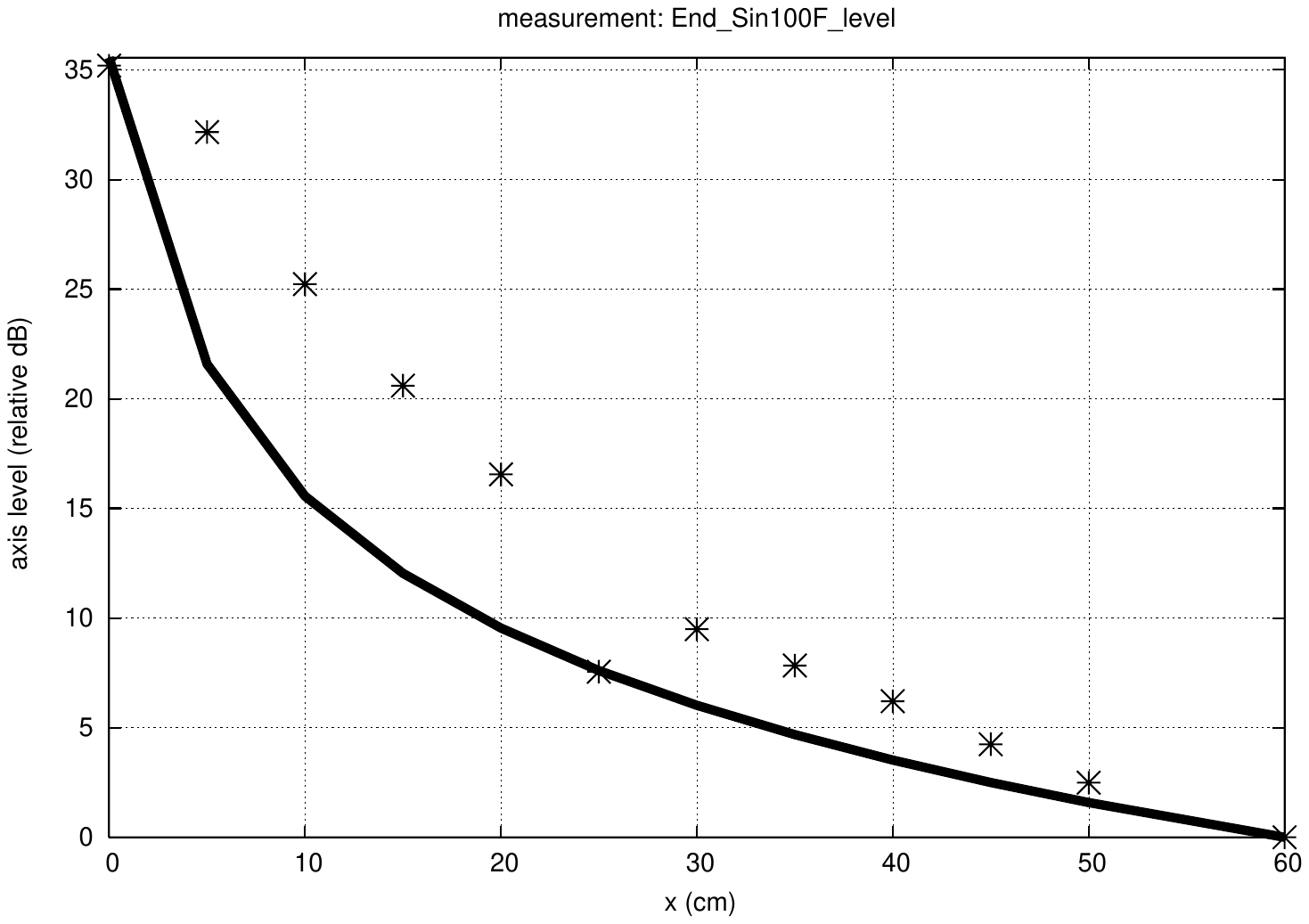}
\end{center}
\caption[32]{Mean level as a function of $x$: experimental points (crosses); theoretical amplification (solid line). Sinusoid at 65~Hz and 120~dB SPL (left), sinusoid at 100~Hz and 130~dB SPL (right).}
\end{figure}

\pa As we have used high mean excitation levels (120 and 130~dB SPL), even with an attenuation of -37.5~dB for 65~Hz sinusoid and of -41.36~dB for 100~Hz sinusoid for $x=100$~cm (compared to $x=0$~cm), it seems normal to be able to hear both sinusoids up to the 1~m distance from the source according to the Fletcher hearing threshold curve for sinusoids.

\pa So, even with a naked loudspeaker, we find similar experimental results for a sinusoid under the cut-off frequency and another one over the cut-off frequency. Thus, there is no destructive interferences between front and back waves under the cut-off frequency for the experiments we made. As we have used high levels, we think that with lower levels, both sinusoids would be audible only for a shorter distance range and we should not forget the fact that we need more power to get a similar level  for the frequencies under the cut-off frequency compared to the one over the cut-off frequency.

\pa For music stimuli, the spectral balance is the result of the mixing process
in studio using at least two stereophonic listening system which are quite different from the naked loudspeaker. And, using a flat amplification, we will not provide more power to the low frequency content when diffusing these music stimuli with a naked loudspeaker. Thus, under the cut-off frequency of the loudspeaker, the mean level will be lowered for the musical content and then may not be audible, according to the diffusion mean level.

\pa To get some more information, we review in the next paragraph what happens for the spectral balance in the low frequency range when the distance from the source varies.

\subsection{Low frequency spectral balance vs distance from the source}
\pa We consider the  evolution of the spectral balance for the frequency range 50-200~Hz using the following mapping (the IDS analyzer permits to define any desired mapping) with filters with impulse responses of 16383 samples length (giving quite stiff filters): 
\begin{itemize}
\item subband 1: 0-50~Hz (not presented here);
\item subband 2: 50-75~Hz ;
\item subband 3: 75-100~Hz ;
\item subband 4: 100-125~Hz ;
\item subband 5: 125-150~Hz ;
\item subband 6: 150-170~Hz ;
\item subband 7: 175-200~Hz ;
\item subband 8: 200-22050~Hz (not presented here).
\end{itemize}

\pa Fig. 33 to 35 present the evolution of the mean weight, according to the distance from the source, for subbands 2 to 7 using the music stimulus and the Endevco pressure sensor.

\begin{figure}[h!t]
\begin{center}
\includegraphics[width=7.5cm]{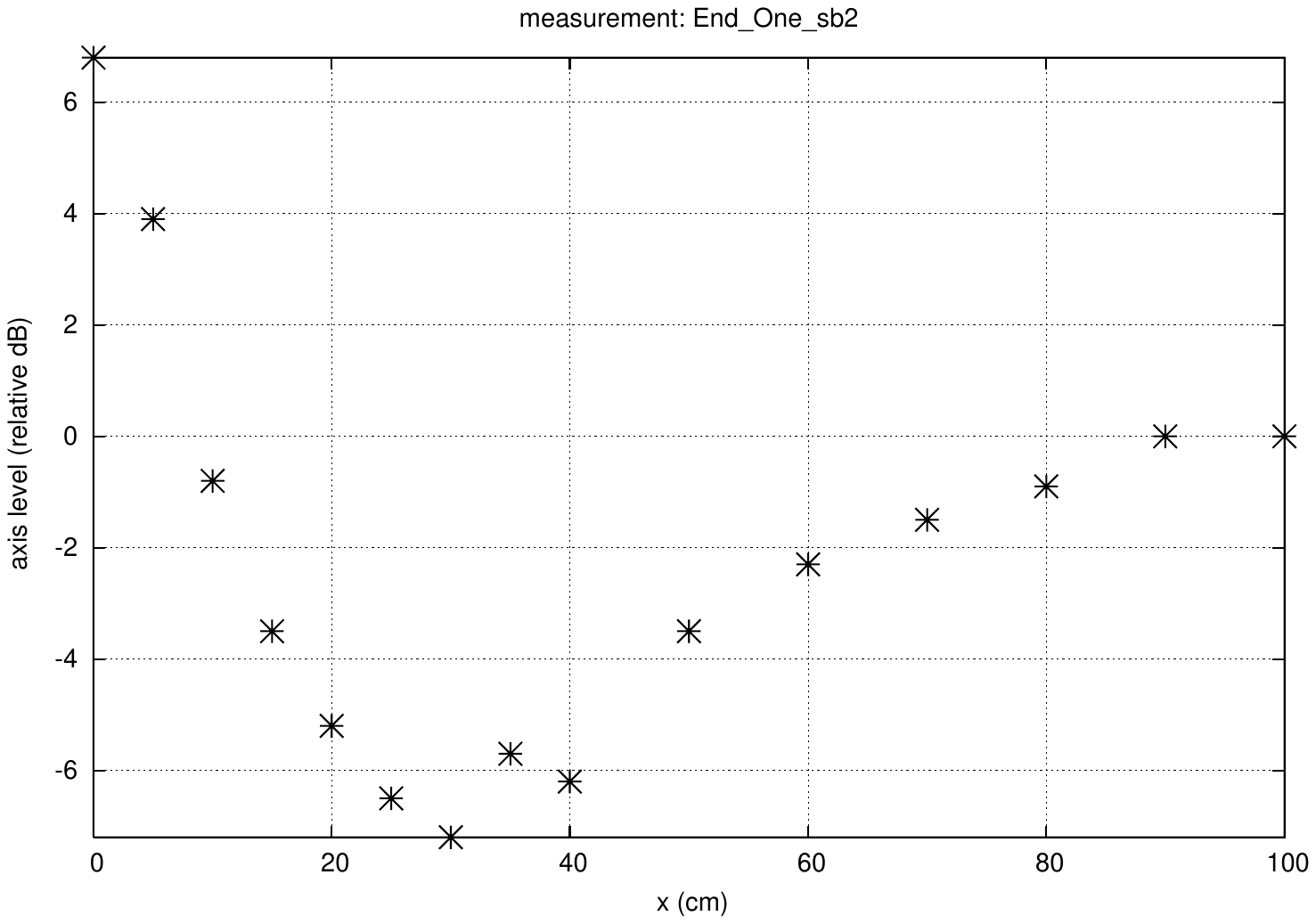}
\includegraphics[width=7.5cm]{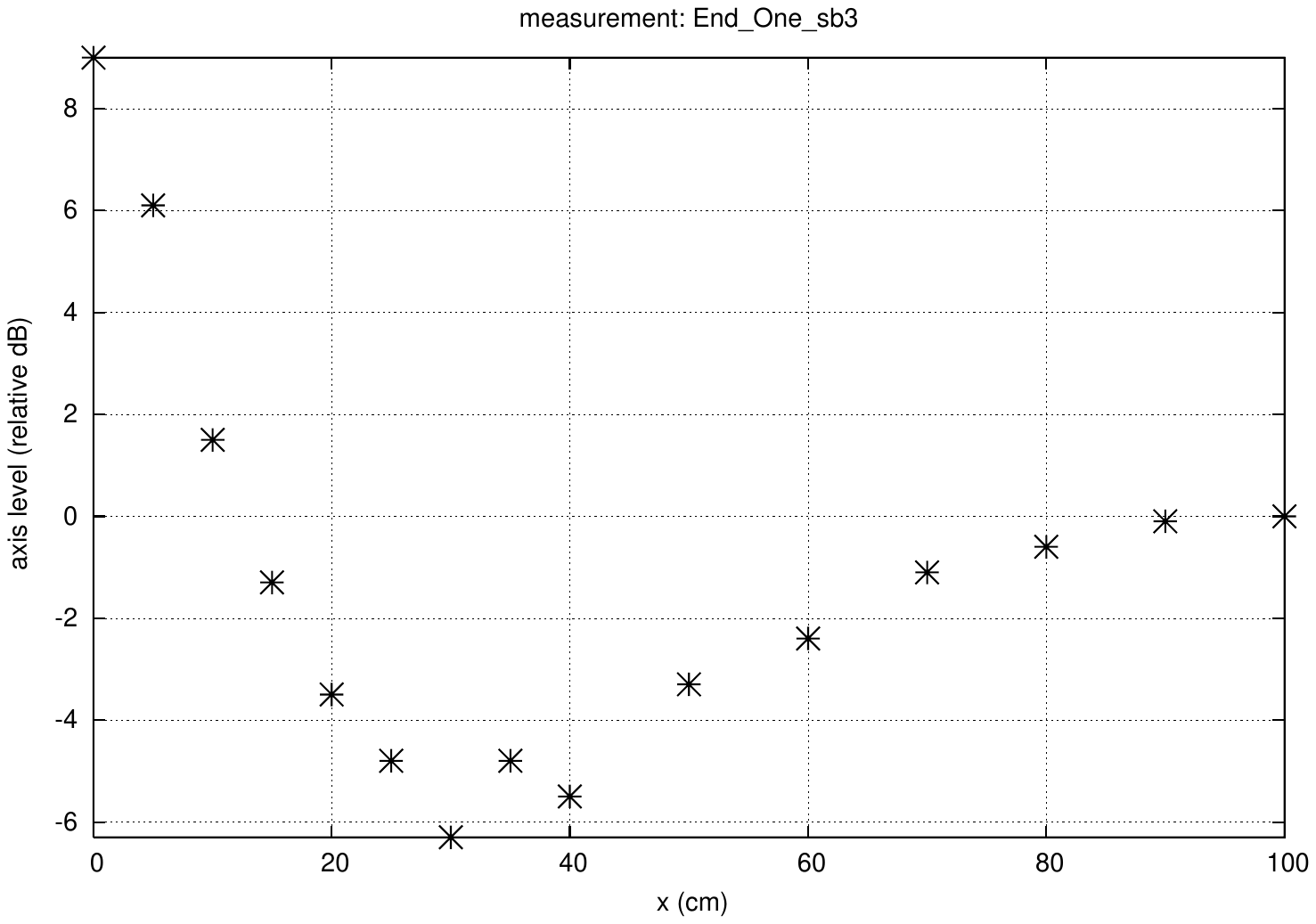}
\end{center}
\caption[33]{Evolution of the weight in the spectral balance as a function of $x$; music and Endevco pressure sensor: subbands 2 (left) and 3 (right).}
\end{figure}

\pa We can first note that the evolution looks seem rather similar for all these subbands with subband 2 corresponding to a frequency range under the cut-off frequency of the naked loudspeaker, the 5 other subbands corresponding to frequency ranges over the cut-off frequency. So, the behavior of each subband according to the distance from the source is equivalent under and over the cut-off frequency which is not coherent with the classical description of the short-circuit phenomenon for a naked loudspeaker. 
  
\pa Listening these 6 series of signals with Sennheiser HD 280 pro headphones and maximal diffusion level, we have observed that we were not able to listen anymore the sound in noise for subband 2 for distances bigger than 20~cm, while we kept on listening a signal (with some noise) up to 1~m for 5 other subbands. But, even if not audible, the weight evolution of subband 2 can be derived from 0 to 1~m. Thus, there is still some signal in subband 2, lost in the noise, which could easily be verified listening normalized versions of signals for subband 2 using the normalization at the same mean level for all the subbands signals, for all the distances from the source.

\begin{figure}[h!t]
\begin{center}
\includegraphics[width=7.5cm]{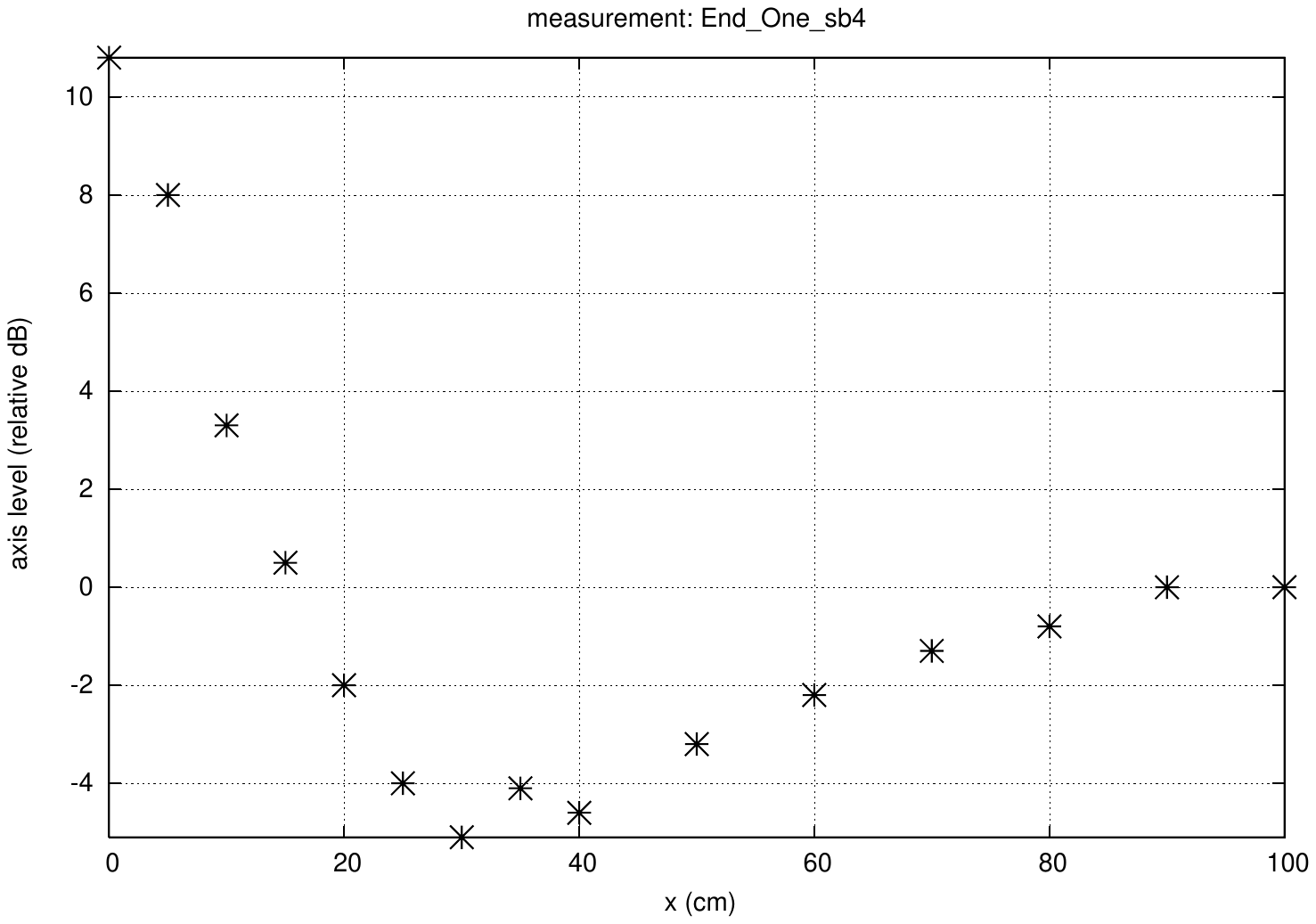}
\includegraphics[width=7.5cm]{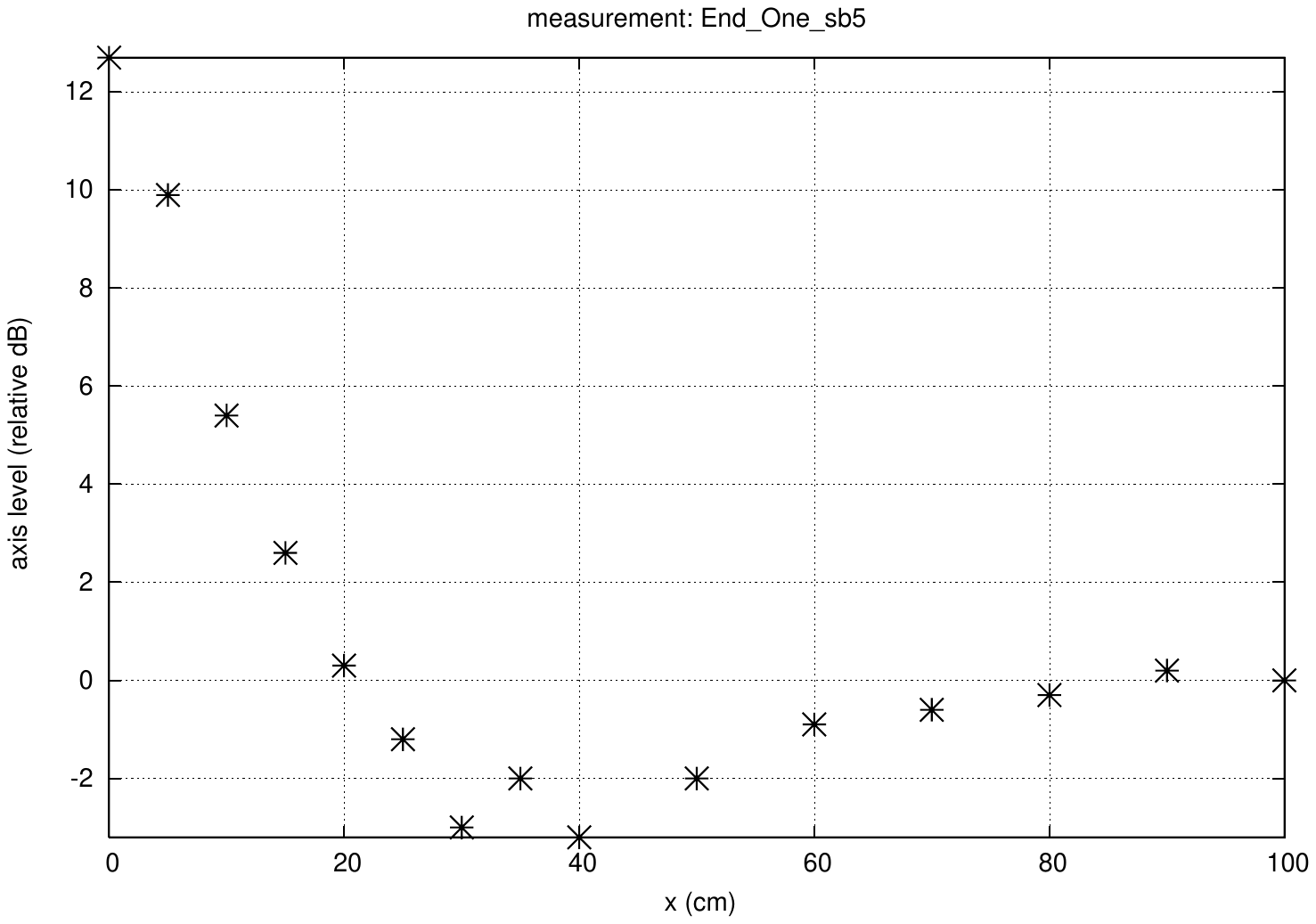}
\end{center}
\caption[34]{Evolution of the weight in the spectral balance as a function of $x$; music and Endevco pressure sensor: subbands 4 (left) and 5 (right).}
\end{figure}

\pa The similar look for all subband weight evolutions may partially be explained considering the role of the noise. We first encounter a decreasing for the weight related to the in-axis attenuation up to around 25/30~cm, with the noise level which increases and a broader noise. Then, we have a strange behavior for the evolution between  30 and 40~cm which can not be really explained even with listening of normalized versions of recordings. After, we encounter a a rise for the weight which may be related to the increase of the noise contribution, and the broadening of its spectrum, with a more or less fast and complete convergence to 0~dB weight.

\pa And, as the subband signals can be listen easily even for the 1~m distance, in the noise, this could mean that when the divergence to the 0~dB weight is rather small, a spherical wave assumption would be locally valid. Indeed, when listening the normalized recordings for all subbands, and especially subbands 6 and 7, it seems difficult to perceive an evolution of the signal spectral balance when we increase the distance from the source: the spectral balance seems invariant when the distance varies in the "converging" distance range. 

\begin{figure}[h!t]
\begin{center}
\includegraphics[width=7.5cm]{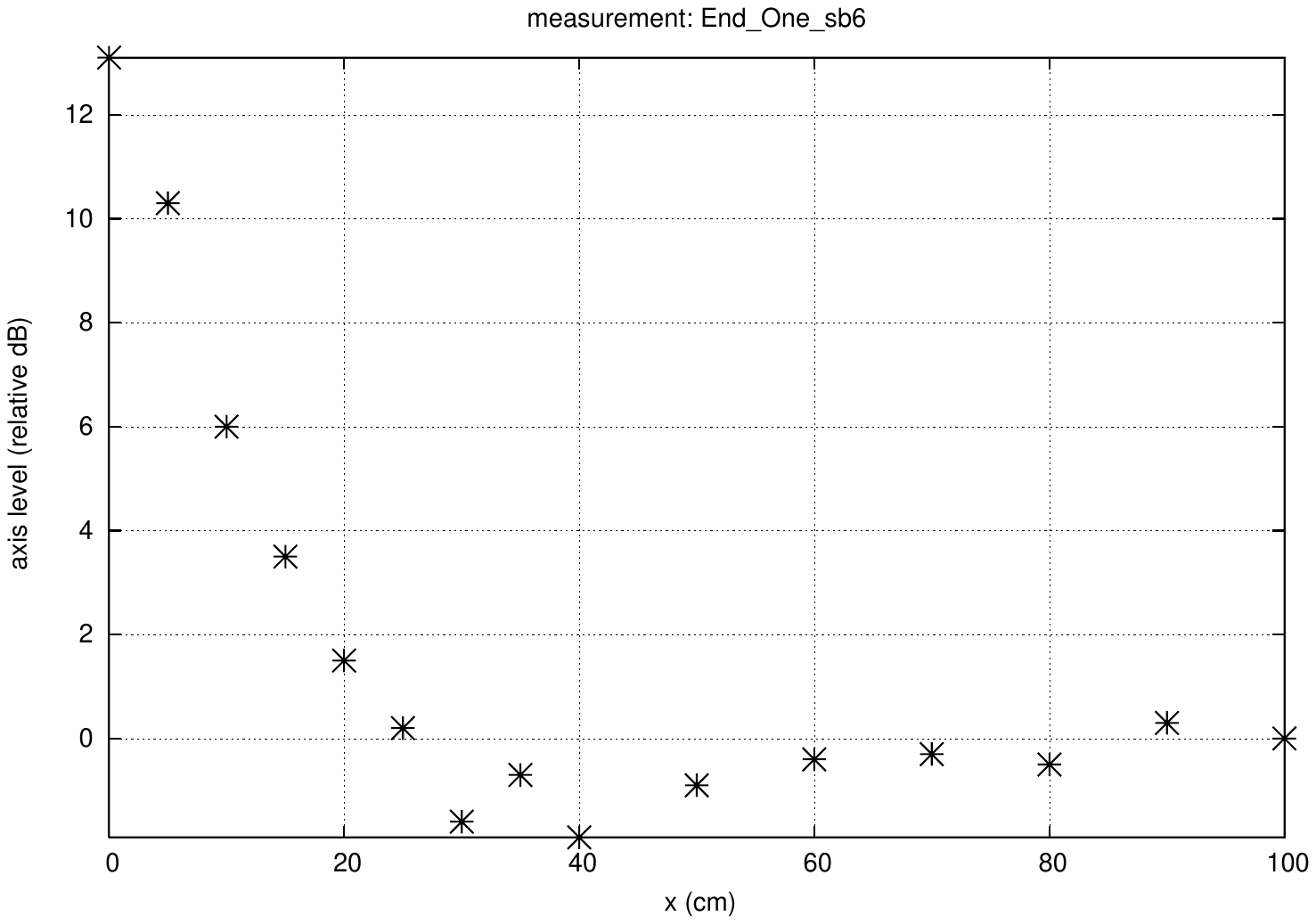}
\includegraphics[width=7.5cm]{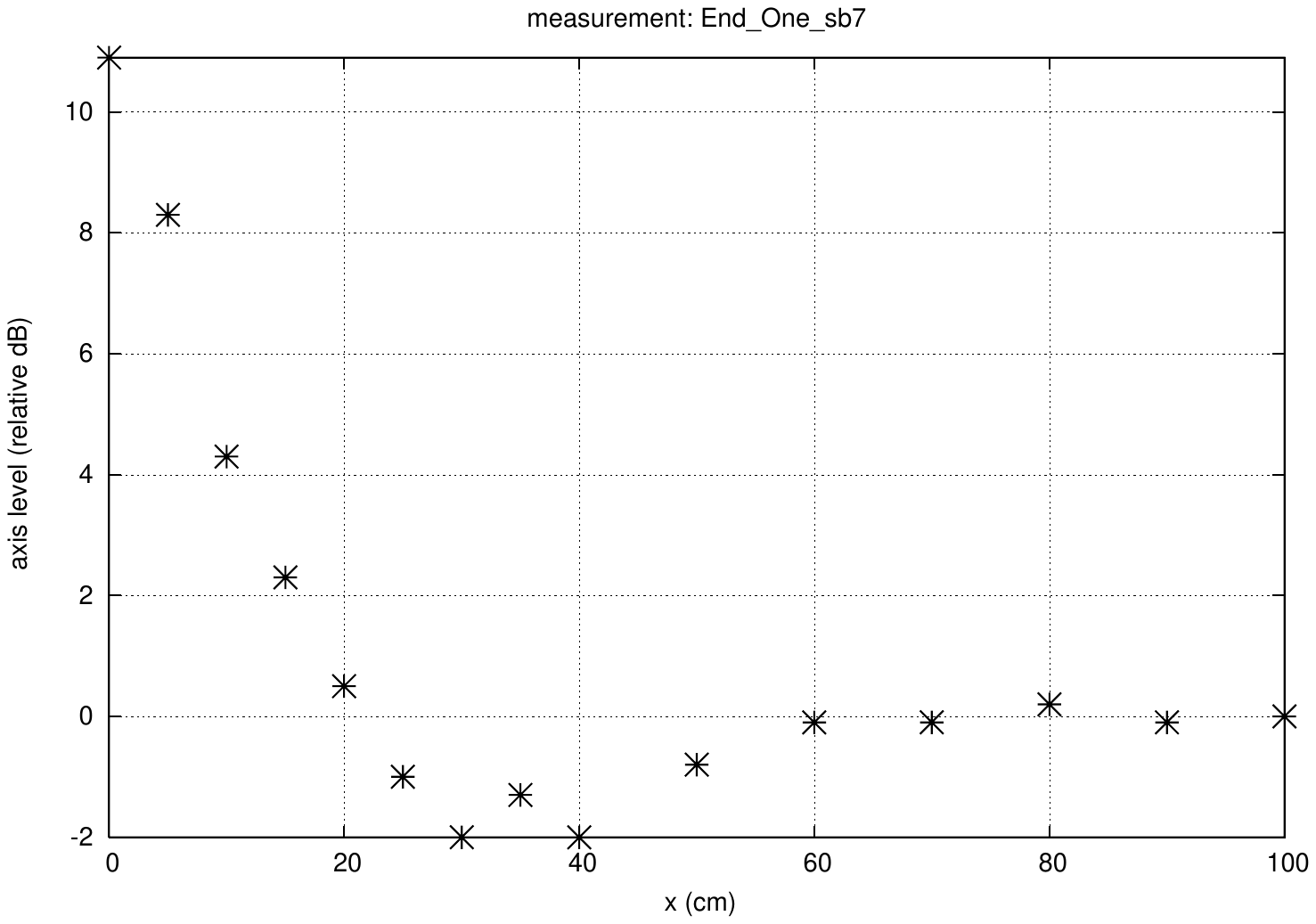}
\end{center}
\caption[35]{Evolution of the weight in the spectral balance as a function of $x$; music and Endevco pressure sensor: subbands 6 (left) and 7 (right).}
\end{figure}

\pa Some new measurements using a pressure sensor with a greater sensitivity and a bigger dynamic (more than 16 bits) DAQ card would permit to get some less noisy signals and then to be able to determine either if the weight of the signal increases following what we called the "re-equalization" process or if this weight keeps on decreasing for distances bigger than 30~cm and finish to converge more or less quickly and completely to the 0~dB reference weight at 1~m.

\pa With our experimental data, we cannot determine what really happens for the last part of the distance range: convergence or not to the 0~dB weight at least for the higher subbands, continuous decreasing of the subband weights or what we encountered with the guitar combo amplifier and called "re-equalization" phenomenon?  

\pa Nevertheless, we can propose an alternative explanation for acoustical phenomena as done in the following paragraph.

\section{Alternative explanation of acoustical phenomena}
\subsection{Case of the closed loudspeaker}
\pa For the moment, we assume that the membrane of the loudspeaker moves as a rigid plate.

\pa When the membrane moves forward like a rigid plate, it creates a local displacement of a thin air layer with a section almost equal to the section of the plate, like a free jet: the particle velocity is assumed to be constant inside the "free jet" and zero outside this "free jet".

\pa Moving particles try to communicate their motion to the closed "at rest" particles in the neighborhood of the "free jet", but this energy transfer takes times, due to inertia, and the motion of the particle at the surface of the jet is progressively curved by this energy transfer, inducing the creation of vortices around the "free jet".

\pa When the particles go forward, the vortices increase and contaminate the core of the "free jet", destroying it by violent and/or fast turbulence but over distance range which may depends on the initial velocity communicated to the air particles by the membrane of the loudspeaker: the "free jet" should go as far away from the membrane as the initial velocity would be important so for big velocities the turbulence and vortices should need time and a significant distance to appear.  

\pa Thus, we can consider that we first have a "free jet" flow area in the proximity of the membrane when it goes forward. Then we have a second area where we find turbulence associated with the dissipation of the "free jet", by growing vortices and turbulence. 

\pa And, after the "turbulent area" we should encounter a "residual flow" area where the particles go with a nearly constant velocity but inside a solid angle associated with the force of the vortices, that is to say which depends also on the initial velocity at the membrane. 

\pa In this third area, the "residual flow" one, we are far away enough from the source and we could demonstrate (to be reported elsewhere using local mass and momentum conservation equations, without any linearization of both equations, in association with adiabatic and perfect gas assumptions) that forward spherical waves model constitutes a limit case, of the local phenomena in this area, for the pressure field (and also the density field). 

\pa Examination of the mathematical description underlines the fact that mass conservation can be formulated with a left term corresponding to a forward spherical wave propagation equation (of first order) and a right term which depends on the first order spatial derivative of the particle velocity, assumed to be small in the third area. It is also quite interesting to say that if we can recognize a forward propagation equation for spherical pressure (or density) waves when looking at the left term of mass conservation, the sound speed corresponds to the particle velocity! This may mean that the sound speed is a limit speed for particle velocity corresponding to the theoretical adiabatic perfect case.

\pa So the particles do not have a quasi zero velocity but should move at high velocity, closed to sound speed with decreasing temporal and spatial variations, as long as there is no obstacle on the particle trajectory. Indeed, the presence of an obstacle would introduce some new dissipation, turbulence and associated local vorticity according to the size of the obstacle. And the dissipation should be consequent and quasi localized in this "turbulent" area.

\pa The observation of the momentum conservation equation permits to observe that it corresponds to an unsteady, compressible and irrotational (no more turbulence and associated vortices; a quasi laminar flow) Bernoulli relation for the scalar velocity potential which means that the particle velocity obeys a flow equation, not  a wave propagation equation. This observation is coherent with the fact that the sound speed found for mass or density conservation, assuming negligible the second right term, is the particle velocity.  

\pa So, in "free field" conditions for the third area, one can eventually consider forward spherical wave model for pressure and density field but waves cannot be a convenient model for the particle velocity which obey a flow equation.

\pa If we consider that the membrane is no longer rigid, the "free jet" area is reduced to a tiny core and/or even would quasi not exist for limit flexible cases: we may directly encounter the "turbulent area" in the proximity of the membrane.

\pa Now, we must consider the case of the backward motion of the membrane and first assume, again, that the membrane is a solid plate.

\pa When going backward, the rigid membrane creates a potential flow for air particles in its closed pro\-ximity associated with a depression. Potential flow, or "drawing", is a more powerful process for air particles and would affect a wide angle range, generating local vortices to "draw" the particle in something like a powerful backward "free jet" sticking at the front of the membrane. 
  
\pa According to the temporal interval between the backward motion to the forward one, the backward motion of the membrane would work against the forward excitation of the \textit{medium}. Indeed, if the membrane goes very fast from the forward motion to the backward one, the forward motion of the "free jet" would be disrupted and would reduce to bursts of short particle jets, which may perhaps look like so-called "shock waves": particles "free jets" would be suddenly released if the membrane moves quickly backward, less violently if the membrane velocity decreases gradually to zero before going backward. So these forward "solid" and backward "potential" motions of the \textit{medium} would be associated with the high frequency content of the excitation voltage. 

\pa If we now consider that the membrane is a little flexible, this membrane flexibility would disrupt the backward motion of the air in the proximity of the membrane and thus disrupt the theoretical potential motion, consisting in a less effective dissipation process for the forward motion of the \textit{medium} in front of the loudspeaker, so for the so-called "free jet" motion. 

\pa We should also point out the fact that, if they exist, the bursts of "free jets" would be associated with very low frequency content in the space region where the "free jet" passes through. Indeed, a "free jet" is a rather coherent motion of particles over a given surface which is directly related to low frequency spatial and temporal variations inside  the "jet". On the contrary, in the "turbulent" area, there will be violent and localized energy transfer from low frequency to high frequency, as classically found in turbulent phenomena.  And, in the "residual flow" area, the magnitude of the temporal and spatial changes would be less and less important and would tend to become negligible: in this case, there would have not enough temporal and spatial flow variations to be perceived as a sound. 

\pa In fact, within this alternative description of acoustical phenomena, we consider that acoustics and sounds correspond to the possible perceptible spatial and temporal fluctuations of the local medium, or to the possible perception of acoustical  flows!

\subsection{Case of the naked loudspeaker}
\pa For the case of the closed loudspeaker, there is a total decoupling between the acoustical phenomena occurring at the front and the back of the membrane.

\pa When we consider the case of the naked loudspeaker, there is no more decoupling between front and back acoustical phenomena so they can interact.  

\pa If we consider a forward motion of the loudspeaker, in front of the membrane we have a "free jet" which will be disrupted and destroyed by vortices and turbulence while at the back of the membrane we have, in the same time, an opposite depression associated with a "potential flow". But, the potential flow at the back is a more powerful mechanism which may favor the creation of vortices in the front of the membrane.

\pa So, for small velocities of the membrane, related to very low frequency content of the excitation signal, the back "potential flow" would correspond to an amplifier for the dissipation mechanisms in front of the membrane. So, (very) low frequency content would suffer from greater dissipation in front of the membrane compared to higher frequencies. 

\pa And, as we need more pressure power to be able to perceive low frequencies, there would not be enough power to hear the frequencies under a cut-off frequency related to the diameter of the membrane: the diameter of the membrane control the coupling or decoupling between front and back of the membrane. Indeed, the time needed for the back "potential flow" to favor the vortex shedding in the front of the  membrane is directly related to the velocity  of the membrane and to its diameter which gives the minimum distance from the core of the back "potential flow" to the core of the front "free jet" which must be disrupted and destroyed. 

\pa So, for low membrane velocities and low frequency excitations, the "free jet" release can be significantly disrupted by the back depression while, the release has already occurred for higher frequencies or velocities when the back depression could disrupt the front flow. And, introducing a finite plane baffle is just a way to realize a stronger decoupling between front and back phenomena as it increases the effective minimum distance separating the back and front flows. In this case, a significant  perturbation of the front flow by the back flow would need much more time and would be only possible for lower velocities of the membrane related to lower excitation frequencies.  

\pa And the combo guitar amplifier look likes a finite baffle case where the baffle has been curved but still behaves as a decoupling mechanism between front and back of the membrane. This means that there will be much more low frequency content with a combo guitar amplifier with a loudspeaker of smaller diameter than with a naked loudspeaker of bigger diameter. 

\pa The inverse configuration between front and back of the membrane is found when the membrane goes backward: at the back we now have a "free jet" like flow and, in front of the membrane, "potential flow" potentially disrupting the back acoustical phenomena. This means that low frequency content (under the cut-off frequency) cannot be emitted with a sufficient power to be audible in the case of the naked loudspeaker.

\pa We have refound a similar description of the phenomena in front of  and at the back of a rigid oscillating plate (alternative "free jet" and "potential flow") with also the idea of the attenuation of the "free jet" by amplification by the "potential flow" in \cite{Bouasse:62} but with neither connection to the acoustical short-circuit phenomenon for naked loudspeaker nor relation to the introduction of a finite plane baffle in order to create a decoupling between back and front of the rigid oscillating plate. And, Millot found this preli\-minary description after having formulated an alternative description of acoustical phenomena within the proximity of a source. Indeed, this description is also valid for any vibrating body or for the output of an airport or of any pipe, for instance (in these former case we effectively have a free jet at the output).

\pa For instance, flow visualizations for free reeds, mouth organ, flute or bass-reflex loudspeaker airport, notably, have been realized by Fabre at the Laboratoire d'Acoustique Musicale from University Paris 6 a few years ago and exhibit free jet and turbulence dissipation after vortex shedding (see  \cite{Fabre:03} for some examples of flow visualizations).

\pa Thus, with our complementary experimental observations and Millot's theoretical works (to be reported elsewhere), we have got some serious clues which favor the research of alternative proposals.

\subsection{Alternative explanation for the proximity effect for directional microphones}
\pa As in the "free jet" jet area we have more low frequency, the bass boost found for subband 1 and 2  with ECM8000 microphone is understandable. Then, entering in the "turbulent flow" area we find fast spatial and temporal energy transfer which correspond to a fast spatial increase of the high frequency content (subband over 5) and a fast decrease of the low frequency content. And, when escaping from the "turbulent flow" area or going further inside it, the fast fluctuations and related high frequency diminish and, at the outside of the "turbulent flow" area, that is to say inside the "residual flow" area, spatial and temporal pressure fluctuations become less and less significant which corresponds to the 0~dB asymptote situation we found, notably for higher subbands (at least 8 to 10), for the subband weight. 

\pa This also may be coherent with the idea of a "re-equalization" phenomenon found within the measurements made with the guitar combo amplifier and notably the ECM8000 pressure sensor, and perhaps refound with the naked loudspeaker and the Endevco pressure sensor.

\pa By the way, this means that there is a natural proximity and distance "re-equalization" phenomenon in the real pressure field.

\pa Using a directional microphone, we find an enhancement of the proximity effect notably as shown in presented experimental results.  And we pointed out that the proximity effect was more important with the bidirectional directivity in the case of the U89i microphone (but more progressive according to the distance from the source compared to the omnidirectional case).

\pa This may be understandable for the following reasons. 

\pa We have already proposed the idea of a "free jet" area in the proximity of the source, related to a bass boost as there is rather dominant (very) low frequency content in a free jet. 

\pa We consider a bidirectional microphone constituted with a single membrane (but it would be similar for a dual diaphragm case as there would be one diaphragm behind the other for a normal incidence angle) whose output signal is proportional to the difference of pressure between front and back sides of this membrane. 

\pa When we put this bidirectional microphone inside the "free jet" area, in the axis of the loudspeaker, we have a situation like the case of a spoon put in a tap water flow. And, one can simply observe, by doing this experiment in his kitchen, that:
\begin{itemize}
\item the water is stopped by the front size of the flow;
\item the water flows around the spoon;
\item there is no water at the back side of the spoon. 
\end{itemize}

\pa For our bidirectional microphone, the situation may be similar. The front size of the membrane stops the "free jet" which may keep on flowing around the membrane without going to the back side, which may stay exposed to "rest" pressure. As, the microphone output signal is proportional to the difference of pressure between both sides of the membrane, we have a reinforcement of the frequency content present in the "free jet" that is to say notably the dominant (very) low frequency content. Thus, the bass boost effect or proximity effect could be much more important than the one already present in the pressure field (without attenuation mechanism added to the design of the bidirectional microphone). And stopping the flow could also generate some very high frequency content.

\pa If, we put now the microphone into the "turbulent flow" area we may observe the decrease of low frequency subbands associated with the increase of the medium and high frequency subbands, which is coherent with the experimental results we have found.

\pa And using a cardioid microphone, the effect would be less pronounced than the one found for a bidirectional microphone but more pronounced than the one for an omnidirectional one. This seems also coherent with our experimental results.

\pa In Table 1, we present the subband weight comparison between the music stimulus and the recor\-dings made for distance $x=100$~cm for each microphone or pressure sensor, and also the mean levels information about each recording. A positive difference for a subband weight means that this subband is more important in the spectral balance of the music stimulus while a negative difference is associated with a stronger weight of this subband for the recording with the microphone or the pressure sensor.

\begin{table*}[h!b!t!]
\begin{center}
\begin{tabular}{rrrrrrrrrrrr}
subband & 1 & 2 & 3 &4 &5 &6 &7 &8 &9 &10 & mean levels\\
\hline
One/Endevco & -6.8 & +17 & +9 & +0.5 & -8.6 & -11.6 & -8.3 & -3.9 & -9.7 & -25.4 & -18.6/-40.9\\
&&&&&&&&&&\\
One/ECM8000 & +8.2 & +3.9 & -2.3 & +3.5 & -0.1 & -3.2 & -5 & -7.1 & -4.7 & -6.8 & -18.6/-50.2\\
One/U89i omni & +6.1 & +2.1 & -1.6 & +1.4 & 0 & -4 & -4.8 & -5.1 & -2.9 & -3.9 & -18.6/-53\\
&&&&&&&&&&\\
One/U89i bidi & +5.4 & +5.1 & -1.7 & +3.2 & -1.6 & -4 & -5 & -9 & -5.8 & -6.2 & -18.6/-55\\
One/U89i cardio & +5.4 & +4 & -1.9 & +2.1 & -1.3 & -4.4 & -5.4 & -7.4 & -4.6 & -5.5 & -18.6/-54.5\\
One/AT2020 & +12.1 & +5.5 & -0.7 & +2.2 & -1.3 & -5.8 & -7.6 & -8.3 & -8.2 & -14 & -18.6/-48.5\\
One/C-2 & +7.3 & +8.9 & -1.6 & +2.1 & -1.8 & -4.9 & -7.9 & -9.8 & -5.8 & -7.3 & -18.6/-55.2\\
\end{tabular}
\caption{Comparisons of the subband weight (relative dB) and of the mean levels (dB FS) between the music stimulus noted "One" and any sensor or microphone for distance $x=100$~cm. The One/Endveco line corresponds to the case of the naked loudspeaker with a recording of music stimulus using the Endevco pressure sensor. The other lines correspond to the case of the combo guitar amplifier with a recording of music stimulus using the other microphones. The 10 subband decomposition corresponds to the one defined in subsection 4.4.}
\end{center}
\label{diff_ids}
\end{table*}

\pa The first line presents the subband weight differences for the naked loudspeaker and the Endevco pressure sensor.  We find that, at a distance from the source equal to 1~m, the pressure field exhibits a lower weight for subbands 2 to 4, but an increase of the subband 1 weight related to the very low frequency so maybe to the presence of a residual flow. The negative difference for subbands 5 to 10 cannot be totally attributed to the presence of noise in the recording, as we refind some similarities within the spectral balances for the other recordings at 1~m with the guitar combo amplifier and the other microphones, notably the ECM8000 omnidirectional one. 

\pa Considering the other lines, we can underline that the sign of each subband weight difference is the same when we change the microphone: subbands 1 and 2 have smaller weight compared to the music stimulus, subband 3 has a bigger weight, subband 4 has systematically a smaller weight and subbands 5 to 10 have bigger weight.

\pa Comparison of ECM8000 microphone, giving an idea of the "real" pressure field, and U89i one with omnidirectional directivity shows that the spectral balance modification of the diffused music stimulus is corrected by the U89i microphone with omnidirectional directivity. The only subband where the difference is bigger for the U89i microphone is the subband 6, and there is only a reinforcement of this subband in spectral balance of 4~dB compared to 3.2 for the ECM8000.

\pa Using the bidirectional directivity, for the U89i microphone, lowers the low frequencies (subbands 1 and 2), the subband 4 and gives more medium and high frequencies (subbands 5 and 7 to 10). The use of the cardioid directivity for the U89i microphone gives an intermediary result except for the subband 3, associated with most of the fundamentals, which exhibits a small increase of weight.

\pa Comparison of the three cardioids microphones  shows that the U89i microphone has a better respect of the low frequency content (subbands 1 and 2), gives more weight to the fundamental frequency range (subband 3) and adds less medium, high and very high frequencies than the two other microphones (subbands 6 to 10). 

\pa One can note that the two other microphones enhance different area of the spectral balance: subband 6, 9 and 10 for AT2020 microphone with quite less very low frequency content (subband 1); subband 2 with much lower weight, subbands 7 and 8 with much more weight fro C-2 microphone.  

\pa And, it seems quite difficult to find a correlation with the influence of the size of the diaphragm as the U89i and C-2 microphones have maybe rather closed spectral balances compared to the AT2020 microphone while the C-2 microphone has a small size and the two other microphones have large diaphragms. 

\pa It is also interesting to compare the ECM8000 and C-2 microphones, which are proposed by the same company, which permits to underline the fact that it seems difficult to find some common trends for these two microphones even if some subbands present "rather closed" weight: subband 1, 3,  4, 9 and 10 if one wants to find some proximity between both microphones...   

\pa But, it seems important to recall the fact that, for all microphones (so not for the Endevco pressure sensor) and even if their magnitude differ from one microphone to the other, all the corrections are in the same sense compared to the music stimulus and, notably, to note that the subbands where the weight variations are the more important are subbands 1, 2 and  8 to 10, that is to say the extremes: low frequency and high or very high frequencies. By comparison, the differences are quite smaller within subbands 3 to 7.

\section{Conclusion and perspectives}
\pa Within this paper, we have shown that it is possible and more informative to use musical stimuli rather than pink noise, or sinusoids, for objective measurements even if we consider global information as global spectral balance and mean level evolutions according to the distance $x$ from the source.

\pa We pointed out the fact that the pressure acoustical field presents a proximity effect and a distance effect which may correspond to a "re-equalization" effect. And we also observed that, when considering both mean level and spectral balance when the distance from the source varies, it is often hard or even impossible to find a distance over which one may consider spherical waves as a fair approximation for the "real" pressure field.

\pa Studying the response of different microphones we observed that  it is possible to design a directional microphone in order to follow the theoretical $1/x $ curve for the whole tested distance range. We also observed that for cardioid microphones, we found subband weight evolutions which correspond to the description of proximity effect found in the literature. And, if the proximity effect can have greater magnitude for bidirectional microphone compared to the others, it was interesting to find out that the subband weight evolutions are faster for the tested omnidirectional microphones which may mean that the evolution of the spectral balance would be more sensible when varying a little the distance from the source, in its closed proximity, with an omnidirectional microphone. 

\pa Thus, as the tested omnidirectional microphone is given as a pressure sensor, designed for acoustical measurements, it would be interesting to verify if the omnidirectional and directional microphones designs for usual recordings (speech, song, music) are rather designed in order to minimize the real spatial and frequency evolutions of the initial real pressure field. And, if as found for the tested microphones, the modifications have the same sense as the ones we have found. 

\pa It would also be quite interesting to study if the most important corrections introduced by a microphone always concern the extreme subbands: 1 and 2 for the low frequency range, 8 to 10 for the high frequency range.

\pa With the experimental results for the study of the short-circuit phenomenon for a naked loudspeaker, we were able to find out that the classical explanation is not valid as the acoustical phenomena are similar under and over the cut-off frequency and that the short-circuit phenomena is due to a stronger attenuation mechanism which does not permit to emit low frequencies with an audible level at large distance from the source.

\pa Using all the collected information, we proposed another explanation for the short-circuit phenomenon for a naked loudspeaker, the so-called proximity effect for the directional microphones and the evolution of the pressure field according to the distance from the source. 

\pa It seems quite important to precise that this alternative model does not rely on waves, as we think that waves would emerge naturally where they make sense if they make sense. Indeed, we explained that there may be three different areas: the "free jet" area, the "turbulent flow" area and finally the "resi\-dual flow" area where one may try to find some validity for forward spherical wave model.

\pa With this proposal, Acoustics is then defined as the possible perception of the spatial and temporal variations of so-called acoustical flows.  

\pa Another measurements campaigns are needed to settle down the results we have collected and these campaigns would be done using a pressure sensor with a bigger sensitivity than the Endevco one, several different musical stimuli and a bigger anechoic chamber.

\pa Using a spatial grid of pressure sensors recorded simultaneously could be a solution to study what happens over the whole surface of a source and, above all, outside the "effective" section of the source: the possible outside of the "free jet" core.    

\pa We performed our experiments for quite high le\-vels at the source but it would be quite interesting to perform the same experiments for several diffusion levels to verify if the size of the three areas, introduced in our proposal,  depends on the diffusion level.

\pa We have used a guitar combo amplifier as source but it would be rather interesting to design and build a closed loudspeaker with a diameter giving a higher cut-off frequency to see if we refind the 
low frequencies and distance "re-equalization" phenomena using a medium loudspeaker.

\pa To test the validity of our assumptions, it also would be necessary to have access to a computers grid to perform numerical simulations with 
sufficient spatial and temporal precisions.

\pa And, at this time, we must precise that our proposal does not permit to derive a simple extension of the broadband microphone model we described in a previous paper. And we do believe that, before trying to describe the inside of a microphone, we need some more information about what happens when we put a given microphone (either small or not) in the pressure field. Indeed,  we found out in a former paper \cite{Millot:06} that the available microphones models are derived assuming that there is no physical presence of the microphone! In fact they rely on the linear combination of the filtering of pressure fields for at least three points.

\section{Acknowledgments}
\pa The authors wish to thank Pr. Yvon Chevalier and his team, from the Laboratoire d'Ingénierie des Systèmes Mécaniques et des Matériaux (LISMMA), ISMEP-Supméca (Saint-Ouen, France), for having permitted us to perform measurements in their anechoic chamber.



\end{document}